\PassOptionsToPackage{table,xcdraw}{xcolor}
\documentclass[sigconf]{acmart}

\usepackage{lipsum} 

\usepackage{acmart-taps}

\usepackage{enumitem}

\usepackage{tabularx}

\usepackage{multirow}

\usepackage{framed}

\usepackage{soul}

\usepackage{colortbl}

\newcommand{\rev}[1] {{#1}}

\AtBeginDocument{%
  }

\copyrightyear{2025}
\acmYear{2025}
\setcopyright{rightsretained} 
\acmConference[DIS '25]{Designing Interactive Systems Conference}{July 5--9, 2025}{Funchal, Portugal}
\acmBooktitle{Designing Interactive Systems Conference (DIS '25), July 5--9, 2025, Funchal, Portugal}
\acmDOI{10.1145/3715336.3735785}
\acmISBN{979-8-4007-1485-6/2025/07}

\definecolor{metaiscolor}{RGB}{251, 230, 162}
\definecolor{socrataiscolor}{RGB}{255, 137, 0}
\definecolor{hephaistuscolor}{RGB}{144, 104, 255}
\definecolor{expertcolor}{RGB}{0, 119, 255}

\setul{0.5ex}{0.3ex}

\aptLtoXcmd{
\long\def\highsocratais#1{\xbox{aptbox}{\XMLaddatt{style}{border-bottom:2px solid \#FF8900;}{#1}}}
\long\def\highhephaistus#1{\xbox{aptbox}{\XMLaddatt{style}{border-bottom:2px solid \#9068FF;;}{#1}}}
\long\def\highexpert#1{\xbox{aptbox}{\XMLaddatt{style}{border-bottom:2px solid \#0077FF;}{#1}}}
}{

\newcommand{\highsocratais}[2][socrataiscolor]{\setulcolor{#1}\ul{#2}}
\newcommand{\highhephaistus}[2][hephaistuscolor]{\setulcolor{#1}\ul{#2}}
\newcommand{\highexpert}[2][expertcolor]{\setulcolor{#1}\ul{#2}}}

\begin{document}

\title{Exploring the Potential of Metacognitive Support Agents for Human-AI Co-Creation}

\author{Frederic Gmeiner}
\email{gmeiner@cmu.edu}
\affiliation{
  \institution{Carnegie Mellon University}
  \city{Pittsburgh}
  \state{PA}
  \country{USA}
}

\author{Kaitao Luo}
\email{kaitaol@andrew.cmu.edu}
\affiliation{
  \institution{Carnegie Mellon University}
  \city{Pittsburgh}
  \state{PA}
  \country{USA}
}

\author{Ye Wang}
\email{ye.wang@autodesk.com}
\affiliation{
  \institution{Autodesk Research}
  \city{San Franscisco}
  \state{CA}
  \country{USA}
}

\author{Kenneth Holstein}
\email{kjholste@cs.cmu.edu}
\affiliation{
  \institution{Carnegie Mellon University}
  \city{Pittsburgh}
  \state{PA}
  \country{USA}
}

\author{Nikolas Martelaro}
\email{nikmart@cmu.edu}
\affiliation{
  \institution{Carnegie Mellon University}
  \city{Pittsburgh}
  \state{PA}
  \country{USA}
}

\renewcommand{\shortauthors}{Gmeiner et al.}

\begin{abstract}
Despite the potential of generative AI (GenAI) design tools to enhance design processes, professionals often struggle to integrate AI into their workflows. 
Fundamental cognitive challenges include the need to specify all design criteria as distinct parameters upfront (intent formulation) and designers’ reduced cognitive involvement in the design process due to cognitive offloading, which can lead to insufficient problem exploration, underspecification, and limited ability to evaluate outcomes.
Motivated by these challenges, we envision novel \textit{metacognitive support agents} that assist designers in working more reflectively with GenAI. 
To explore this vision, we conducted \rev{exploratory prototyping through a} Wizard of Oz elicitation study with 20 mechanical designers probing multiple metacognitive support strategies. 
We found that agent-supported users created more feasible designs than non-supported users, with differing impacts between support strategies. 
Based on these findings, we discuss opportunities and tradeoffs of metacognitive support agents and considerations for future AI-based design tools. 
\end{abstract}



\maketitle

\section{Introduction}

Generative AI (GenAI) models offer increasing capabilities in supporting design workflows by generating images \cite{saharia_photorealistic_2022},
videos \cite{ho_imagen_2022}, 
or complex mechanical parts \cite{formlabs_generative_2020, yang_simulearn_2020}. 
In mechanical design, working with AI allows designers to co-create designs that would be extremely tedious or even infeasible without AI support, such as reducing the weight of an electric wheelchair component \cite{formlabs_generative_2020} or generating parts using emerging manufacturing processes \cite{yang_simulearn_2020} with 3D geometric GenAI solvers. 
However, despite the growing promise of AI design tools to augment design processes, professionals often struggle to effectively integrate AI into their workflows \cite{gmeiner_exploring_2023a, zhang_cautionary_2021}.
GenAI demands new (computational) workflows that require designers to work differently than they are used to or trained in \cite{woodbury_elements_2010a, menges_computational_2011, gmeiner_exploring_2023a, tankelevitch_metacognitive_2023}.
Current GenAI-supported workflows pose a set of unique cognitive challenges, including:

\begin{enumerate}
\item [(1)]\textbf{Intent Formulation}: Designers have to specify all design criteria necessary for generating feasible parts as distinct parameters upfront instead of iteratively modeling, testing, and visualizing a part’s 3D geometry \cite{gmeiner_exploring_2023a, subramonyam_bridging_2024}. 
This is a particular challenge for GenAI systems with lengthy and expensive inference times, such as complex 3D geometric solvers (e.g., \cite{autodesk_fusion_2020}).

\item [(2)]\textbf{Problem Exploration}: 
To tackle design problems sufficiently, designers must thoroughly think through and consider many aspects, but GenAI workflow automation can reduce cognitive engagement and foster overreliance due to "cognitive offloading," making problem exploration more challenging \cite{zamfirescu-pereira_why_2023, zhang_cautionary_2021, lee_trust_2004}.

\item  [(3)] \textbf{Outcome Evaluation}: Designers are also required to evaluate generated designs according to the problem, but when their problem understanding is limited due to cognitive offloading, they won't be able to effectively evaluate and refine generated
designs \cite{zhang_cautionary_2021, tankelevitch_metacognitive_2023}.
\end{enumerate}

Motivated by such cognitive challenges, we explore interaction patterns and interfaces to support professionals in more effectively working with AI-driven design tools. 
In this work, we follow an exploratory \rev{prototyping approach \cite{zamfirescu-pereira_fake_2021}} to explore the potential of voice-based agents that support designers’ \textit{metacognition} \cite{flavell_metacognition_1979}\footnote{\textit{Metacognition} refers to mental processes of thinking about one’s own thinking, enabling individuals to regulate and improve their cognitive strategies by reflecting on their decisions and problem-solving approaches.} 
while working on a manufacturing design task in a 3D GenAI CAD tool, where the designer specifies their goals as parameters and geometry within a graphical CAD interface. 
Broadly, we ask: \textit{What interfaces and interaction patterns can support designers in better thinking through and formulating design problems, and evaluating generated outcomes, when working with GenAI-based design tools?}

Informed by theories and findings from human-AI interaction, learning sciences, and the study of design processes, \rev{we engaged in exploratory prototyping \cite{zamfirescu-pereira_fake_2021}} to explore a design space for metacognitive support agents through three different design probes \cite{boehner_how_2007b} and observe how each influences designers' processes and outcomes in a GenAI-based design task. 
\rev{In this prototyping process,} we used the “Wizard of Oz” (WoZ) technique \cite{dahlback_wizard_1993, zamfirescu-pereira_fake_2021}, in which a human operator controlled the agent probes \rev{in a flexible manner but within certain probe-dependent constraints.}
Each probe followed a different support approach: 
(1) \textit{SocratAIs} asks reflective questions to prompt deeper reflection-in-action and (2) \textit{HephAIstus} prompts task planning and diagramming supported by suggestions for design strategies and software operation.
\rev{While the first author enacted these two agents, we also included (3) external experts in mechanical and generative design from Autodesk to act as wizards in some sessions, who we invited to provide \textit{their own} interpretations of metacognitive support strategies in a freeform manner.}

Since CAD-based work is highly visual-spatial, we explore all support agents through voice modality to reduce cognitive load. 
Inspired by the concept of \textit{"think-aloud computing"} \cite{krosnick_thinkaloud_2021}, we prompt designers to verbalize their thoughts while working on the design task to foster deeper reflection-in-action and to elicit their knowledge and situational intentions for the support agents.

We conducted a formative study with 20 trained mechanical engineers new to working with generative AI systems.
The designers were supported by one agent probe (or received no support in a control condition) while working on a realistic mechanical design task in the "Generative Design" extension of the commercial CAD software Autodesk Fusion 360 \cite{autodesk_fusion_2020}.

By comparing the design processes, outcome quality, and participant post-task interviews through video interaction and thematic analysis, we investigate the following research questions:

\begin{itemize}[font=\bfseries, align=left]
    \item[RQ1] \textit{How do different agent support strategies impact the design process?}
    \item[RQ2] \textit{What are the perceived benefits and challenges of metacognitive support agents?}
 \end{itemize}

Overall, we found that agent-supported users created more feasible designs than unsupported users. 
Most users actively engaged with and appreciated the agent's support in helping them think through the design task and operate the software.
We also identified that different agent strategies had different impacts.
For example, question-asking strategies that prompted mental simulations or visualizations through sketching helped designers with intent formulation and problem exploration regarding the part's mechanical loads.
However, we also observed that asking questions alone was less impactful when users had solidified incorrect assumptions\rev{, and that sometimes agent support could lead to additional overreliance}.
Finally, our findings provide insight into design trade-offs and differences in user preferences for metacognitive support agents.

We conclude by discussing design implications for future metacognitive support agents for GenAI-based design tools.
While our paper explores support for mechanical design tasks, we discuss how our findings may generalize to other GenAI design activities.
In sum, this paper makes three main contributions:
\begin{enumerate}
    \item Opening a design space for metacognitive think-aloud support agent interfaces for computational design tasks;

    \item Sharing empirical insights into how designers interact with metacognitive think-aloud support interfaces in the context of GenAI-based manufacturing design workflows;

    \item Proposing design considerations for future metacognitive support interfaces for GenAI-assisted design tasks.
\end{enumerate}

\section{Related Work}

\subsection{Challenges of AI-Assisted Design Workflows} \label{related-work:AI-challenges}

Many GenAI design tools operate as black boxes---designers specify objectives and then examine one or more generated designs. 
This poses key barriers to iterative trial-and-error design workflows, especially for GenAI systems with lengthy and expensive inference times, such as geometric solvers  (e.g., \cite{autodesk_fusion_2020}). 
Therefore, research has explored the design of systems that facilitate faster, more interactive design exploration paired with computational design techniques    \cite{kazi_dreamsketch_2017, chen_forte_2018, davis_drawing_2015,zaman_gemni_2015,matejka_dream_2018,kim_interaxis_2016}.
However, recent research has also identified several unique cognitive challenges professionals face when using AI-based design tools: 

\textbf{Intent formulation:} GenAI tools demand designers to specify all design criteria required for generating feasible parts upfront, shifting focus to careful upfront planning of design requirements and formulating design intents in distinct parameters instead of iteratively modeling, testing, and visualizing a part’s 3D geometry \cite{gmeiner_exploring_2023a, subramonyam_bridging_2024}.
This design process demands a shift in designers' attitudes, skills, and mental processes compared with traditional (CAD modeling) practices \cite{woodbury_elements_2010a, menges_computational_2011}.

\textbf{Problem exploration:} Design typically requires designers to think carefully through many different facets (explicit and implicit) to tackle design problems sufficiently. 
However, GenAI workflow automation can foster reduced cognitive involvement in the design process and overreliance due to "cognitive offloading" \cite{risko_cognitive_2016, tankelevitch_metacognitive_2023}, making it more challenging to explore and define design problems adequately \cite{zamfirescu-pereira_why_2023, zhang_cautionary_2021, lee_trust_2004}.
For example, empirical research found that designers in geometric "traditional" modeling environments engaged more in semantic-level actions, leading to unexpected discoveries and diverse design solutions. In contrast, those in parametric environments followed a top-down process with fewer exploration and goal changes \cite{cetintunger_comparison_2020}.

\textbf{Outcome evaluation:} Designers need to assess generated designs in relation to the design problem at hand. 
However, if their understanding of the problem is limited due to AI-imposed cognitive offloading, their capacity to effectively evaluate and refine the generated designs will also be constrained \cite{zhang_cautionary_2021, tankelevitch_metacognitive_2023}.

Motivated by such challenges, recent research highlights the need to rethink parametric design tools and develop systems that better support designers in parametric modeling and computational thinking \cite{veuskens_identifying_2022}.
Similarly, other recent work has emphasized better support for the metacognitive challenges imposed by GenAI-based workflows \cite{tankelevitch_metacognitive_2023}. 
\textit{Metacognition} \cite{flavell_metacognition_1979} involves reflecting on and regulating one’s own thinking to improve decision-making and problem-solving strategies.
For GenAI workflows, Tankelevitch et al. \cite{tankelevitch_metacognitive_2023} highlight three critical phases that demand more explicit metacognitive support: 
(1) "prompting" GenAI (formulating inputs), (2) evaluating GenAI outputs, and (3) deciding on if and how to incorporate GenAI into one's workflow best.  

Motivated and building atop prior work identifying (meta)cognitive challenges of GenAI-assisted design workflows, we explore novel support interfaces to help users work better with GenAI-assisted workflows. 
In the next sections, we will review metacognitive support strategies from learning science and design, and then highlight the role of asking questions in design and problem-solving as a distinct metacognitive support strategy.

\begin{figure*}[t]
  \centering
  \includegraphics[width=\linewidth]{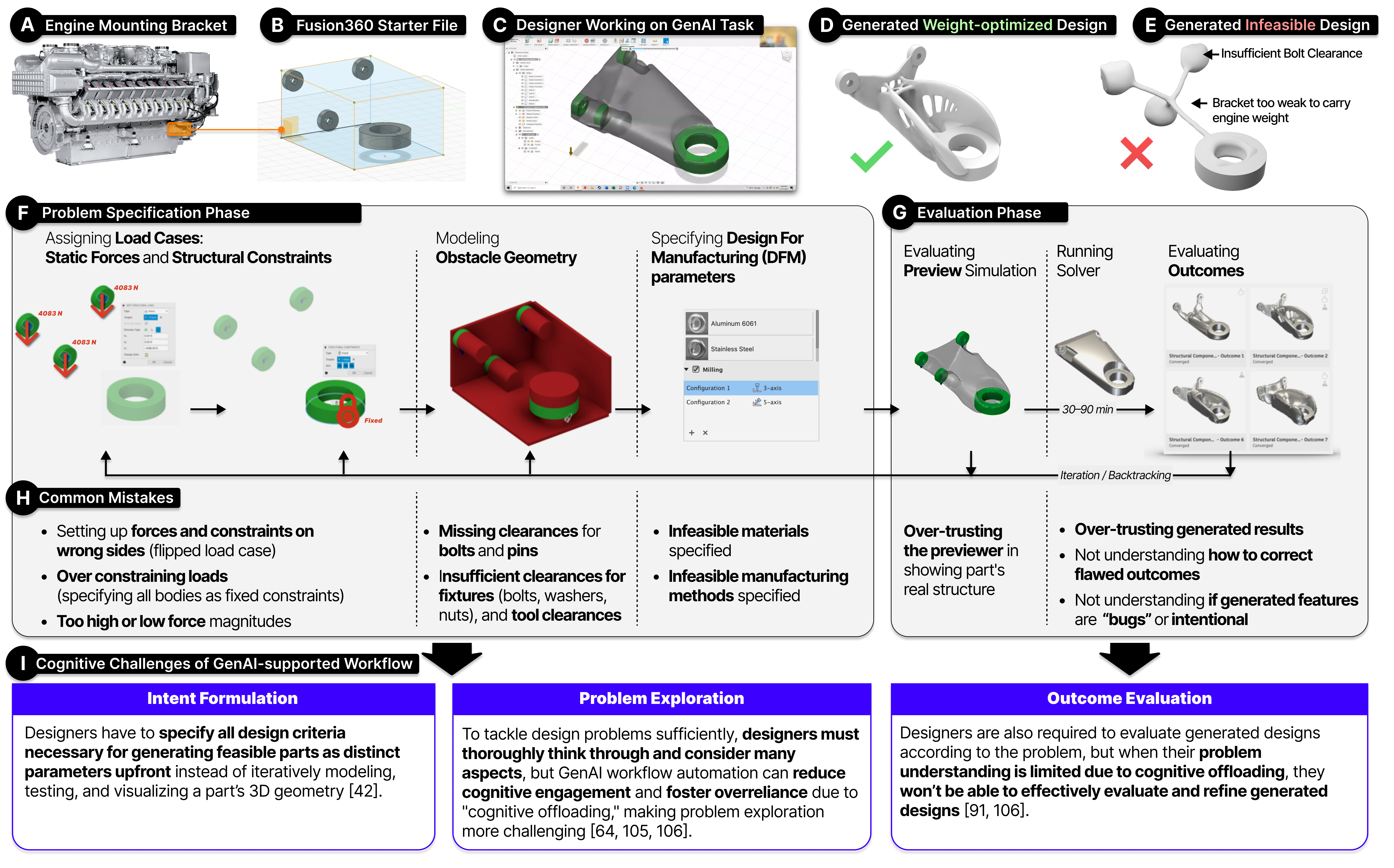}
  \caption{ Overview of the Fusion 360 design task (A-E), workflow (F-G), common user mistakes (H), and cognitive challenges (I). The task involves (A) designing an engine bracket that connects the engine to a damper. (B) A starter file containing connection holes and bounding dimensions is provided to the users to initiate the design in (C) Fusion 360. The user is prompted to create (D) a viable design while minimizing weight and avoiding (E) infeasible features. (F–G) The workflow involves six steps, and based on the AI system's solutions, the user may iterate the design by adjusting the design constraints and criteria to produce new solutions. Task taken from \cite{gmeiner_exploring_2023a}. (Image A: © Rolls-Royce Power Systems)}
  
  \Description{The figure illustrates the Fusion 360 design task and workflow for creating an engine mounting bracket. It is divided into multiple panels. Panel A shows an image of an engine mounting bracket connecting an engine to a damper. Panel B presents a 3D starter file containing predefined connection holes and bounding dimensions for initiating the design task. Panel C provides a screenshot of a Generative AI-based design workspace in Fusion 360 where users specify design parameters. Panel D highlights a weight-optimized design as a successful outcome, while Panel E shows an infeasible design with annotations indicating issues such as insufficient bolt clearance and inadequate strength to support engine weight. Panel F outlines the “Problem Specification Phase,” detailing steps such as assigning static forces and constraints, modeling obstacle geometry, and specifying design-for-manufacturing parameters, with visual examples of these actions. Panel G describes the “Evaluation Phase,” which includes steps for previewing simulations, running the solver, and evaluating design outcomes, with an emphasis on iteration to refine constraints and rerun the solver. Panel H summarizes common user mistakes, such as over-constraining load cases, incorrectly setting forces, and failing to provide sufficient clearances for fasteners or tools. Panel I breaks down cognitive challenges in the Generative AI workflow, categorized into “Intent Formulation,” “Problem Exploration,” and “Outcome Evaluation,” which highlight the difficulties of specifying criteria upfront, exploring problems thoroughly, and effectively evaluating generated designs. The figure uses annotations and visual cues, such as red crosses and green check marks, to differentiate between correct and incorrect design practices.}

    \label{fig:fusion-task}
 
\end{figure*}

\subsection{Metacognitive (Design) Support Strategies}

In the cognitive and learning sciences, metacognitive support has been shown to play a crucial role in enhancing problem-solving abilities by enabling individuals to reflect upon and actively regulate their own cognitive processes \cite{hacker_metacognition_1998, ku_metacognitive_2010}. 
\textit{Self-regulated learning (SRL)} is closely tied to metacognition and involves studying and supporting learners’ ability to manage and direct their learning processes through metacognitive skills like planning, monitoring, and evaluating their actions, typically occurring in distinct cyclical phases \cite{vaughan_metacognition_2022, panadero_review_2017, heirweg_mine_2020, bannert_process_2014}.
Prior research has identified effective metacognitive support strategies such as "self-explanation", where prompting individuals to articulate their reasoning and underlying assumptions to themselves supports them in clarifying and organizing their own understanding \cite{vanlehn_model_1992}.   
This process, often in combination with think-aloud-style verbalizations of thoughts, promotes the integration of new information with prior knowledge, fostering critical thinking and cognitive engagement in a task \cite{hausmann_effect_2010, hacker_metacognition_1998}.  
Research has studied metacognitive support strategies and interactive systems to promote reflection and problem-solving in various contexts, including software debugging \cite{tamang_comparative_2021, deliema_debugging_2019, ko_designing_2004,kulesza_principles_2015,parreira_robot_2023}, data analysis \cite{drosos_its_2024}, learning computational skills \cite{chaudhury_exploring_2023} and exploratory learning \cite{chase_development_2015}.

Design research increasingly emphasizes the central role of metacognitive monitoring and control processes for design activities. For example, 
Ball and Christensen \cite[p.~49]{ball_advancing_2019} explicitly draw parallels between metacognitive processes and prior design theories, such as the role of \textit{"reflection in and on action"} in design practice \cite{schon_reflective_1983, dove_argument_2016a, sharmin_reflect_2011}.
    Furthermore, research has found evidence for the importance of metacognition for learning and mastering design skills \cite{pontificiauniversidadjaveriana_metacognition_2019, plumb_measuring_2018, kavousi_role_2020, kurt_improving_2017}.
Building on this understanding of metacognition in design, the following section explores how questioning strategies can foster metacognitive engagement, critical thinking, and deeper cognitive exploration while supporting designers in tackling complex design challenges.

\subsection{The Role of Asking Questions in Design and Problem-Solving}

Questioning can support thinking and foster deeper cognitive engagement during problem-solving. 
In educational contexts, deep-level reasoning questions (e.g., questions probing underlying principles or causal relationships) or inquiry-based prompts (e.g., prompts encouraging student-led questioning and investigation) have been shown to enhance learning outcomes by stimulating critical thinking and deeper exploration of complex concepts \cite{dillon_classification_1984, graesser_question_1994, tawfik_role_2020, craig_deeplevelreasoningquestion_2006, berger_more_2014}.
A specific strategy is the \textit{Socratic Method}, which employs guided open-ended questioning to stimulate critical thinking and reflection \cite{elder_thinkers_2016}. 
This approach has been used effectively across various fields, such as in healthcare education to develop critical thinking skills among students \cite{ho_thinking_2023}, programming to aid novice debuggers in identifying and resolving code issues \cite{alshaikh_experiments_2020, al-hossami_socratic_2023, wilson_socratic_1987, ko_designing_2004}, supporting academic career development \cite{park_thinking_2024}, and in creativity research to foster co-creativity between humans \cite{stenning_socratic_2016}.

Similarly, asking questions plays a central role in guiding designers' thinking in exploring and refining ideas by challenging their assumptions as they work through complex, evolving problems \cite{eris_effective_2004}. 
When tackling ill-defined problems, designers must navigate the solution and problem spaces simultaneously, often using abductive reasoning to reframe problems and synthesize new insights and possibilities \cite{kolko_abductive_2010, dorst_creativity_2001, cross_designerly_2006}. 
Research analyzing design team communication has shown the crucial role of question asking for problem-solving \cite{aurisicchio_characterising_2007}, idea generation \cite{Coimbra_inflection_2016}, design reviews \cite{cardoso_question_2014}, and design studio education \cite{hurst_comparing_2023}.
Other work has explored ways of supporting designers through targeted questioning to enhance the design process by helping them articulate product requirements \cite{wang_asking_2009}, stimulate idea generation \cite{royo_guiding_2021}, and highlight awareness of bias in designerly thinking \cite{price_asking_2022}.

Eris developed a taxonomy of questions asked during design teamwork \cite{eris_effective_2004}, building on prior taxonomies by Lehnert \cite{lehnert_process_1978} and Graesser \cite{graesser_question_1994}. 
Eris's taxonomy outlines three types of questions for design: 
1. \textit{Low-level questions} for clarification, 2. \textit{Deep reasoning questions} for causal explanations, and 3. \textit{Generative design questions} for exploring alternative solutions.
Research showed that student teams asking more Deep Reasoning and Generative Design questions achieved more innovative design outcomes \cite{eris_effective_2004}.

\subsection{Multimodal and Collaborative CAD Systems }

Current computer-aided design (CAD) tools primarily rely on WIMP (Windows, Icons, Menus, Pointer) interfaces, using pointer movements and keystrokes as input \cite{niu_multimodal_2022} to modify geometry in visual-spatial interfaces. 
However, multimodal inputs for CAD work, such as gestures or speech in combination with WIMP \cite{khan_gesture_2019}, can offer advantages, as experimentally demonstrated by Ren et al. \cite{ren_experimental_2000}. 
This combination---leveraging the split-attention \cite{ayres_splitattention_2005} and the modality effect \cite{castro_modality_2020}---allows users to process visual and auditory information simultaneously, which can reduce cognitive load and enhance performance in complex tasks.

Similarly, user research studies often utilize the concurrent think-aloud protocol \cite{kuusela_comparison_2000} to understand participants’ thoughts and actions by encouraging them to speak about their thoughts as they perform a task. 
Concurrent verbalization can offer rich insights into users' knowledge and intents while only slightly increasing cognitive load during complex tasks \cite{vandenhaak_retrospective_2003}, such as annotating existing CAD models through speech \cite{plumed_voicebased_2021}.
Inspired by the think-aloud protocol, Krosnick et al. \cite{krosnick_thinkaloud_2021} propose the interaction paradigm of \textit{think-aloud computing} where computer users are encouraged to speak while working on a CAD design task to transcribe and capture rich knowledge with relatively low effort in real-time. 

Other empirical research shows how conversation and real-time support during CAD sessions can improve problem-solving. 
For example, revealing unique communication patterns that make multi-disciplinary engineering design work more effective \cite{roy_designing_2024} or systems for supporting CAD users by connecting them with human CAD experts in real-time \cite{chilana_supporting_2018, joshi_micromentor_2020} or automatically provide context-sensitive learning resources \cite{matejka_ambient_2011}.

In this study, we aim to explore voice-based agent support interfaces and think-aloud user interactions for augmenting GenAI-driven CAD workflows by enabling low-effort continuous speech-based user intent and context-elicitation.

\begin{table*}[t]
  \centering
    \caption{Overview of agent probes' support strategies and behaviors.}
  \Description{Table 1 provides an overview of the support strategies and behaviors of three agent probes: SocratAIs, HephAIstus, and Expert. The table is structured into rows detailing their main strategies, support principles, response to user queries, modalities, and examples of interactions. SocratAIs focuses on asking questions to prompt self-explanation and reflection, responding to user queries with questions only, and communicates solely through voice. An example interaction includes the question, “What might be reasons why the GenAI system generated the shapes this way?” HephAIstus employs planning, sketching, and suggesting strategies by supporting users through planning activities and proposing design strategies or workflows. It answers factual user questions and uses multiple modalities, including voice, link sharing, and screen annotations. An example provided is, “Can you walk me through your load cases and constraints by sketching a free-body diagram? I shared a link to a board for you to sketch on in the chat.” The Expert probe is categorized as expert-determined, using strategies and responses to support users through voice and screen annotations, though specific strategies and responses are not further detailed in this table. This table highlights distinct approaches across agent probes in providing design task support.}
  \label{tab:probes}
  \includegraphics[width=\linewidth]{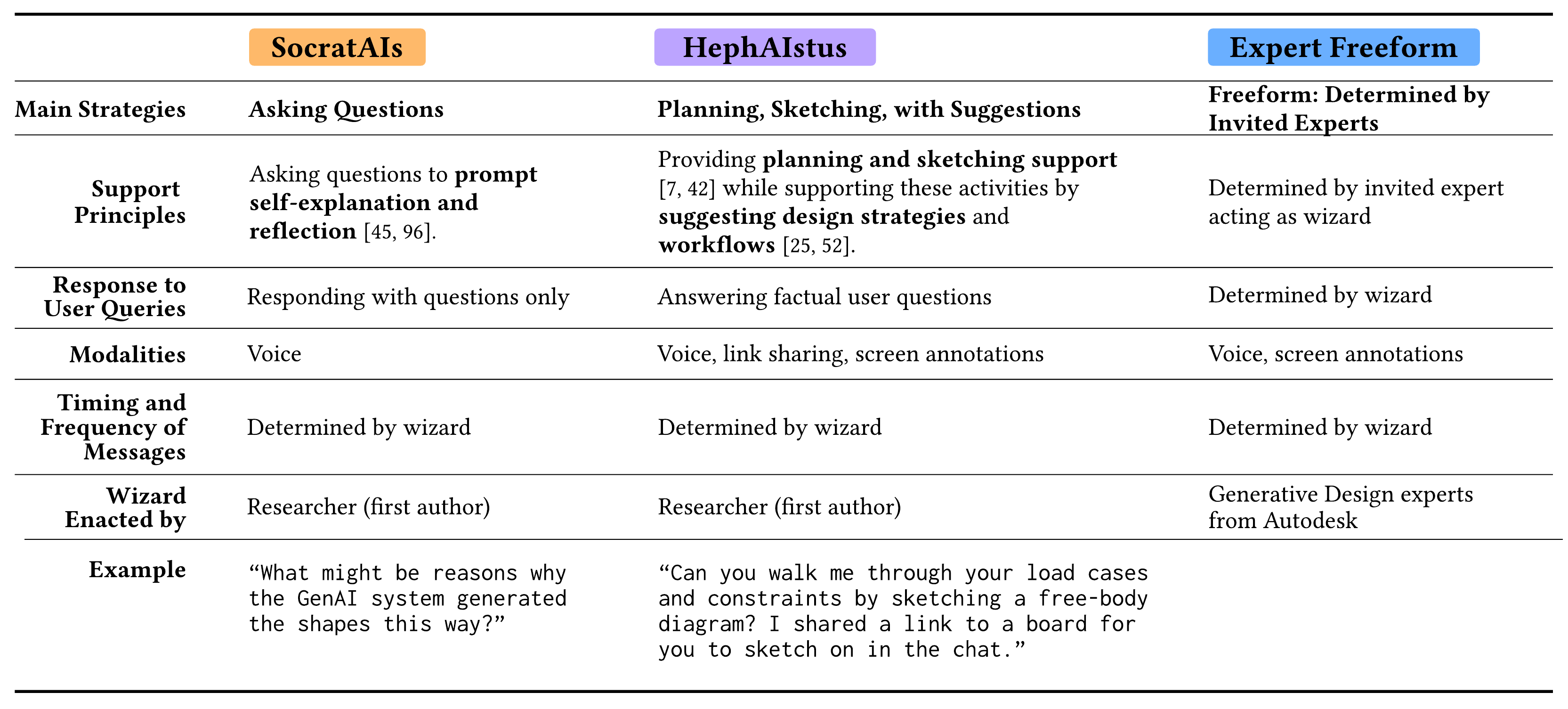}

\end{table*}

\section{Case Study: Challenges of AI-Assisted Mechanical Design Tasks in Fusion360 Generative Design} \label{case-study}

Our work aims to develop support agents that help designers overcome the cognitive challenges they face when working with AI-assisted design tools. 
\rev{We replicate our previous} study of mechanical designers working with AI assistance \cite{gmeiner_exploring_2023a}, which found that designers working with AI for the first time often failed to create feasible mechanical parts despite being familiar with the design tasks and CAD tools in general. 
In that study, designers worked with the "Generative Design" feature of Autodesk Fusion360 \cite{autodesk_fusion_2020}, which helps designers create lightweight and strong parts through topology optimization and genetic algorithms \cite{matejka_dream_2018}. 
In the task (Figure \ref{fig:fusion-task}A--E), the designer is asked to design a material-efficient and structurally sound engine mounting bracket by considering the optimal manufacturing and material combination from a large pool of possibilities. 
While designing mounting brackets is common for mechanical engineers, optimizing designs for different manufacturing methods and materials is difficult without simulation and AI support. 
Traditionally, engineers build a part and then gradually remove or add material based on structural analysis to derive a weight-optimized part. 
Exploring different manufacturing options is necessary for every material and manufacturing process---a time-consuming and tedious task. 
In contrast, Generative Design can automatically generate many options based on specified requirements, which the designer can explore and choose from.

Concretely, in Autodesk Generative Design, designers specify the \textit{structural loads} a part has to hold, the 
 GenAI solver's \textit{obstacle geometry} (part areas that must remain free of material, such as clearances for bolt holes), and the \textit{material and manufacturing properties} (Figure \ref{fig:fusion-task} F). 
Designers then optionally request a preview simulation before running the solver and then evaluate many AI-generated solutions to identify viable designs (Figure \ref{fig:fusion-task} G). 
If no outcomes are deemed satisfactory, designers might iterate the design by adjusting the input criteria. 

In our prior study \cite{gmeiner_exploring_2023a}, we observed that \textbf{while most designers learned to specify some of these input parameters successfully over time, many failed to correctly specify structural loads and obstacle geometry for sufficient part clearances.} 
As a result, most of the submitted designs were unfeasible because they were either too heavy or weak, larger than the allowable, or had insufficient clearances around bolt holes, preventing the bracket from being mounted.  

Figure \ref{fig:fusion-task} H lists common errors that occur during the task of designing a ship engine mounting bracket using Generative Design related to insufficiently \textbf{specified loads}, \textbf{obstacle geometry},  \textbf{materials and manufacturing options (DFM)}, and during \textbf{outcome evaluation}. These observed common mistakes relate to key cognitive GenAI workflow challenges of \textit{intent formulation}, \textit{problem exploration}, and \textit{outcome evaluation} (see Figure \ref{fig:fusion-task} I and Section \ref{related-work:AI-challenges}). In the following section, we describe how we probe different support strategies aimed at helping designers overcome these challenges.

\section{Constructing Support Agent Design Probes}

Our motivation was to \rev{prototype,} study and compare different metacognitive support strategies in the context of GenAI-supported design tasks.
Previous studies have shown that existing support resources and strategies (such as help menus, online forums, or video tutorials) seem to be ineffective in helping designers overcome the cognitive challenges involved in working with GenAI \cite{gmeiner_exploring_2023a}. 
Thus, we speculated that metacognitive support strategies, such as asking reflective questions to prompt self-explanation or planning and sketching activities, might be more effective.
Inspired by previous work and findings on metacognitive support, we therefore asked: 

\begin{enumerate}
    \item[1)] \textit{What if a support agent simply asked questions? Could this in itself be enough to promote productive reflection-in-action and improve human-AI co-creation?}
    \item[2)] \textit{What if an agent prompted designers to plan and sketch while supporting these activities through suggestions?}
\end{enumerate}

To explore these metacognitive support strategies prior to developing functional AI-based agent systems, we \rev{engaged in exploratory prototyping \cite{zamfirescu-pereira_fake_2021} and} constructed two different agents as design probes \cite{boehner_how_2007b}: 
\rev{(1) \textit{SocratAIs}, a Socratic agent that asks reflective questions to prompt deeper reflection-in-action, and (2) \textit{HephAIstus} that prompts task planning and diagramming supported by suggestions for design strategies and software operation (see Table \ref{tab:probes}).
These two agent probes were enacted by the first author using an exploratory \textit{"Wizard of Oz"} \cite{dahlback_wizard_1993, zamfirescu-pereira_fake_2021} approach.
\rev{The wizard followed guidelines to adhere to the general rules for each agent while also having flexibility over when to send messages and the exact phrasing of messages given the in-the-moment context of each participant's session (see \ref{study:woz-setup} for details).}}

We hypothesized that each of these support strategies could be effective in supporting designers on their own, but also likely in combination.    
However, as a start, we wanted to investigate how certain strategies would work in isolation to better understand their impact, benefits, and tradeoffs on designers' metacognition and design process. 

In addition to these two probes, to move beyond our research team's assumptions, we also asked:
\begin{enumerate}
    \item[3)] \textit{How would human experts in generative design support designers new to working with GenAI in this task?}
\end{enumerate}
\rev{To answer this question, we also invited (3) external experts from Autodesk to act as wizards during some sessions, to observe and compare their natural strategies of supporting other designers in this task (freeform).} 
The following sections describe our three agent probes \highsocratais{\textit{SocratAIs}}, \highhephaistus{\textit{HephAIstus}}, and \highexpert{\textit{Expert-Freeform}} in more detail. 


\subsection{\textbf{\highsocratais{\textit{SocratAIs}} Probe – Asking Questions}}

Inspired by previous research on prompting self-explanation \cite{vanlehn_model_1992,hacker_metacognition_1998}, \textit{\highsocratais{SocratAIs}} proactively asks users questions as they complete a design task to \textbf{prompt self-explanation and reflection on their design decisions}.
Specifically, this agent follows a \textbf{Socratic questioning approach} to support designers' metacognition by constructing questions relevant to the phase of a design task that a designer is currently working on, %
such as specifying the part's loads, obstacle geometry, manufacturing considerations, or evaluating outcomes (see Appendix for agent guide). 
In line with a Socratic questioning approach, \highsocratais{\textit{SocratAIs}} \textbf{only responds to user requests with further questions} and refuses to provide direct answers.

\subsection{\textbf{\highhephaistus{\textit{HephAIstus}} Probe – Planning, Sketching, with Suggestions}}

To explore our second question (\textit{what if an agent prompted designers to plan and sketch while supporting these activities through suggestions?}), we constructed \highhephaistus{\textit{HephAIstus}} \textit{(referencing Hephaestus, the Greek god of craftsmanship)}. 
This agent provides metacognitive support in the form of \textbf{planning and sketching} and by supporting these activities with suggestions around design strategies and tool operation.
Inspired by prior research on the benefits of externalization activities in design \cite{babapour_roles_2015,gmeiner_exploring_2023a}, the agent offers deliberate planning and sketching activities parallel to the CAD workspace to help users think through the design problem more strategically and visually. 
For planning, the agent suggests the user engage in a \textbf{project planning activity} by sharing a pre-generated text document outlining critical project-relevant aspects with the user \rev{(see Appendix \ref{appendix-hephaistus} for an example document)}. 
This strategy aims to encourage users to think through the design task more deeply before switching to the CAD interface. 

For the sketching activity, the agent suggests that the designer \textbf{sketch out load case-relevant forces and constraints as a \textit{free-body diagram}\footnote{Free body diagrams are common mechanical visual representations to illustrate the forces acting on physical objects in a given situation, helping to simplify complex mechanical problems and reason about its structure.}} by sharing a link to a 2D drawing canvas containing the side and top view of the bracket as a starting point \rev{(see Appendix Figure \ref{fig:appendix_sketching_board} for an example)}.

To support these planning and sketching activities, the agent proactively offers suggestions for the design task and software operation, inspired by work on supporting software learning and work processes \cite{chilana_supporting_2018, joshi_micromentor_2020}. 
This entails providing alternative design options, highlighting overlooked software features, notifying about unintentional execution errors, or recommending tools and techniques to improve the overall design process.

In contrast to \highsocratais{\textit{SocratAIs'}} question-asking approach, \highhephaistus{\textit{HephAIstus}} responds to user queries with direct answers, similar to chat assistants such as ChatGPT. 
Lastly, the agent can visually highlight areas on the user's screen to direct their attention to what the agent is talking about.

\subsection{\textbf{\highexpert{\textit{Expert-Freeform}} – Support from an External Generative Design Expert}} \label{ExpertSupportAgent}

Lastly, we explored how human experts attempt to support designers’ metacognition in this task, \rev{when provided agency over how to do so}. We invited experts in mechanical and generative design from Autodesk (the maker of Fusion360)\rev{---who were not a part of our research team---}to serve as wizards, allowing them to provide their own interpretation of metacognitive support (freeform).
For these sessions, we recruited wizards from Autodesk's employee pool via internal mailing lists and snowball sampling (see Appendix Table \ref{tab:experts}).
Participants ranged between 27 and 47 years of age, with mechanical design experience between 3 and 15+ years. 
All experts had high self-rated proficiency in Fusion360 Generative Design, were closely involved with its development, and had substantial experience training others to use the tool.  

The expert wizards were instructed to support the other designer in working with Generative Design and the design task by controlling the voice agent. 
\rev{We refrained from explicitly telling them to follow a specific support strategy, and instead,} they were asked to provide real-time support in their preferred way, so long as they did not directly instruct the other designer on what to do.

\subsection{\textbf{Common Agent Capabilities}}

Besides the differing support strategies outlined above, all support agent probes shared the common capabilities:
\begin{itemize}
    \item the agent possesses (non-exhaustive) knowledge of additive manufacturing and generative design tasks 
    \item the agent has access to the users' screen and think-aloud speech in real-time;
    \item the agent can identify inconsistencies between the requirements stated in the design brief and the GenAI parameters specified by the designer by comparing the design brief and screen activities  (e.g., detecting over/under-constrained load cases, infeasible material combinations, or wrong force setup)\footnote{Such
requirements could be explicit nature (e.g., the force the bracket needs to hold) or implicit features, such as bolt clearances, which were
not explicitly mentioned in the design brief};
    \item the agent can send voice messages to the user and (in \highhephaistus{\textit{Heph\-AI\-stus}} and \highexpert{\textit{Expert-Freeform}} cases) annotate the user's screen and share links via chat.
\end{itemize}

\section{Study Design}

To elicit the impacts, potential benefits, and drawbacks of the support agent probes, we conducted a formative \rev{between-subjects} study with trained mechanical designers new to working with generative AI. 
Each designer was supported by a different agent probe (facilitated by a human operator in the background) while working with the Autodesk Generative Design tool to design a ship engine mount~\footnote{See Section \ref{case-study} for a description of this task, previously used in  \cite{gmeiner_exploring_2023a}.}.

We used an \rev{exploratory} “Wizard of Oz” (WoZ) prototyping approach \cite{zamfirescu-pereira_fake_2021}, where a human operator controlled the voice agent probes in the background to simulate different support strategies. 
\rev{Instead of only following strict predefined rules, the wizards had certain degrees of freedom in enacting the agent probes' support strategies to explore broader design possibilities and implications in response to emerging situations during user sessions \cite{zamfirescu-pereira_fake_2021} (see \ref{study:woz-setup}).}

While working on the task, designers were asked to think aloud to elicit their cognitive processes (e.g., mental models \cite{chi_laboratory_2006}, learning \cite{young_direct_2009}) and knowledge and intents so that they could be used by the (WoZ-controlled) voice-based support agents. 
Participants worked between 31 and 99 minutes, then submitted their designs and completed a semi-structured interview to reflect on their experience working with the support agent.

We collected the following data:
\begin{itemize}
\item Video, screen, and audio recordings with machine-generated transcripts of the agent-supported think-aloud design sessions
\item Audio recordings and machine-generated transcripts of the post-task interviews 
\item 3D designs created during the think-aloud sessions
\item Log files with timestamps of all human-facilitated agent messages 
\end{itemize}

\subsection{Participants}

We recruited 20 designers (aged 20 to 42 \textit{(M = 26.1, SD = 5.9))} with mechanical engineering backgrounds from engineering departments of North American universities and through the Upwork freelance hiring platform\footnote{http://www.upwork.com} (see Appendix Table \ref{tab:participants}). 
Participants had between one and ten years of Mechanical Design experience and between zero to ten years of industry experience, as determined via a screening questionnaire. 
All participants had at least two years of experience using CAD and Autodesk Fusion360 but no experience working with the Generative Design extension.
We recruited participants familiar with Fusion360 so that they could focus on learning to work with the AI-driven Generative Design feature rather than learning the CAD tool’s user interface.
Participants included a mix of undergraduates, graduate students, and professional engineers.
Before the study, all participants signed a consent form approved by our institution’s IRB. Participants were compensated 20 USD per hour.

\subsection{Design Task and System}

Participants were instructed to design a light and strong engine mounting bracket with Autodesk Fusion360's \cite{autodesk_fusion_2020} “Generative Design” feature (see Section \ref{case-study} and Figure \ref{fig:fusion-task} for a detailed description). 
Since we adopted the task from an existing study \cite{gmeiner_exploring_2023a}, we verified the suitability of the task for our study by first piloting it with mechanical engineers from our institution and an external user proficient with Generative Design, all of whom successfully completed the task without receiving any support.

\subsection{Procedure}

The study was structured into four phases, split into two sessions:

\textbf{1) Onboarding (30 minutes):} After an introduction to the study, participants received a hands-on tutorial demonstrating Fusion360 Generative Design's core functionalities through a step-by-step example design task.

\textbf{2) Design Task - Part 1 (up to 70 minutes):} 
After onboarding, participants were introduced to the design brief, task, and starter file containing predefined geometric constraints. 
They were also told that a virtual AI agent would support them during the design task (except for the members of the No Support group).
Sessions were conducted over video conference (Zoom) with audio, screen, and video recording. 

Participants worked while sharing their screens and thinking aloud, with research team members following the video call remotely and operating the support agent. 
Participants were allowed to use any available support resources, such as internal Autodesk help files, external video tutorials, or online user forums.

Participants worked until they completed specifying the generative design inputs. 
They then started the Generative Design extension's solver, which completed the first session. 
Since the solver required 30 minutes of runtime, participants took a 30-minute break and then returned for the second session. 

\textbf{3) Design Task - Part 2 (up to 30 minutes):}
After the solver finished, participants evaluated the generated designs. 
If satisfied with the results, they could directly select three designs. Otherwise, they could re-adjust the design criteria and restart the solver, in which case they would return to evaluating the generated designs after the exit interview and select their final designs.

\textbf{4) Exit Interview and Debriefing (20 min): }
After task completion, participants participated in a semi-structured remote interview with a research team member. 
Participants were asked to reflect on their experience working with the Generative Design extension, the think-aloud activity, and the agent support (see the Appendix for interview protocol). 
Additionally, after the expert-facilitated sessions, we interviewed the experts to gain further insight into their support strategies and challenges they perceived.
The interview was audio and video recorded and the interviewer took notes. 
At the end of the inteview, participants were debriefed about how humans had actually controlled the AI support agents.

\begin{figure*}[t]
  \centering
  \includegraphics[width=\linewidth]{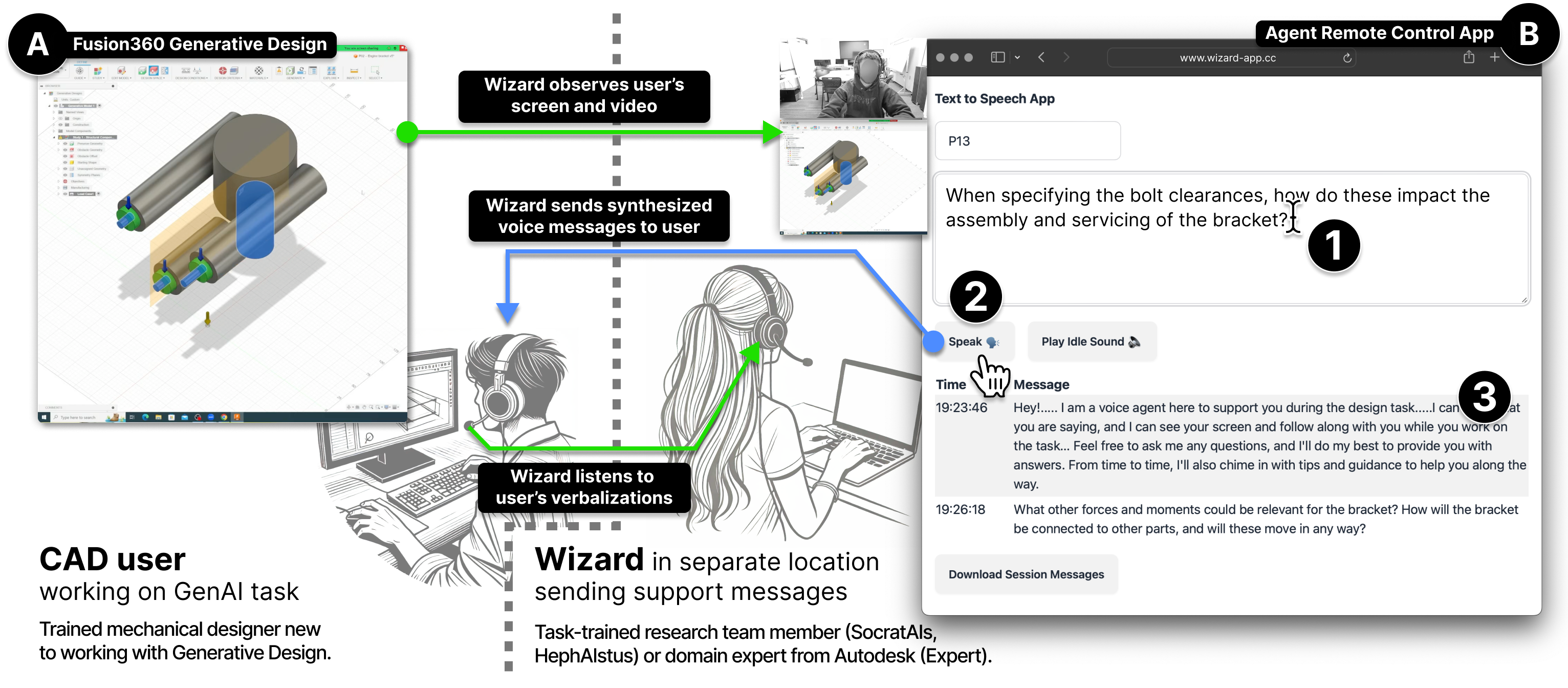}
  \caption{Process diagram of Wizard of Oz setup. The remotely located wizard (right) followed the designer’s actions (left) by listening to their verbalizations and observing their screen and webcam stream. 
  Using a web interface, the wizard could (1) type messages and send these as (2) synthesized voice messages to the user as agent messages. 
  (3) All agent messages were logged with timestamps. }
  
  \Description{The figure illustrates the Wizard of Oz setup used to facilitate the study. It depicts two main components: the CAD user working on a Generative AI-based design task and the wizard providing support remotely. Panel A shows the CAD user, a trained mechanical designer, interacting with Fusion 360 Generative Design software. The user shares their screen and webcam stream with the wizard. The wizard, located remotely, observes the user’s actions and listens to their verbalizations to provide real-time support. Panel B shows the wizard’s remote control app interface, which includes a text-to-speech tool where the wizard types messages. These messages are sent as synthesized voice messages to the user, indicated by the “Speak” button and message log with timestamps. The process emphasizes the wizard’s role in sending prompts, answering questions, and guiding the user without directly instructing them, enabling dynamic interaction in a simulated AI-agent environment.}

    \label{fig:agent-interface}
 
\end{figure*}

\subsection{Wizard of Oz Setup} \label{study:woz-setup}

\rev{Overall, we followed an exploratory Wizard of Oz prototyping approach \cite{zamfirescu-pereira_fake_2021} as a design space exploration where wizards would have some flexibility in enacting the agent probes. 
This allowed us to make meaningful comparisons between the support strategies (\textit{asking questions} vs. \textit{planning and sketching support} vs. \textit{expert freeform}) while also giving wizards flexibility on how to enact the different agent probes in detail (such as the exact message timing and phrasing).
Below, we detail the instructions given to wizards and the study setting. }

\subsubsection{\rev{Wizard Details}}

The \highsocratais{\textit{SocratAIs}}, and \highhephaistus{\textit{HephAIstus}} agents were facilitated by the first author with experience in mechanical engineering, Fusion360, and generative gesign
\footnote{\rev{In some sessions, a second research team member with experience in mechanical engineering and generative design was co-present, verbally supporting the wizard.}}.

\rev{This wizard followed these general guidelines: 
\begin{enumerate}
  \item[1)] Follow the designer’s verbalizations and screen actions and pay close attention to the task-specific design steps and challenges as outlined in Section \ref{case-study}.
    \item[2)] Pay close attention to inconsistencies between the requirements stated in the design brief and the input parameters set by the designer\footnote{Such requirements could be explicit (e.g., the force the bracket needs to hold) or implicit features, such as bolt clearances, which were not explicitly mentioned in the design brief.}.
  \item[3)] Never directly tell the participant what to do, but rather provide supportive questions, hints, or suggestions (depending on the enacted agent type). 
  \item[4)] You are free to send messages whenever and how often you consider it helpful to the designer. However, pay special attention to moments in which designers transition between design sub-tasks (such as from specifying obstacle geometry to specifying loads), as well as when designers show hesitation or use hedging expressions.
  \item[5)] You are free to formulate the messages in a way you consider to be most helpful, while adhering to the agent's support strategy (e.g., only asking questions).
\end{enumerate}
}

\rev{
For the \highexpert{\textit{Expert-Freeform}} agent wizards, we did not provide specific guidelines, but only instructed them to provide support in their preferred way, so long as they did not directly instruct the supported designer on what to do (see Section \ref{ExpertSupportAgent}).  
}

\subsubsection{\rev{Setting}}

For all sessions, participants and wizards were in separate locations during the design task, and communication between the wizards and participants was established via Zoom video conferencing software (see Figure \ref{fig:agent-interface}). 
Although most sessions were co-located, with participants and wizards in separate but neighboring rooms of our research lab, eight sessions were conducted remotely.
In the lab sessions, participants completed the task on a computer workstation running Fusion360 with the Generative Design extension. 
Remote participants were provided access to a web-based computer\footnote{using Paperspace} with the same setup for remote sessions.

Participants shared their screens via Zoom and wore an audio headset during the task to capture their verbalizations and ensure they could hear the agent’s voice.  
The wizard joined the same video call using a generic name (‘Agent’) with a deactivated webcam to follow the participant’s screen actions and verbalizations.
In addition, the wizard could generate and send agent voice messages using a self-developed web control interface (see Figure \ref{fig:agent-interface} right). 
For the \highexpert{\textit{Expert-Freeform}} sessions, in addition to the researcher and the designer, the external task expert from Autodesk anonymously attended the conference call in the background and remote-controlled the voice agent via the web interface. 

The agent control interface was developed in React.js and uses Google’s text-to-speech API to generate the agent voice from the wizard-typed text (see Figure \ref{fig:agent-interface}). 
The interface features a button to toggle the playback of an idle sound cue to sonically indicate an ‘agent is processing’ state to the participant. 
Additionally, the tool logs all generated messages with a timestamp and session ID exportable in JSON format. 
We used audio-routing software\footnote{https://existential.audio/blackhole/} on the wizard’s computer to inject the generated agent speech audio into the video call.
To mitigate possible gender bias effects, we deliberately selected a gender-ambiguous voice option for the agent, following suggested best practices from prior research \cite{tolmeijer_female_2021}.

\subsection{Measures and Analysis}

To gain insight into our research questions, 
\textbf{RQ1} \textit{How do different support strategies impact the design process?} and 
\textbf{RQ2} \textit{What are the perceived benefits and challenges of metacognitive support agents?}, we evaluated the design outcomes and analyzed \textasciitilde19 hours of think-aloud videos and \textasciitilde6 hours of interview recordings using a combination of video interaction analysis and reflexive thematic analysis.

\subsubsection{\textbf{Design outcome evaluation}}
We evaluated the design outcome feasibility by checking the submitted engine brackets against the requirements in the design brief, rating across five criteria, each yielding one point:
\begin{enumerate}
    \item The structural soundness was validated using finite element analysis (FEA).
    \item The feasible load case setup was checked in their Fusion360 project file.
    \item The optimized mass was not extremely light or heavy.
    \item The part had feasible fastener clearance (i.e., clear bolt holes)
    \item The part’s mass and volume fit within the acceptable bounding dimensions.
\end{enumerate} 
The GenAI solver generated around 20 designs, and to compensate for possible variability in the generated outcomes, we asked participants to choose three feasible parts from which we then selected the highest-scoring part as their final design.

\subsubsection{\textbf{Video interaction analysis \rev{to determine agent impact}}} \label{video-analysis}
We used \textit{video interaction analysis} \cite{baumer_comparing_2011} of the think-aloud recordings to understand how agent support impacted participants’ design process.
\rev{Specifically, to determine the impact of the agents' messages on the design process, we analyzed whether participants considered new design aspects after receiving a message based on their verbal reflections or concrete actions.}
The think-aloud video and transcript data were equally distributed among three researchers who applied the following coding procedure:

\rev{\textbf{1) Tracking GenAI input specifications:}} First, the coders tracked participants’ interactions with the Generative Design features relevant to the design task and documented whether the actions would produce satisfactory outcomes. 
Specifically, they tracked how participants specified \textit{(1) structural loads} (forces), \textit{(2) mechanical constraints}, and the obstacle geometry feature to control the bracket's \textit{(3) bolt} and \textit{(4) dampener pin clearances}, and \textit{(5) overall size}. 

\rev{\textbf{2) Coding message impact:}} 
Second, the coders tracked \rev{the impact of the agent's messages on the design process:} 
For each agent message, they coded if the message had an observable impact on the participant considering a new aspect related to the design task, which needed to be apparent from the designer’s verbalizations or actions \textit{(coded with ‘none,’ ‘weak,’ or ‘strong’)}.
For the design assistant agent probe (HephAIstus), they also tracked users’ direct messages to the agent and if agent messages were generally observably helpful to the user \textit{(yes/no)}.  

\rev{\textbf{3) Coding planning and sketching interactions:}} 
Third, the coders tracked when the agent sent the planning sheet or the free-body diagram sketching board and when the designers interacted with these. 

Between these coding sessions, the researchers met frequently to discuss edge cases and ensure consistency in their coding practices. 
From this data, we then created time-series event plots for each session with \texttt{R} and \texttt{ggplot2} to visually identify patterns (see Appendix Figure \ref{fig:appendix_timelines}).
In addition, we created \textit{summary videos} for each participant, highlighting all situations featuring agent messages or other interesting designer-agent interactions (please see the video figure in the supplementary material for an example).

\subsubsection{\textbf{Reflexive thematic analysis}}

To understand participants’ attitudes toward the agent support, we performed a \textit{reflexive thematic analysis} \cite{braun_reflecting_2019} of the interview data (transcripts). 
We followed an iterative inductive coding process and generated themes through affinity diagramming. We used ATLAS.ti to analyze transcripts, audio, video, and Miro for affinity diagramming. 

First, the first author coded the interview transcript data utilizing both a \textit{semantic} (what people said) and \textit{latent} (our interpretations of the data) coding strategy. 
Next, the research team collectively identified initial codes and themes.
Based on the time series plots from the video interaction analysis and the summary videos, we then associated the participant statements from the interviews with specific situations in the design sessions to cross-validate the impact of agent messages and identify additional qualitative themes and interaction patterns. 
We iteratively reviewed and revised codes and themes until we identified a stable network of coherent and rich themes.

\begin{table*}[t]
  \centering
    \caption{ Table summarizing participants' outcome design scores across five criteria (checkmarks) and process statistics by support group. \rev{Normalized Message Frequency represents the number of agent-initiated messages divided by the session duration. 
    Note that the number of agent-initiated messages is lower than the total agent messages for the HephAIstus and Expert-Freeform groups since these exclude agent messages in response to user-initiated queries.} }
    
  \Description{Table 2 summarizes the task outcomes and processes of participants across four support conditions: SocratAIs, HephAIstus, Expert-Facilitated, and No Support. Each column represents individual participants within the respective support group, with rows capturing their performance in five task criteria: Passing Structural Analysis (FEA), Correct Load Setup, Mass Optimization, Feasible Fastener Clearances, and Feasible Package Size. Checkmarks indicate successful completion of each criterion.

The table also includes average outcome design scores for each group, with SocratAIs achieving the highest average score of 4.0, followed by Expert-Facilitated at 3.4, HephAIstus at 3.2, and No Support at 1.0. Process metrics such as the number of agent messages, user messages, and triggered new considerations are recorded, showing that HephAIstus had the highest average number of agent messages (39), while SocratAIs led in triggering new considerations (6.0). The duration of tasks, in minutes, is also detailed, with HephAIstus having the longest average duration at 81.7 minutes and Expert-Facilitated the shortest at 42.5 minutes.

This table provides a comprehensive comparison of task outcomes and user-agent interaction processes across different support conditions, highlighting the effectiveness of agent-assisted strategies relative to the No Support condition.}
  \label{tab:results-summary-table}
  \includegraphics[width=\linewidth]{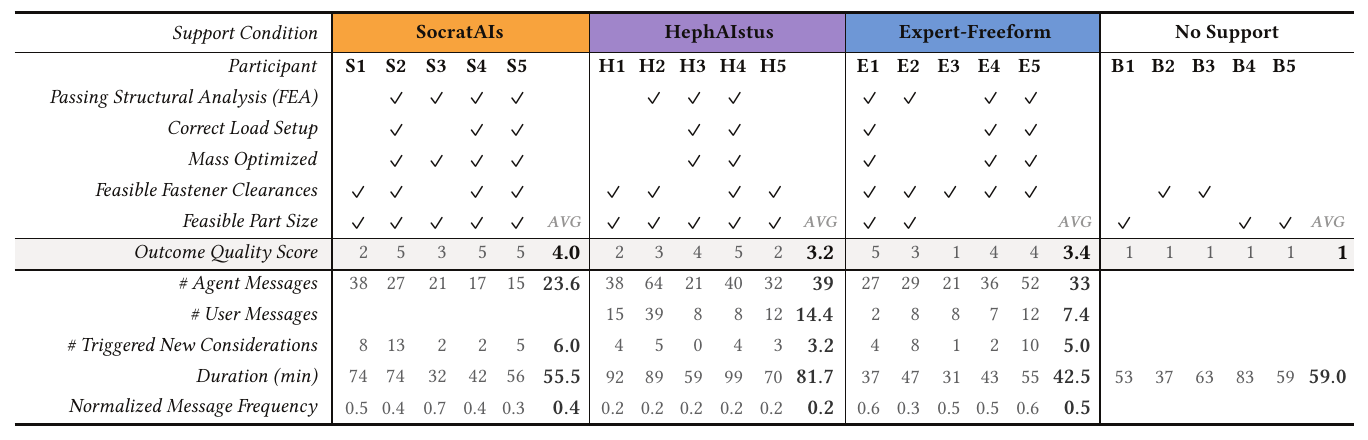}

\end{table*}

\section{Findings}

All participants completed the task without abandoning it. 
Overall, most agent-supported designers overcame more GenAI workflow challenges and produced more feasible designs than unsupported designers (Table \ref{tab:results-summary-table}). 
However, different agent strategies impacted the design process in different ways. 
Most designers saw benefits in agent support, but we also elicited various trade-offs and differing preferences for support interactions.
\rev{In the following sections, we present our findings on the impact of different support agent strategies on the design process, along with the perceived benefits and challenges associated with these.}

\subsection{Impact of different support strategies on design process (RQ1)}

\aptLtoX[graphic=no,type=html]{\begin{figure}
        \centering
        \includegraphics[width=\linewidth]{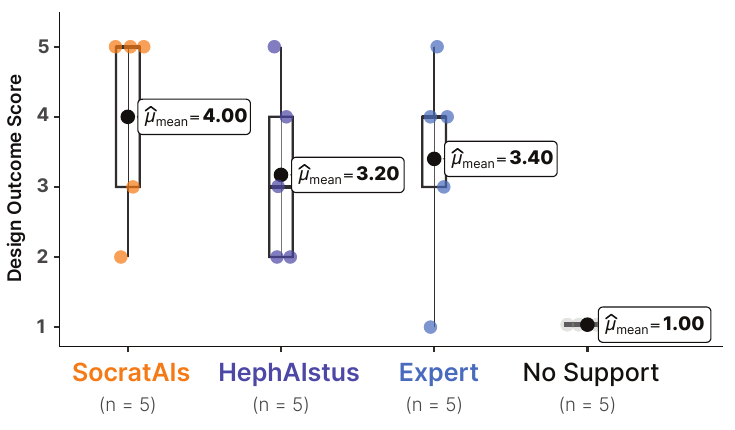}
        \caption{Plot showing the design outcome scores between agent-supported groups and no support.}
        \Description{The figure illustrates a plot comparing design outcome scores across four groups: SocratAIs, HephAIstus, Expert, and No Support, each with five participants. The Y-axis represents the design outcome scores ranging from 1 to 5, and the X-axis shows the group labels. Data points for each group are displayed as colored dots, with error bars indicating the range of scores. SocratAIs has the highest mean score of 4.00, followed by the Expert group with a mean score of 3.40, and HephAIstus with a mean score of 3.20. The No Support group has the lowest mean score of 1.00, with no variability across its participants. Each group’s mean score is prominently labeled next to its respective cluster of points. This plot highlights the performance differences between agent-supported groups and the control group without support.}
        \label{fig:design-outcomes-score-plot}
    \end{figure}
    \hfill
    \begin{figure}
        \centering
        \includegraphics[width=\linewidth]{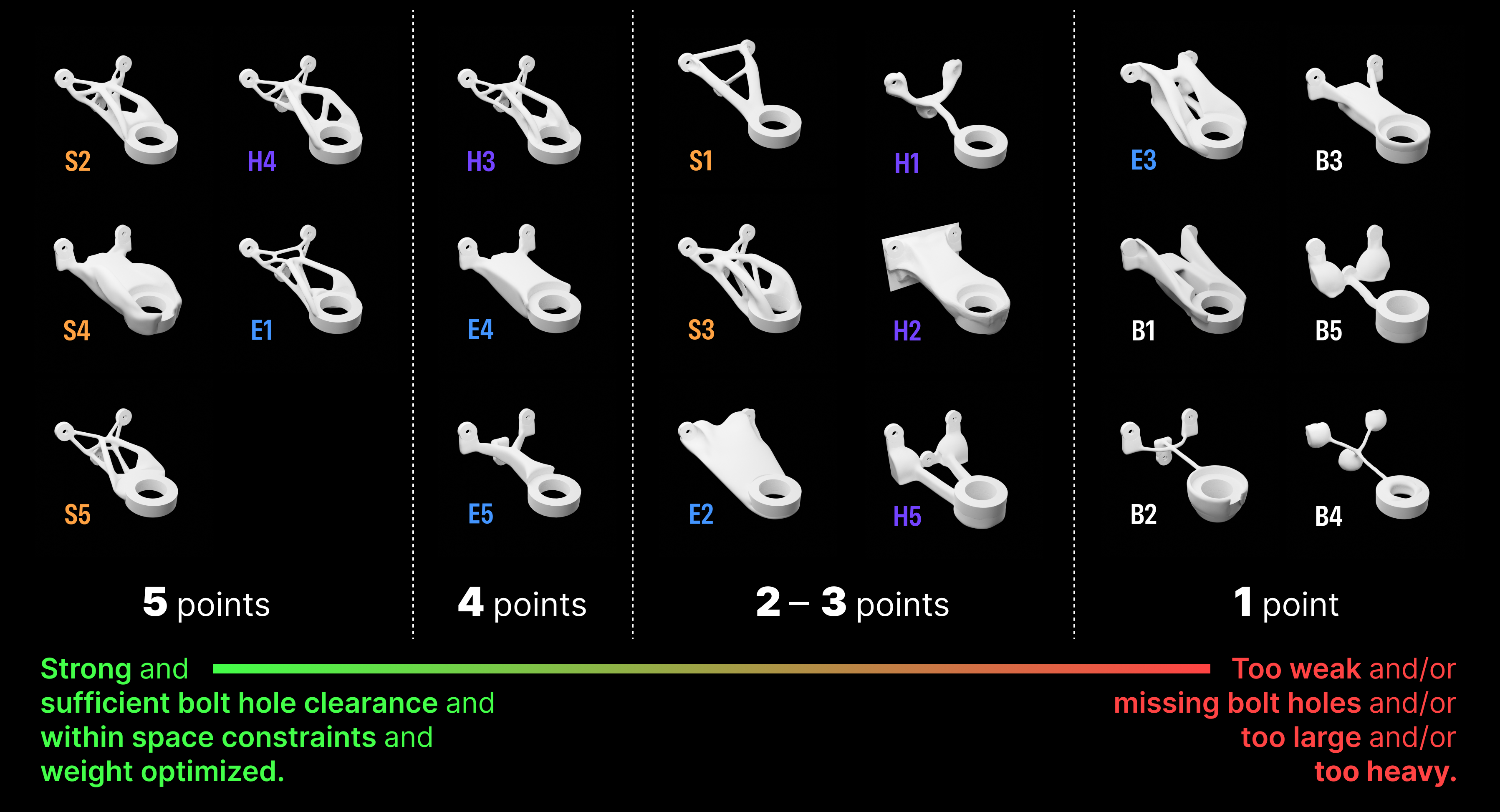}
        \caption{Overview of engine bracket designs created by participants grouped by quality score (1–5). White IDs indicate participants from the \textit{No Support} group (all one point).}
        \Description{The figure shows engine bracket designs created by participants, categorized by quality scores from 1 to 5. Each column represents a score group, with higher scores on the left and lower scores on the right. Designs scored 5 points are labeled with IDs such as S2, H4, and S5, and are characterized as strong, having sufficient bolt hole clearance, and being optimized for weight within space constraints. Designs scored 4 points, including H3 and E4, exhibit similar features but with minor deficiencies. Designs with scores of 2 to 3 points, such as S1 and H2, show issues like insufficient bolt clearance or weight optimization. Designs scored 1 point, labeled with white IDs such as B3 and B5, are from the No Support group and are marked as too weak, missing bolt holes, or being too large or heavy. The gradient below the designs highlights the criteria from strong and optimized to weak and deficient. White labels distinguish participants in the No Support group, whose designs uniformly scored 1 point.}
        \label{fig:brackets_designs}
\end{figure}}{\begin{figure*}[t]
    \centering
    \begin{minipage}[t]{0.48\linewidth}
        \centering
        \includegraphics[width=\linewidth]{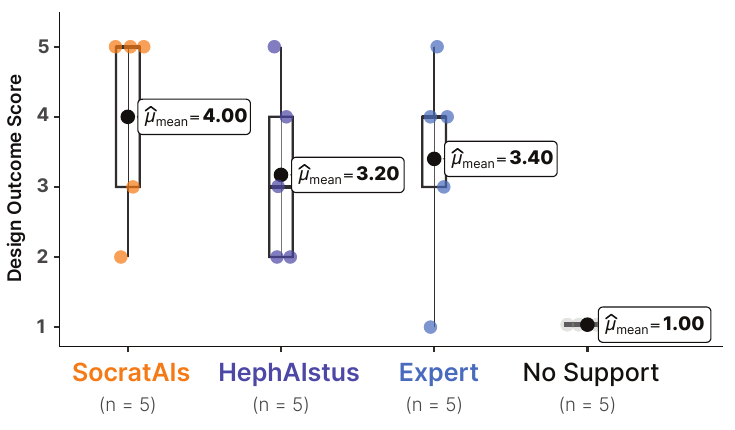}
        \caption{Plot showing the design outcome scores between agent-supported groups and no support.}
        \Description{The figure illustrates a plot comparing design outcome scores across four groups: SocratAIs, HephAIstus, Expert, and No Support, each with five participants. The Y-axis represents the design outcome scores ranging from 1 to 5, and the X-axis shows the group labels. Data points for each group are displayed as colored dots, with error bars indicating the range of scores. SocratAIs has the highest mean score of 4.00, followed by the Expert group with a mean score of 3.40, and HephAIstus with a mean score of 3.20. The No Support group has the lowest mean score of 1.00, with no variability across its participants. Each group’s mean score is prominently labeled next to its respective cluster of points. This plot highlights the performance differences between agent-supported groups and the control group without support.}
        \label{fig:design-outcomes-score-plot}
    \end{minipage}
    \hfill
    \begin{minipage}[t]{0.48\linewidth}
        \centering
        \includegraphics[width=\linewidth]{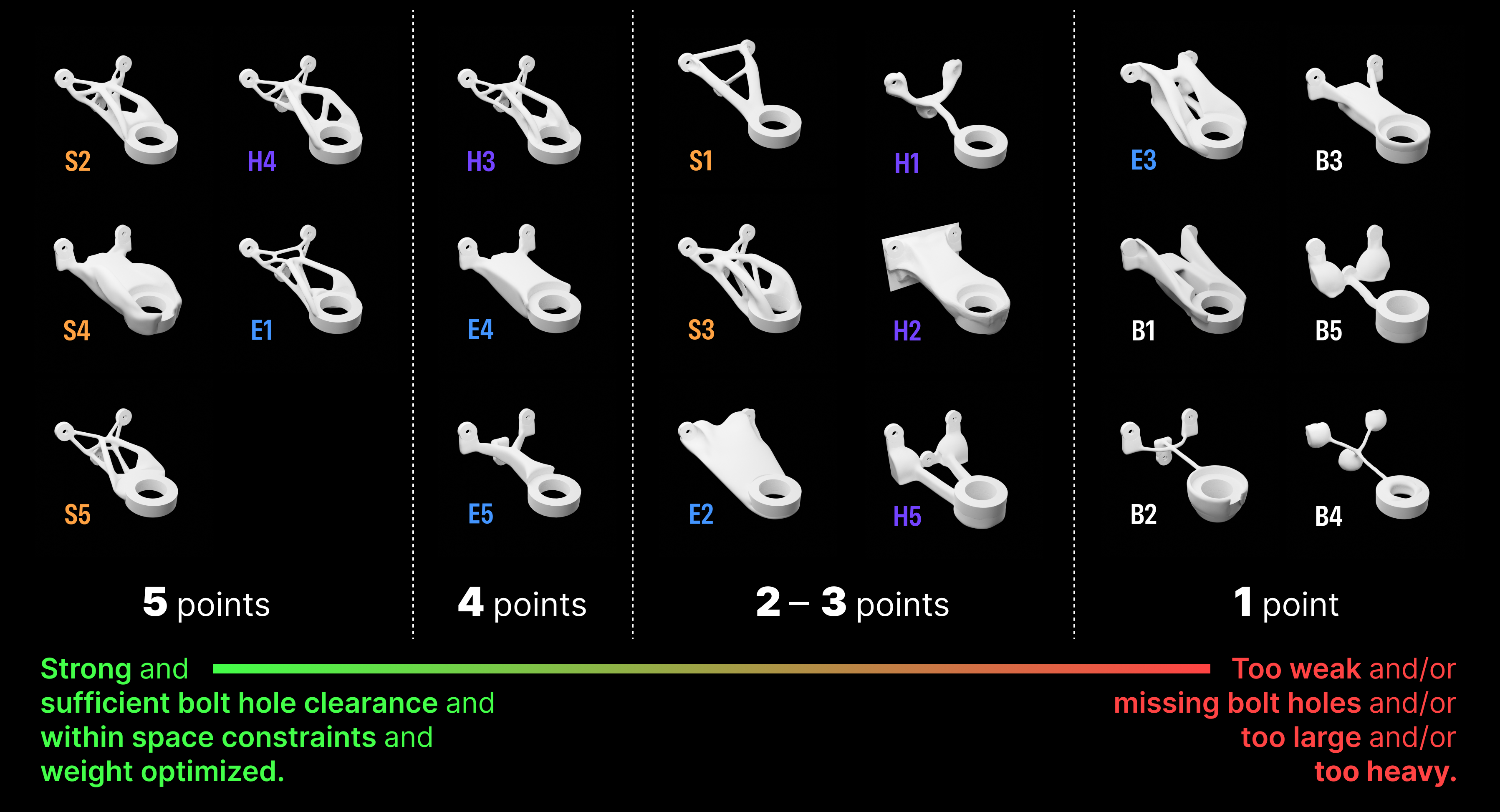}
        \caption{Overview of engine bracket designs created by participants grouped by quality score (1–5). White IDs indicate participants from the \textit{No Support} group (all one point).}
        \Description{The figure shows engine bracket designs created by participants, categorized by quality scores from 1 to 5. Each column represents a score group, with higher scores on the left and lower scores on the right. Designs scored 5 points are labeled with IDs such as S2, H4, and S5, and are characterized as strong, having sufficient bolt hole clearance, and being optimized for weight within space constraints. Designs scored 4 points, including H3 and E4, exhibit similar features but with minor deficiencies. Designs with scores of 2 to 3 points, such as S1 and H2, show issues like insufficient bolt clearance or weight optimization. Designs scored 1 point, labeled with white IDs such as B3 and B5, are from the No Support group and are marked as too weak, missing bolt holes, or being too large or heavy. The gradient below the designs highlights the criteria from strong and optimized to weak and deficient. White labels distinguish participants in the No Support group, whose designs uniformly scored 1 point.}
        \label{fig:brackets_designs}
    \end{minipage}
\end{figure*}}

\begin{figure}[t]
  \centering
  \includegraphics[width=\linewidth]{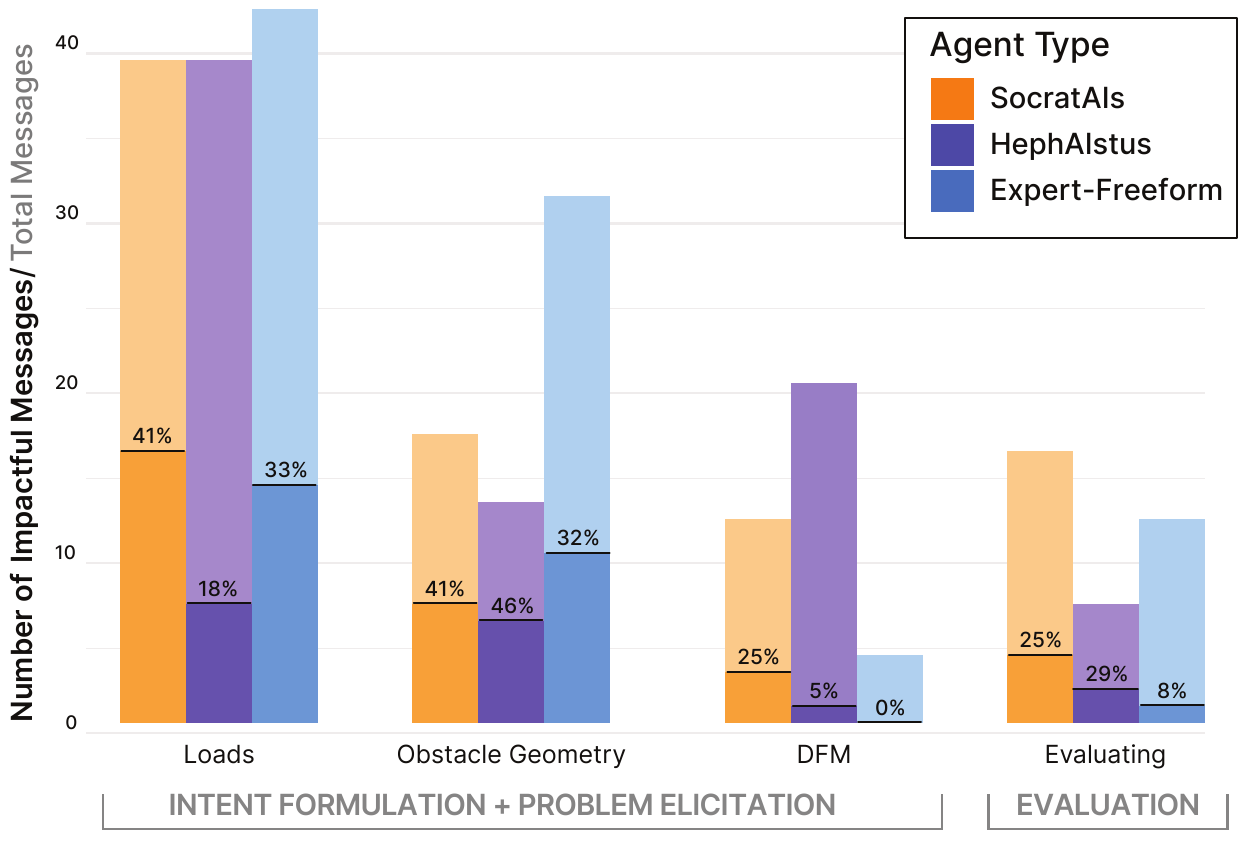}
  \caption{Plot illustrating the total number of messages per support topic category and agent type (unsaturated colored bars) and the percentage of observable impactful messages that triggered observable new considerations (saturated areas).}
  \Description{The figure presents a bar chart illustrating the total number of messages across four support topic categories—Loads, Obstacle Geometry, DFM (Design for Manufacturing), and Evaluating—segmented by agent type: SocratAIs, HephAIstus, and Expert. The unsaturated bars represent the total number of messages sent by each agent type, while the saturated areas indicate the percentage of impactful messages that triggered observable new considerations. For the Loads category, SocratAIs had 41\% impactful messages, Expert had 33\%, and HephAIstus had 18\%. For Obstacle Geometry, HephAIstus led with 46\% impactful messages, followed by SocratAIs at 41\%, and Expert at 32\%. In the DFM category, SocratAIs had 25\% impactful messages, HephAIstus had 5\%, and Expert had no impactful messages. In the Evaluating category, SocratAIs had 25\%, HephAIstus had 29\%, and Expert had 8\% impactful messages. This chart highlights the varying effectiveness of each agent type in generating impactful support across different design task categories.}
  \label{fig:messages-impact-plot}
\end{figure}

\subsubsection{Design Outcome \rev{Comparison}}
Overall, participants with support produced notably higher-quality parts (see Table \ref{tab:results-summary-table} and Figure \ref{fig:design-outcomes-score-plot} and \ref{fig:brackets_designs}), with an average outcome score of $M=3.5$ ($SD=1.4$), compared to participants with no support who had an average outcomes score of $M=1.0$ ($SD=0.0$).
All participants in the \textit{No Support} group incorrectly specified the bracket’s load case, and consequently, no final design passed the structural analysis (see designs with white labels in Figure \ref{fig:brackets_designs}). 
Additionally, brackets created in the \textit{No Support} group had inaccurately specified obstacle geometry, resulting in infeasible fastener clearances or material exceeding the required package size. \rev{These low-quality outcomes match our prior study's results in the same unsupported task \cite{gmeiner_exploring_2023a}.}

In contrast, while the outcome quality varied within and across the agent-supported groups, the majority of supported designers created brackets that passed the structural analysis and had feasible fastener clearances while staying within the required space limitations. 
Between supported groups, the average design score varied slightly, with \highsocratais{\textit{SocratAIs}}-supported users having the highest number of designs fulfilling the load and spatial requirements.
\rev{While the small sample size per group prevents us from drawing conclusions about statistically significant differences between the support conditions, the consistent gap between supported and unsupported users points to clear benefits of having agent support}.

\subsubsection{\rev{Comparison of} Agent Message Frequency and Impact (across conditions)} \label{findings-message-impact}

\rev{In terms of \textbf{number of messages},} agents sent between 15 and 64 messages per session, with the highest group average being $M=39\rev{(SD=15.8)}$ in the \highhephaistus{\textit{HephAIstus}} group, $M=33 \rev{(SD=11.9)}$ in the \highexpert{\textit{Expert-Freeform}} group, and $M=23.6 \rev{(SD=9.3)}$ in the \highsocratais{\textit{SocratAIs}} group (see Table \ref{tab:results-summary-table}). 
\rev{However, these counts of the number of messages sent by \highhephaistus{\textit{HephAIstus}} and \highexpert{\textit{Expert-Freeform}} also include responses to user-initiated queries and therefore are naturally higher than for the \highsocratais{\textit{SocratAIs}} group. 
To gain a \textbf{normalized comparison of message frequency between conditions}, we calculated the number of messages initiated by the agents divided by the session duration (agent-initiated messages/session duration), which revealed a similar message frequency per minute across agent groups of $M=0.4 (SD=0.15)$ for \highsocratais{\textit{SocratAIs}}, $M=0.2 (SD=0.03)$ for \highhephaistus{\textit{HephAIstus}}, and $M=0.5 (SD=0.14)$ for \highexpert{\textit{Expert-Freeform}}.}

\textbf{Regarding the messages' impact on the design process,} the number of messages that triggered observable new considerations (i.e., impactful messages, as defined in \ref{video-analysis}) ranged from zero to 13, with means ranging between $M=6.0 \rev{(SD=4.6)}$ \highsocratais{\textit{SocratAIs}}, $M=5.0 \rev{(SD=3.9)}$ \highexpert{\textit{Expert-Freeform}}, and $M=3.2 \rev{(SD=1.9)}$ \highhephaistus{\textit{HephAIstus}}.
\rev{Interestingly, although the \highhephaistus{\textit{HephAIstus}} group had fewer messages that triggered observable new considerations compared to the other two groups, their final design outcomes were comparable. 
This suggests that the planning and sketching activities prompted by the \highhephaistus{\textit{HephAIstus}} agent may have supported productive design reasoning, even when fewer individual messages were coded as impactful.}

\rev{\textbf{Analyzing the messages' topics across all agent groups,}} most agent messages concerned \textit{intent formulation} and \textit{problem exploration} in the \textit{Loads}, \textit{Obstacle Geometry} and \textit{DFM} categories followed by messages supporting \textit{Evaluation} (see Figure \ref{fig:messages-impact-plot}). 
\rev{Further analyzing the messages' topics in terms of their impact on considering new design-relevant aspects,} the highest number of impacts had messages supporting intent formulation and problem exploration: \textit{Obstacle Geometry} (between 32\% to 46\% across groups) and \textit{Loads} (41\% \highsocratais{\textit{SocratAIs}} and 33\% \highexpert{\textit{Expert-Freeform}}). 

In terms of the \textbf{contrast between the number of messages and impact on design considerations}, agents varied drastically between the support aim topics:
\highhephaistus{\textit{HephAIstus'}} \textit{Loads} and \textit{DFM} categories had only half or a third of the impact (18\% and 5\%) as \highsocratais{\textit{SocratAIs}} while having a similar or larger number of total messages.
Similarly, the \highexpert{\textit{Expert-Freeform's}} messages supporting \textit{Evaluating} had only half of the impact (8\%) as the other support agents, while having a similar number of total messages.
In contrast, for the \textit{Obstacle Geometry} and \textit{Evaluating} categories, messages in the \highhephaistus{\textit{HephAIstus}} group had the highest consideration impact while having the lowest number of messages compared to the other groups (46\% and 29\%).

\begin{figure*}[t]
  \centering
  \includegraphics[width=\linewidth]{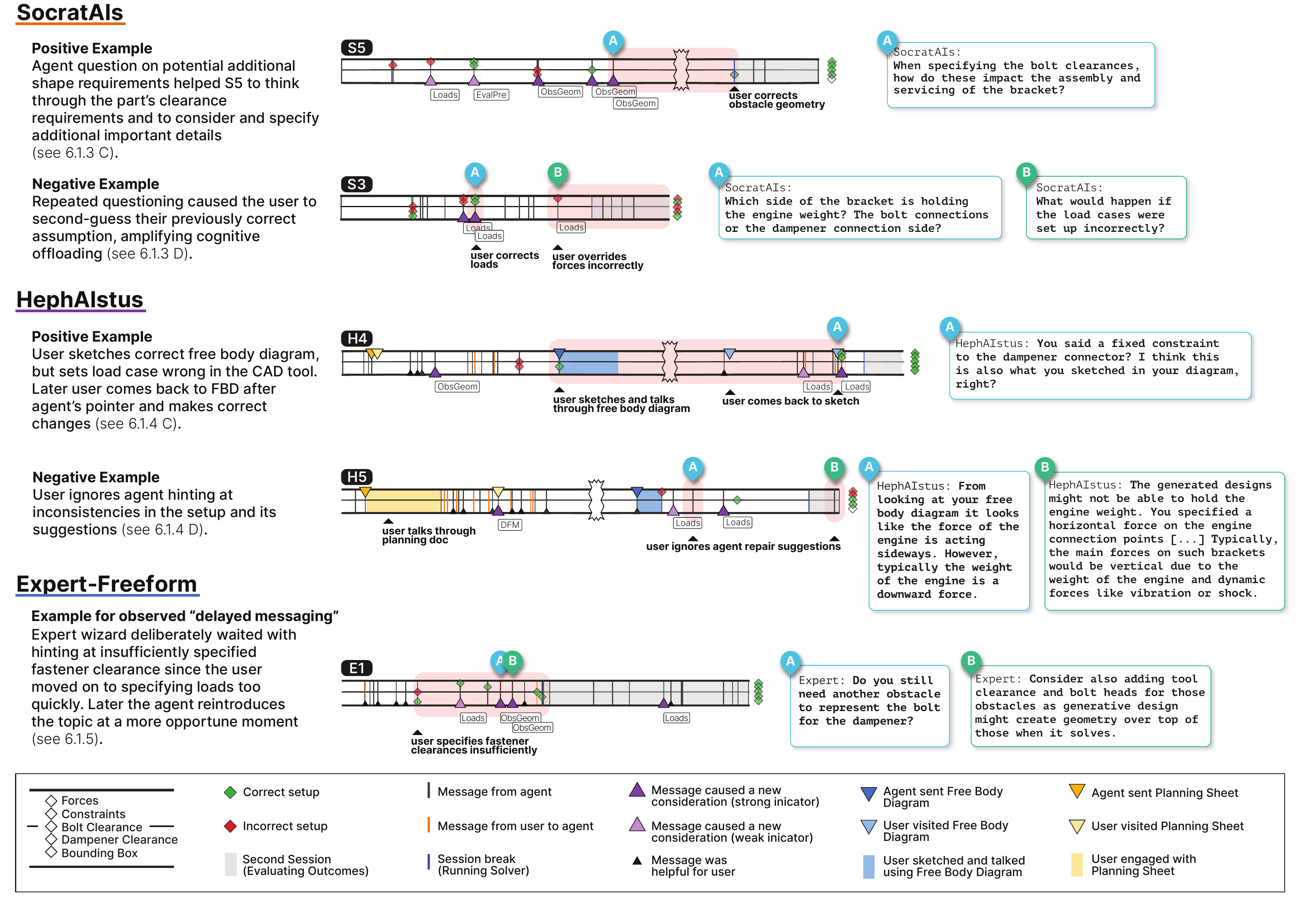}
  \caption{Timeline excerpts visualizing participant and agent interactions throughout the design task; timelines are divided into lanes, each showing (in)correct GenAI input specifications (diamond shapes) for 
  (1) forces, (2) constraints, (3) bolt and, (4) dampener clearances, (5) bounding box  (from top to bottom); black and orange vertical lines represent exchanged agent and user messages with purple and black triangles indicating an observable impact on the design process.}
  
  \Description{The figure presents timeline excerpts visualizing participant and agent interactions throughout the design task, with examples categorized by the type of agent (SocratAIs, HephAIstus, Expert). Each timeline is divided into lanes representing (in)correct GenAI input specifications for forces, constraints, bolt clearance, dampener clearance, and bounding box. Diamond shapes indicate (in)correct input setups, while vertical lines show agent and user messages. Purple and black triangles mark messages that caused new considerations or were helpful.

For SocratAIs, a positive example shows how a question about bolt clearances prompted a participant (S5) to think through assembly requirements and adjust their design. A negative example highlights repeated questioning that led to cognitive offloading and flawed outcomes for S3. For HephAIstus, a positive example demonstrates how a free-body diagram prompted H4 to correct their load case setup. A negative example illustrates how H5 ignored direct repair suggestions from the agent. For the Expert agent, a delayed messaging strategy is shown where the agent waited to highlight fastener clearance issues until the user (E1) was ready to revisit the topic, ultimately helping refine the design.

These timelines showcase different interaction dynamics and the varying impacts of agent strategies on addressing design challenges.}

    \label{fig:timelines}
 
\end{figure*}

\subsubsection{\highsocratais{SocratAIs'} Effects on the Design Process} \hfill
\label{findings-socratais}

{
\raggedright
\textbf{A) User-Agent Interaction Dynamics:}
}
Overall, participants paused their think-aloud verbalizations when listening to agent messages (118/118 questions). 
However, depending on the user's situation, they responded differently: 
immediately responding to the agent's questions by giving an answer (61/118 questions); finishing their line of thought and sub-task before replying (28/118); pausing to think and reflect silently before verbally replying (13/118); or directly responded with simple acknowledgments after thinking for a while in silence, such as \textit{"Yeah, you are right,"} or \textit{"That's a good point,"} even if the message was not a direct suggestion but an open-ended question (11/118); or providing no response (5/118). 
Some participants (2/5) asked the agent a question at the beginning of the session, but they stopped asking the agent more questions afterward, recognizing that it would not provide an answer but reply to user requests only with questions.
\\ \\
{\raggedright
\textbf{\rev{B) Agent Impact on Overcoming GenAI-Related Challenges:}}
}
\\
\textbf{\highsocratais{\textit{SocratAIs}} had mixed impacts on helping designers overcome design challenges across participant sessions} (see example timelines in Figure \ref{fig:timelines} and Appendix Figure \ref{fig:appendix_timelines} for all sessions). 
In some sessions (2/5), agent messages had an observable strong impact on helping designers overcome design challenges (S2, S5). 
Meanwhile, in other sessions (2/5), we could only observe weak evidence of impact, where some agent messages had a positive impact, but overall, designers were unable to overcome most major challenges (S1, S3). 
In one session (S4), the designer created feasible outcomes without facing major challenges or showing observable agent impact, but \highsocratais{\textit{SocratAIs}} helped them to consider additional design-task-related factors.
\\ \\
{
\raggedright
\textbf{C) \highsocratais{\textit{SocratAIs}}' Positive Effects:}
} 
\\
In most sessions (S2, S3, S4, S5), \textbf{reflective agent questions helped users with more precise intent formulation and problem exploration.} 
For example, an agent question probing reflection on potential additional shape requirements of the part helped S5 think through the part's clearance requirements and consider and specify additional important details, leading to a feasible bracket design:
\begin{itemize}
\setlength{\leftskip}{1cm}
\setlength{\rightskip}{1cm}
    \ttfamily
  \item[\highsocratais{\textit{SocratAIs}}:] When specifying the bolt clearances, how do these impact the assembly and servicing of the bracket?
  \item[S5:] [Looks at the preview simulation] So this part won't be serviceable [...] you'd need enough clearance for the socket. So I will go back to edit model..." [user adds more obstacle geometry].
\end{itemize}
(see also timeline S5 in Figure \ref{fig:timelines}). 
Later, S5 reflected on the helpfulness of the agent's message:
\textit{"It was asking something about how the obstacle geometry affects the serviceability of the part. That was basically telling me that I needed to leave some clearance for tools and for maintenance. That was very helpful"} (S5).

We also noticed in several cases that \textbf{questions were especially effective} in helping designers formulate intent and specify the design problem more accurately \textbf{when prompting users to mentally simulate the real-world aspects of the bracket}. For example, S3 had mistakenly modeled the load case reversed to the real-world situation ("flipped" load case), causing the solver to build a structurally unstable bracket. 
An agent question prompting them to reflect on the part's function from a real-world perspective (\texttt{Which side of the bracket is holding the engine weight? The bolt connections or the dampener connection side?}) helped the user update their mental model and load case.

In two sessions (S2, S5), \textbf{reflective questions during the preview and outcome evaluation phases helped designers better evaluate and correct faulty designs.}
For example, S2 generated brackets based on incorrectly specified loads (forces and constraints assigned to the wrong sides). 
While evaluating the design, the user noticed the structurally weak parts but was unsure about the cause.
An agent question probing deeper thought about the GenAI solver's mechanism  caused the designer to realize their flawed mental model and correct the load specification:
\begin{itemize}
\setlength{\leftskip}{1cm}
\setlength{\rightskip}{1cm}
    \ttfamily
  \item[\highsocratais{\textit{SocratAIs}}:] What might be the reasons why the solver generated the shapes this way?
  \item[S2:] [user thinks] Yeah, because it did not consider building material between these three [bolt connections] ... maybe because of the forces...  [user checks forces again] It should be acting downward based on the weight of the engine [user corrects load case].
\end{itemize}

\textbf{For confident users, questions helped in problem exploration by considering additional design factors.}

In one session, S4 did not encounter major challenges, and while the agents also asked similar questions as in the other sessions, these also had no observable impact on supporting critical cognitive challenges.
However, agent questions (such as "\texttt{Considering the engine’s environment, what alternative or additional assumptions might we make about the load cases?}") helped the user to consider additional relevant factors, such as suitable materials for the bracket's maritime environment or additional forces resulting from ship movements.
Later, S4 reflected in the interview:
\textit{"I think it was pretty useful [...] it prompted me about the assumptions about the loading. And I was like: 'oh yeah, this is on a ship, so it has to survive lateral loads and not just ground loads'." (S4)
}
\\ \\
{\raggedright
\textbf{D) \highsocratais{\textit{SocratAIs}}' Negative Effects:}
} \\
In one session (S3), 
\textbf{repeated questioning amplified cognitive offloading, leading to flawed results.}
At first, \highsocratais{\textit{SocratAIs}}' questions during the initial setup phase helped S3 correct their overconstrained load case. 
However, a later question \texttt{(What would happen if the load cases were set up incorrectly?)} made them doubt their initially correctly specified load case setting and change it for the worse (see S3 in Figure \ref{fig:timelines}).

In another case, \textbf{questions were less impactful in correcting a user's solidified wrong assumptions}.
S1 had incorrectly set up a load case with fixed constraints and forces assigned to the same geometry, which would cancel out the impact of the force on the bracket's structure in the solver. 
An agent question (\texttt{Can you walk me through your intention of assigning a force and a fixed constraint to the same geometry?}) caused S1 to provide an explanation of their reasoning, eliciting their incorrect assumptions. 
But, instead of realizing misconceptions, their explanations reinforced their assumptions, preventing them from correcting issues.

\subsubsection{\highhephaistus{\textit{HephAIstus}}' Effects on the Design Process}  \hfill
\label{findings-hephaistus}

{\raggedright
\textbf{A) User-Agent Interaction Dynamics:}
} \\
Besides proactive support, \highhephaistus{\textit{Heph\-AI\-stus}} also responded to user requests like a voice-based chat assistant such as Alexa or Siri, and users addressed the agent between 8 to 39 (M=14.4) times per session (see Table \ref{tab:results-summary-table} and orange vertical lines in Figures \ref{fig:timelines} and \ref{fig:appendix_timelines}). 
These user-initiated requests included asking the agent about manufacturing-related facts, such as material properties, or asking for help with load case-related tasks, such as calculating forces.
Some users requested confirmation or feedback from the agent on the design process (\textit{"Am I missing anything, agent?"}) or requested guidance (\textit{"Ok, what's next?"}). 
Overall, these sessions were characterized by phases of active back-and-forth conversations between the user and agent, as apparent from the clusters of dense orange and black vertical lines in the event timelines in Figures \ref{fig:timelines} and \ref{fig:appendix_timelines}. The proactive messages initiated by the agent included suggestions regarding Fusion360 operation (such as specific tools within the Generative Design extension) and overcoming design challenges (such as reminding users to use the preview or pointing out mismatches between the design brief and their setup). 
\\ \\
{\raggedright
\textbf{\rev{B) Agent Impact on Overcoming GenAI-Related Challenges:}}
}
\\
Overall, we found that \textbf{\highhephaistus{\textit{HephAIstus}} had mixed impacts on helping designers overcome design challenges across participant sessions.} 
In 1/5 sessions, the agent messages had an observable strong impact on helping designers overcome design challenges (H4), while in 3/5 sessions, weak impacts on overcoming design challenges were observable (H1, H2, H5). 
In 1/5 sessions, a designer created almost feasible outcomes without facing major challenges or observable agent impact (H3), but it was clearly observable that the agent helped the user operate the software more effectively.
\\ \\
{\raggedright
\textbf{C) \highhephaistus{\textit{HephAIstus}}' Positive Effects:}
} \\
\textbf{The agent-provided sketching board helped some designers in loads-related problem specification and intent formulation.}
When the agent prompted designers to explain the bracket's load case by sketching a free-body diagram (FBD) and sharing a link to a prepared drawing board, all users followed the link and used the board to sketch out diagrams while verbalizing their thoughts (see blue triangles and highlighted passages in Figure \ref{fig:timelines} and \ref{fig:appendix_timelines} and example board in Appendix). 
In several cases, the sketched free-body diagram served as a conversational anchor and reference point between the user and the agent. 
For example, while H4 had first sketched an FBD with feasible load cases, they then incorrectly specified the load case in Fusion360. 
Later, the agent pointed out the inconsistency between the sketched FBD and the load case setup in the CAD tool, which led the designer to correct the input specification (see H4 in Figure \ref{fig:timelines}).

\textbf{All designers explored the agent’s project planning sheet, but few revisited it during the session.}
While 2/5 users quickly went through the document at the beginning, 3/5 users (H2, H3, H5) spent between 3–10 minutes in the document, utilizing its provided structure to talk through, reflect on and plan the design process step by step before starting to work in Fusion360 (see yellow triangles and highlighted passages in Figure \ref{fig:timelines} and \ref{fig:appendix_timelines}). 
For example, the planning sheet supported H5 in reflecting on and exploring suitable material options while using the document to add notes about different material characteristics (see H5 in Figure \ref{fig:timelines}). 
However, only two users revisited the document later in the session (H4, H5).

We also observed that in many cases, \textbf{proactive agent suggestions reminded designers about overlooked steps,  unintentional execution errors (slips), or software features.} 
For example, the agent reminded H2 to run a preview simulation to better assess bolt clearances before starting the solver (\texttt{You might run a preview simulation at a later point to evaluate the clearances before starting the solver}).
This message led H2 to run a preview and realize insufficient bolt clearance. 
\\ \\ 
{\raggedright
\textbf{D) \highhephaistus{\textit{HephAIstus}}' Negative Effects:}
} \\
While 52\% of agent messages had an observable impact on helping the user work on the task and operating the software (101/196 messages, see black triangles in Figure \ref{fig:timelines}), only a fraction (0.06\%) directly helped overcome cognitive design challenges by considering new design-task relevant aspects (16/196 messages, purple triangles in Figure \ref{fig:appendix_timelines}).
We also observed that \textbf{directly pointing out inconsistencies in the users' setup only helped some users correct existing issues.} 
For example, in two cases (H1, H5), designers repeatedly failed to correct ill-defined load case setups despite the agent directly pointing these out and providing concrete suggestions for correcting them (see H5 in Figure \ref{fig:timelines}).
In those situations, participants decided not to follow the agent's suggestions and instead followed their own (partially incorrect) intuition.

\subsubsection{\rev{\highexpert{Expert-Freefrom} Observed Support Strategies}} \label{findings-expert-support-strategies} \hfill

From analyzing the session videos and post-task interviews with the expert facilitators, we identified several support strategies that helped designers similar to our support agent probes. 

We observed that \textbf{experts frequently supported designerly thinking and metacognition while also highlighting overlooked design issues to guide users}.
Similar to \highhephaistus{\textit{HephAIstus}}, experts proactively highlighted potential issues, such as missing or misrepresented GenAI parameters, to ensure critical considerations were addressed early.

Additionally, we observed that \textbf{some experts used a question-asking strategy} to support users in intent formulation, problem exploration, and outcome evaluation similar to \highsocratais{\textit{SocratAIs}}. 

Some \textbf{experts also deliberately delayed messages} when the user moved on to a different sub-task too quickly and waited to reintroduce the topic later at a more opportune moment (see E1 in Figure \ref{fig:timelines}).

Lastly, we observed that \textbf{experts also frequently supported users in navigating CAD software features}, helping them by offering guidance on Generative Design functions, recommending workflow optimizations, helping with calculations such as load distribution, and providing help to locate tools and options as needed.

\subsection{Perceived benefits and challenges of metacognitive support agents (RQ2)}

Overall, participants appreciated the support from the agents.
S2 stated that \textit{"it's doing a good job [...] by assisting you throughout the whole design work"}  and H1 noted a perceived efficiency gain: \textit{"I feel like without [the agent], [...] it would have definitely taken a longer amount of time."}
Besides positive aspects, participants highlighted trade-offs and challenges, which we present in the following sections, organized around the different support strategies and general agent interactions.

\subsubsection{User Feedback on \highsocratais{\textit{SocratAIs}} (Question-Asking)} \label{findings-user-feedback-socratais}\hfill

\textbf{Participants consistently found the agents’ questions valuable, particularly those that prompted reflection on key design aspects}, such as load cases and clearances. 
These questions helped refine GenAI input specifications by encouraging them to reconsider functional details and correct initial assumptions, as one participant noted, \textit{“The questions [...] helped me reflect and go back over my train of thought and see, ‘Am I missing something? Does this look like I’m doing what I’m supposed to do?’”} (S5).

Participants also reported that questions encouraged them to slow down and critically evaluate their thinking, much like a professor would in a one-on-one setting, as S1 explained: \textit{"when [..] you're just sitting down designing by yourself, you don't often run through those things. So having someone to stop you and say, 'Why do you think that works?' is a good check every now and then"} (S1).

\textbf{Some users found agent questions redundant but preferred them over missing important steps.}
For example, one participant noted that while \textit{"25\% of the questions actually helped,"} the rest pulled their attention away from the current task (S1). 
Others mentioned that the agents sometimes asked questions about actions they were already performing, which felt unnecessary as they were already thinking through those steps. 
However, participants acknowledged that redundancy was preferable to missing something important and found the frequency of questions appropriate.

\textbf{Some found the questions more useful for problem exploration and intent formulation:}
\textit{"It did a good job [during] the initial part of the setting up of the design"} (S3). 
\textbf{Others saw more benefit in supporting outcome evaluation}, particularly when analyzing and comparing designs: 
\textit{"[Questions during outcome evaluation phase were more] valuable because when you're comparing this many designs, it's good to be reminded of what's most important to compare and prioritize"} (S2).

\subsubsection{User Feedback on \highhephaistus{\textit{HephAIstus}} (Planning, Sketching, with Suggestions)} \label{findings-user-feedback-hephaistus} \hfill

\textbf{The agent-provided project planning sheet was generally perceived as helpful}, giving users a structured way to approach their tasks and a document to guide their process, as this participant stated: \textit{"To show you an actual work plan from the design to the actual fabrication and production of the piece is good. [It helps you to] separate [the design process] into different steps and how we're going to work from here"} (H2).

\textbf{Designers found that sketching helped them visually think through the design problem:} 
\textit{"Sketching a free body diagram was definitely helpful. I mean, it was just good to see before I had set it up in Fusion, sort of my plan for where the loads and constraints were gonna go"} (H3).
Several participants suggested that the sketching feature could be improved by making it more interactive---for example, by providing real-time calculations, augmenting sketches with force vectors, and offering a library of example diagrams---to support thinking through a part's design requirements.

\textbf{Many participants valued the agent's proactive suggestions, and
when the agent pointed out possible inconsistencies in their load specifications,}
as H1 described:
\textit{"I wasn't sure why [the solver] was generating so thin [parts] and having the agent explain to me, 'hey, it's probably because of the constraints that you set, you have canceling loads, you should not do that.' That was good feedback to modify the constraints"} (H1).

\textbf{Designers also liked when \highhephaistus{\textit{HephAIstus}} helped them catch slips and correct mismatches in real-time.}
For example, one user appreciated a hint that pointed out a mismatch between the force requirements in the design brief and the load setup:
\textit{“I thought it was extremely helpful. I [mis]read the instructions with the weight capacity [...] I’m glad I received a prompt to make sure that the weight distribution was accurate because it knew there was a difference between the weight the bracket was supposed to hold and the engine’s total weight. I was impressed that the agent provided me that prompt to check and make that design change”} (H1).

Additionally, \textbf{participants also found the agent more efficient than searching online or sifting through video tutorials for answers. }
Being able to ask the agent questions directly about specific software functions or design issues saved time and allowed them to stay focused on the task without interrupting their workflow:
\textit{"I didn't have to stop what I was doing [...] to go to Google and find information"} (H3).

\textbf{The fact that the agent would also annotate the screen and highlight relevant interface elements was widely appreciated for reducing the need to search for tool functions manually}, as one participant noted: \textit{“when I had doubts about specific functions in Fusion, I asked the agent and it was very helpful in that. And the fact that it would highlight the word to click, that was very useful”} (H2).

\subsubsection{Reflections and Suggestions of the \highexpert{\textit{Expert-Freeform}} Wizards} \label{findings-expert-reflections} \hfill

In the post-task wizard interview (after supporting a designer), the task experts reflected on their support strategies and challenges. 
Some \textbf{experts emphasized the difference between operating Fusion360 and thinking through design problems, underscoring that while both of these tasks are equally important, they often require two distinct "mindsets":}
\textit{"I was thinking of it from a 'how do I use Fusion' standpoint. [But it's ] kind of like two parts of the brain: One is like, 'I know where the buttons are, I know the workflow,' and there is like, really creative problem-solving"} (Expert 4).

Furthermore, some also suggested that \textbf{agent-initiated “design reviews” during outcome evaluation could help users in critical evaluation of GenAI outcomes}.
Similarly, others suggested to \textbf{introducing deliberate "checkpoints" or reflection phases between design task steps} (e.g., when the user transitions from specifying loads to obstacle geometry), which could allow for better scaffolding of (metacognitive) support throughout the process.

Several \textbf{experts also saw the potential to pre-structure workflows by offering early guidance on problem exploration and setting up generative design inputs.} 
Some highlighted the value of “preemptive” planning activities, similar to \highhephaistus{\textit{HephAIstus'}} planning sheet---suggesting that preparing a clear design plan before switching to the GenAI tool could improve the overall design process.

\textbf{Some experts deliberately focused on supporting users in outcome evaluation} and also suggested that designers run a solver preview early to obtain visual feedback, helping users quickly assess if their setup was correct.
Another observed recurring strategy was that during outcome evaluation, \highexpert{\textit{Expert-Freeform}} agents suggested looking back to realize flawed load specifications, as this expert described in the interview:
\textit{"If we get really hefty results like blocky stuff, then as the agent, I can say, 'Do these parts seem over-designed? 
Let's look back at our load cases!' And then we can recognize, 'okay, we applied that load to the full load to every entity'"} (Expert 1).

Lastly, experts also suggested instead of directly speaking agent messages, to \textbf{annotate screen elements to signal available feedback from the agent}, for example, by circling critical parts, to provide users opportunities to initiate a conversation with the agent when desired by clicking on these highlighted regions.

\subsubsection{Feedback on General Interactions With Agent Probes } \label{findings-feedback-general-agent-interactions}
\rev{Here we summarize participants' feedback on general aspects of interacting with the support agents.}
\textbf{Participants generally appreciated the voice modality of the agent, finding it faster and more efficient than typing}, as H2 described, \textit{"talking is more time efficient because in chat I'll first have to explain the issue and it will take longer and be less clear"} (H2).

\textbf{Others highlighted the benefit of voice-based interactions for complex tasks that required creative or visual exploration:}
\textit{"In a software like this, I could definitely see use in [voice]. Text can be convenient, [but] you might miss it. Having a voice is helpful [for] anything that requires an exploratory or creative process"} (H1).
However, some users noted that \textbf{voice interaction could be impractical in shared or public workspaces}, such as offices or labs, and would prefer an additional text-based alternative in such environments.

\textbf{Users also shared differing opinions about the overall benefits of the question-asking and support suggestion strategies.} 
Designers in the \highsocratais{\textit{SocratAIs}} group generally valued open-ended questions rather than providing direct answers, as it encouraged them to think critically about their design decisions and allowed for flexibility in their approach, as S5 shared: 
\textit{"I don't think it should have pulled me straight up the answer. I think it was better to tell me, 'Hey, you should think about this,' because not every part is going to be maintained the same way or serviced the same way [...] By asking you a more open-ended question, it pointed you in a direction, but it also left open the possibility to ignore it"} (S5).
However, a few users also suggested that \textbf{more direct educational scaffolding with explanations could be more helpful for less experienced users}, while other, more experienced users could prefer shorter prompts to ensure that their workflow stayed on track. 
Some participants in the \highhephaistus{\textit{HephAIstus}} group also emphasized that to fully rely on the agent, they would need to trust its understanding of complex design contexts and ensure that its recommendations are accurate, especially for critical engineering tasks.

\begin{table*}[t]
\caption{Overview of design considerations and key learnings.}
\Description{ This table is machine-readable}
\rowcolors{2}{gray!15}{white} %
\begin{tabular}{rp{12.4cm}p{1.3cm}}
\toprule
\multicolumn{2}{l}{\textbf{Design Considerations / Key Learnings}} 
    & \textbf{Seen in}                                                \\

\midrule
\multicolumn{3}{l}{\textit{Opportunities for agent-based \textbf{metacognitive support}}} \\
\midrule

A1 & Cueing users with \textbf{thought-provoking open-ended questions} can help with \textit{intent formulation}, \textit{problem exploration}, and \textit{outcome evaluation} in GenAI-assisted design tasks.                             & \ref{findings-message-impact}, \ref{findings-socratais}, \ref{findings-expert-support-strategies} \\

A2 & Prompting \textbf{mental simulations through questions and sketching} can assist designers in thinking through design problems and more accurately formulating intents and specifying GenAI model inputs (supporting \textit{intent formulation} and \textit{problem exploration}).                                                                                  & \ref{findings-socratais}, \ref{findings-hephaistus}   \\

A3 & Offering metacognitive support in \textbf{key moments} of GenAI-based design processes can enhance cognitive engagement, for example, by offering users \textbf{agent-driven “design review sessions” during part evaluation} or introducing \textbf{dedicated "reflection checkpoints"} when transitioning between subtasks.                                                                        & \ref{findings-socratais}, \ref{findings-expert-reflections}  \\

A4 & Giving users \textbf{control over the type of metacognitive support} depending on their needs and experience level.                                                                           & \ref{findings-feedback-general-agent-interactions}  \\

A5 & Providing designers \textbf{custom-generated user-editable design checklists} to support planning and reflection of design decisions.                                                                         & \ref{findings-hephaistus}, \ref{findings-user-feedback-hephaistus}   \\

\midrule
 \multicolumn{3}{l}{\textit{Opportunities for agent-based \textbf{CAD support}}} \\
\midrule

B1 & Offering \textbf{suggestions for design decision and tool operation} in combination with \textbf{metacognitive support} to help \textbf{improve users' tool fluency} and \textbf{overcome cognitive GenAI workflow challenges}. & \ref{findings-expert-support-strategies}, \ref{findings-expert-reflections}   \\

B2 & Enabling users to \textbf{verbally request support from agents} can help to maintain focus and reduce context-switching in complex and visual-heavy CAD tasks.                                                                            & \ref{findings-hephaistus}, \ref{findings-user-feedback-hephaistus}   \\

B3 & In addition to \textbf{voice agent feedback}, utilizing visual \textbf{screen annotations and text} can reduce cognitive load.                                                                          & \ref{findings-user-feedback-hephaistus}   \\

B4 & Agents that follow user behavior over time offer the potential for \textbf{proactively providing reminders, hinting at inconsistencies, and suggestions for metacognitive support, tool operation, and design task considerations}.                                                                           & \ref{findings-user-feedback-hephaistus}, \ref{findings-expert-support-strategies}  \\

B5 & \textbf{Visually signaling available agent feedback} for users to optionally engage in can reduce task interruptions.                                                                           & \ref{findings-expert-reflections} \\

\bottomrule
\end{tabular}
\label{tab:design_consoderations}
\end{table*}

\section{Discussion}

In the following sections, we discuss our findings and their implications for metacognitive design support systems and agent-based CAD support while highlighting key learnings, design considerations, and open questions for future GenAI design support systems (see Table \ref{tab:design_consoderations}).

\subsection{Toward Metacognitive Design Support Systems}

Our findings indicate that agent-facilitated metacognitive support can play a positive role in helping designers overcome the cognitive challenges of GenAI workflows: 
Designers receiving some form of support often had improved design outcomes compared to those without assistance\footnote{ 
Our result in the unsupported group aligns with previous findings using the same task with a similar population \cite{gmeiner_exploring_2023a}, and statistical tests also showed no significant differences in population characteristics between supported and unsupported groups in our study.}.
In our \rev{exploratory prototyping} study, we categorized support strategies into distinct agent probes to reveal nuanced benefits and tradeoffs, but also saw that none of our agents served as a one-size-fits-all solution.
This suggests that \textit{combining multiple strategies may ultimately prove more effective in practice}, and future work should explore systems with blended approaches. 

\rev{Below we reflect on the findings of this study and highlight design considerations for future metacognitive GenAI support systems (Table \ref{tab:design_consoderations}).}
Regarding specific metacognitive support strategies, our findings indicate that \textbf{(A1) cueing users with thought-provoking open-ended questions can help with intent formulation, problem exploration, and outcome evaluation, leading to improved AI-generated outcomes} (see \ref{findings-message-impact}---\ref{findings-socratais}, \ref{findings-expert-support-strategies}).
These findings align with prior evidence on the crucial role of questions within design processes \cite{eris_effective_2004, cardoso_question_2014, price_asking_2022}.
Similar to other recent work \cite{danry_dont_2023, park_thinking_2024}, our findings also emphasize AI agents’ possible role as facilitators that can stimulate users' critical thinking, which challenges the common notion of GenAI systems as ``oracles'' that only provide definitive (but possibly inaccurate) solutions or answers.  
However, going further, a challenge will lie in determining when ``asking'' versus ``telling'' the user would be most appropriate. 
Further investigations could draw on principles from learning science, suggesting metacognitive processing may only be effective if preceded by adequate knowledge or initial instruction \cite{berry_relationship_1984, hacker_metacognition_1998}.

Also, we saw that asking questions alone can have limitations: 
In our study, questions were less effective at challenging ingrained incorrect assumptions, indicating that guidance beyond questioning may sometimes be required, especially when users hold deep misconceptions.
Similarly, repeated questioning also presented a dual effect: while it often helped users to repair flawed inputs, it sometimes led to over-reliance, with designers thinking the AI might know something more than them or that the AI is right, rather than engaging in deeper reflection. 
This risk of dependency aligns with other findings on in-action feedback during design tasks, where excessive guidance was observed to diminish self-reflection and critical evaluation \cite{e_when_2024, zhang_cautionary_2021}. 
Future work should, therefore, explore \textit{when} and \textit{how} metacognitive support systems could provide assistance without increasing automation reliance. 

From a technical perspective, recent advancements in natural language processing (NLP) have enabled automated generation of Socratic questions for teaching math  \cite{shridhar_automatic_2022} or debugging 
 \cite{al-hossami_socratic_2023} and generating domain-specific educational questions by pre-training LLMs \cite{bulathwela_scalable_2023}.
Building atop such technical foundations, future research should explore design task-specific question generation to prompt designers’ self-reflection and critical thinking aligned to specific design domains. 
Furthermore, inspired by emerging process mining techniques focused on metacognition and self-regulated learning phases \cite{abdelshiheed_leveraging_2023, borchers_using_2024, zhang_using_2024}, future work could investigate ways to further optimize systems, for example by tailoring prompts to designers’ specific situational metacognitive needs, such as \textit{intent formulation}, \textit{problem exploration} or \textit{outcome evaluation} phases. 

Regarding specific metacognitive strategies, our analysis indicates that \textbf{(A2) prompting mental simulations through questions and sketching can assist designers to think through and more accurately formulate intents and specify GenAI model inputs} (see \ref{findings-socratais}, \ref{findings-hephaistus}). 
These findings align with previous research on design cognition, suggesting that mental simulation presents a vital metacognitive process in design activities \cite{ball_analogical_2009, ball_advancing_2019}. 
In addition, we saw that providing distinct support for thought externalization and visualization through sketching (a known cognitive amplification strategy in design \cite{babapour_roles_2015}) helped designers more carefully think through input specifications for the GenAI solver, thus improving designers' intent formulation and problem specification. 
Overall, metacognitive agent support might potentially be helpful for many design processes, whether GenAI-supported or not.
Based on these findings, future work should explore further metacognitive support mechanisms relevant to design within and outside of GenAI-assisted tasks, such as prompting mental simulations through questions or guiding users in gradually sketching and eliciting relevant input criteria.

While many questions may not have directly helped designers\footnote{\rev{On average only 6 out of 23 questions of \highsocratais{\textit{SocratAIs}}' sessions had an observable positive impact on the design process where the designer considered a new relevant aspect after receiving the message (see Table \ref{tab:results-summary-table}).}}, we saw that the \textit{right} reflective question at the \textit{right} time can have a significant impact on the design process. 
However, anticipating and catching the right moment can be tricky, but some situations seemed to be more opportune than others. 
For example, during the GenAI preview and outcome evaluation phases, reflective questions helped designers in assessing and correcting generated parts by linking back components' structural errors to insufficiently specified model inputs (see \ref{findings-socratais}). 
Likewise, some external expert wizards and users also emphasized that support during evaluation phases would be especially useful (see \ref{findings-expert-reflections} and \ref{findings-user-feedback-socratais}). 
Building atop this learning, future systems could \textbf{ (A3) offer users agent-driven “design review sessions” during part evaluation} (similar to reviews in design education or professional collaborations \cite{goldschmidt_design_2010, oh_theoretical_2013}) \textbf{and
introducing dedicated "reflection checkpoints" between GenAI setup steps} (e.g., when the user transitions from specifying loads to obstacle geometry) to better scaffold (metacognitive) support throughout GenAI-based design processes (see \ref{findings-expert-reflections}).

Our findings also revealed differences in user preferences between more confident and inexperienced designers regarding questions versus suggestions (see \ref{findings-feedback-general-agent-interactions}), highlighting the desire for systems to give users \textbf{(A4) control over the type of metacognitive support depending on their needs and experience level.}

Lastly, future metacognitive support systems could provide \textbf{(A5) custom-generated user-editable design checklists to support planning and scaffolding of design decisions} (see \ref{findings-hephaistus}, \ref{findings-user-feedback-hephaistus}).

\subsection{Opportunities for Agent-based CAD Support}
\rev{While our study primarily focused on supporting designers working with GenAI within CAD environments, the findings also revealed interesting insights and opportunities for designing agent-based CAD support systems that can complement metacognitive strategies.} 
For example, in our study, \highhephaistus{\textit{HephAIstus}}---in addition to its metacognitive planning and sketching support---had a positive impact on helping users work on the design task and software operation. 
However, \highhephaistus{\textit{HephAIstus}}' suggestions helped designers less to overcome GenAI-related cognitive challenges than \highsocratais{\textit{SocratAIs}}' questions.
Especially for supporting \textit{intent formulation} and \textit{problem exploration} related to load cases, \highsocratais{\textit{SocratAIs}} was twice as effective as \highhephaistus{\textit{HephAIstus}} (see Section \ref{findings-message-impact} and Figure \ref{fig:messages-impact-plot}), indicating that for intent formulation and problem exploration, questions paired with planning and sketching support might be more effective than suggestions. 
Consequently, we conclude that \textbf{(B1) metacognitive support through reflective questions, planning, and sketching is equally crucial for effectively supporting designers in GenAI tasks as providing suggestions for design decision and tool operation }\footnote{This is also indicated by the comparable quality of outcomes across the different agent-supported groups.} (see \ref{findings-expert-support-strategies}, \ref{findings-expert-reflections}). 

Furthermore, based on the insights derived from \highhephaistus{\textit{HephAIstus}} and \highexpert{\textit{Expert-Freeform}}, we see various opportunities for systems providing real-time support for design tasks and software operation, along with metacognitive support in CAD and GenAI design workflows.  
Notably, we see opportunities for \textbf{(B2) enabling users to directly request information and metacognitive support from agents verbally while working on a task}, which seemed to have helped users maintain focus while reducing context-switching (see \ref{findings-hephaistus}, \ref{findings-user-feedback-hephaistus}).

Similarly, users also appraised the agent’s \textbf{(B3) visual tool guidance by directly highlighting relevant GUI elements} (see \ref{findings-user-feedback-hephaistus}).
Recent NLP advancements in speech processing and synthesis \cite{mehrish_review_2023, triantafyllopoulos_overview_2023}, as well as the increasing ability of multimodal AI models to visually understand and operate software GUIs \cite{niu_screenagent_2024, hong_cogagent_2023}, provide promising foundations for future research to explore such multimodal conversational support agents further.

Additionally, by capturing and responding to user behavior, verbalizations, and (screen) context over time, \textbf{(B4) agents can proactively provide reminders, hinting at inconsistencies and suggestions for metacognitive support, tool operation, and design task considerations} (see \ref{findings-user-feedback-hephaistus}, \ref{findings-expert-support-strategies}). 
With multimodal LLM’s increasing context windows, such contextual longitudinal support seems to become increasingly feasible.

Lastly, instead of support agents always directly verbalizing their messages, future support systems could instead \textbf{(B5) visually signal available agent feedback for users to engage in if and when desired}, which could reduce task interruptions (see \ref{findings-expert-reflections}).

\subsection{Limitations}
We highlight the following limitations:
\rev{The study followed an exploratory prototyping approach \cite{zamfirescu-pereira_fake_2021} that enabled us to compare the different agents’ metacognitive strategies while allowing flexibility in how support was delivered (e.g., message timing and phrasing).
As a result, our design insights are partially shaped by the individual wizards (the first author and four external experts), and repeating the study with different wizards may yield slightly different outcomes.
Furthermore, to analyze the impact of agents on the design process, we used video interaction analysis to identify moments when participants visibly considered new, relevant aspects in response to agent messages. While this yielded valuable findings, future work could incorporate additional user-agent interaction dynamics to further surface complementary insights.
}
\rev{In terms of population}, our participants represent only a subset of engineering designers. 
While all participants had relevant training in design and experience with 3D CAD software, many had limited industry exposure. 
To address this imbalance, we included five professionals with more extensive industry experience.
Most of these professional users were part of the expert-facilitated agents, which might have biased the results. 
However, we disregarded this potential bias since the observed behaviors were similar across all supported groups. 
Furthermore, as the participants in our study were self-selected, they were likely interested in or receptive to GenAI systems. 
This openness to AI-supported work may have influenced some of our findings.
Additionally, although we aimed to ensure the design tasks felt realistic, participants knew they were part of a research study and that their designs wouldn’t be produced. 
They might have invested more time learning the tool and thinking through the problem to create practical designs in a real-world setting.

\section{Conclusion}
While GenAI tools promise to enhance design processes, many professionals struggle to work effectively with AI. 
Key challenges include specifying all design criteria upfront (intent formulation) and reduced cognitive engagement due to cognitive offloading, which can limit problem exploration and outcome evaluation. 
To address this, we explored metacognitive support agents in a Wizard of Oz user study. 
Our findings show that users with agent support developed more viable designs, though outcomes varied depending on support strategy. 
While designers recognized the benefits of such assistance, we also uncovered trade-offs and differing user preferences. 
Based on these results, we highlight opportunities and trade-offs of metacognitive support agents and implications for AI-based design tools. 
While this work explores metacognitive support agents for GenAI-assisted mechanical part creation, the findings and design considerations offer promising avenues for research in other AI-assisted workflows and insights for developing new support techniques for AI-based design applications.

\begin{acks}
We thank all study participants, Autodesk experts, and the research assistants Anna Xu and Claire Malella for supporting this work.
We would also like to thank Aniket Kittur, Christopher McComb, and Nur Yildirim for providing early feedback on a draft of this manuscript and the reviewers of this submission for their constructive suggestions.
This material is based upon work supported by the National Science Foundation under Grant No. \#2118924 Supporting Designers in Learning to Co-create with AI for Complex Computational Design Tasks.
\end{acks}

\bibliographystyle{ACM-Reference-Format}

\begin{thebibliography}{108}


\ifx \showCODEN    \undefined \def \showCODEN     #1{\unskip}     \fi
\ifx \showDOI      \undefined \def \showDOI       #1{#1}\fi
\ifx \showISBNx    \undefined \def \showISBNx     #1{\unskip}     \fi
\ifx \showISBNxiii \undefined \def \showISBNxiii  #1{\unskip}     \fi
\ifx \showISSN     \undefined \def \showISSN      #1{\unskip}     \fi
\ifx \showLCCN     \undefined \def \showLCCN      #1{\unskip}     \fi
\ifx \shownote     \undefined \def \shownote      #1{#1}          \fi
\ifx \showarticletitle \undefined \def \showarticletitle #1{#1}   \fi
\ifx \showURL      \undefined \def \showURL       {\relax}        \fi
\providecommand\bibfield[2]{#2}
\providecommand\bibinfo[2]{#2}
\providecommand\natexlab[1]{#1}
\providecommand\showeprint[2][]{arXiv:#2}

\bibitem[Abdelshiheed et~al\mbox{.}(2023)]%
        {abdelshiheed_leveraging_2023}
\bibfield{author}{\bibinfo{person}{Mark Abdelshiheed}, \bibinfo{person}{John~Wesley Hostetter}, \bibinfo{person}{Tiffany Barnes}, {and} \bibinfo{person}{Min Chi}.} \bibinfo{year}{2023}\natexlab{}.
\newblock \showarticletitle{Leveraging {{Deep Reinforcement Learning}} for {{Metacognitive Interventions Across Intelligent Tutoring Systems}}}.
\newblock In \bibinfo{booktitle}{\emph{Artificial {{Intelligence}} in {{Education}}}}, \bibfield{editor}{\bibinfo{person}{Ning Wang}, \bibinfo{person}{Genaro {Rebolledo-Mendez}}, \bibinfo{person}{Noboru Matsuda}, \bibinfo{person}{Olga~C. Santos}, {and} \bibinfo{person}{Vania Dimitrova}} (Eds.). Vol.~\bibinfo{volume}{13916}. \bibinfo{publisher}{Springer Nature Switzerland}, \bibinfo{address}{Cham}, \bibinfo{pages}{291--303}.
\newblock
\showISBNx{978-3-031-36271-2 978-3-031-36272-9}
\urldef\tempurl%
\url{https://doi.org/10.1007/978-3-031-36272-9_24}
\showDOI{\tempurl}


\bibitem[{Al-Hossami} et~al\mbox{.}(2023)]%
        {al-hossami_socratic_2023}
\bibfield{author}{\bibinfo{person}{Erfan {Al-Hossami}}, \bibinfo{person}{Razvan Bunescu}, \bibinfo{person}{Ryan Teehan}, \bibinfo{person}{Laurel Powell}, \bibinfo{person}{Khyati Mahajan}, {and} \bibinfo{person}{Mohsen Dorodchi}.} \bibinfo{year}{2023}\natexlab{}.
\newblock \showarticletitle{Socratic {{Questioning}} of {{Novice Debuggers}}: {{A Benchmark Dataset}} and {{Preliminary Evaluations}}}. In \bibinfo{booktitle}{\emph{Proceedings of the 18th {{Workshop}} on {{Innovative Use}} of {{NLP}} for {{Building Educational Applications}} ({{BEA}} 2023)}}. \bibinfo{publisher}{Association for Computational Linguistics}, \bibinfo{address}{Toronto, Canada}, \bibinfo{pages}{709--726}.
\newblock
\urldef\tempurl%
\url{https://doi.org/10.18653/v1/2023.bea-1.57}
\showDOI{\tempurl}


\bibitem[Alshaikh et~al\mbox{.}(2020)]%
        {alshaikh_experiments_2020}
\bibfield{author}{\bibinfo{person}{Zeyad Alshaikh}, \bibinfo{person}{L. Tamang}, {and} \bibinfo{person}{V. Rus}.} \bibinfo{year}{2020}\natexlab{}.
\newblock \showarticletitle{Experiments with a {{Socratic Intelligent Tutoring System}} for {{Source Code Understanding}}}. In \bibinfo{booktitle}{\emph{The {{Florida AI Research Society}}}}.
\newblock


\bibitem[Aurisicchio et~al\mbox{.}(2007)]%
        {aurisicchio_characterising_2007}
\bibfield{author}{\bibinfo{person}{Marco Aurisicchio}, \bibinfo{person}{Rob~H Bracewell}, \bibinfo{person}{Ken~M Wallace}, {et~al\mbox{.}}} \bibinfo{year}{2007}\natexlab{}.
\newblock \showarticletitle{Characterising Design Questions That Involve Reasoning}. In \bibinfo{booktitle}{\emph{{{DS}} 42: {{Proceedings}} of {{ICED}} 2007, the 16th {{International Conference}} on {{Engineering Design}}, {{Paris}}, {{France}}, 28.-31.07. 2007}}.
\newblock


\bibitem[{Autodesk}(2020)]%
        {autodesk_fusion_2020}
\bibfield{author}{\bibinfo{person}{{Autodesk}}.} \bibinfo{year}{2020}\natexlab{}.
\newblock \bibinfo{title}{Fusion 360 {{Generative Design}}}.
\newblock \bibinfo{howpublished}{https://www.autodesk.com/solutions/generative-design/manufacturing}.
\newblock


\bibitem[Ayres and Sweller(2005)]%
        {ayres_splitattention_2005}
\bibfield{author}{\bibinfo{person}{Paul Ayres} {and} \bibinfo{person}{John Sweller}.} \bibinfo{year}{2005}\natexlab{}.
\newblock \showarticletitle{The {{Split-Attention Principle}} in {{Multimedia Learning}}}.
\newblock In \bibinfo{booktitle}{\emph{The {{Cambridge Handbook}} of {{Multimedia Learning}}} (\bibinfo{edition}{1} ed.)}, \bibfield{editor}{\bibinfo{person}{Richard Mayer}} (Ed.). \bibinfo{publisher}{Cambridge University Press}, \bibinfo{pages}{135--146}.
\newblock
\showISBNx{978-0-521-83873-3 978-0-521-54751-2 978-0-511-81681-9}
\urldef\tempurl%
\url{https://doi.org/10.1017/CBO9780511816819.009}
\showDOI{\tempurl}


\bibitem[Babapour(2015)]%
        {babapour_roles_2015}
\bibfield{author}{\bibinfo{person}{Maral Babapour}.} \bibinfo{year}{2015}\natexlab{}.
\newblock \showarticletitle{Roles of Externalisation Activities in the Design Process}.
\newblock \bibinfo{journal}{\emph{Swedish Design Research Journal}}  \bibinfo{volume}{2014} (\bibinfo{date}{May} \bibinfo{year}{2015}).
\newblock
\urldef\tempurl%
\url{https://doi.org/10.3384/svid.2000-964x.14134}
\showDOI{\tempurl}


\bibitem[Ball and Christensen(2009)]%
        {ball_analogical_2009}
\bibfield{author}{\bibinfo{person}{Linden~J. Ball} {and} \bibinfo{person}{Bo~T. Christensen}.} \bibinfo{year}{2009}\natexlab{}.
\newblock \showarticletitle{Analogical Reasoning and Mental Simulation in Design: Two Strategies Linked to Uncertainty Resolution}.
\newblock \bibinfo{journal}{\emph{Design Studies}} \bibinfo{volume}{30}, \bibinfo{number}{2} (\bibinfo{date}{March} \bibinfo{year}{2009}), \bibinfo{pages}{169--186}.
\newblock
\showISSN{0142694X}
\urldef\tempurl%
\url{https://doi.org/10.1016/j.destud.2008.12.005}
\showDOI{\tempurl}


\bibitem[Ball and Christensen(2019)]%
        {ball_advancing_2019}
\bibfield{author}{\bibinfo{person}{Linden~J. Ball} {and} \bibinfo{person}{Bo~T. Christensen}.} \bibinfo{year}{2019}\natexlab{}.
\newblock \showarticletitle{Advancing an Understanding of Design Cognition and Design Metacognition: {{Progress}} and Prospects}.
\newblock \bibinfo{journal}{\emph{Design Studies}}  \bibinfo{volume}{65} (\bibinfo{date}{Nov.} \bibinfo{year}{2019}), \bibinfo{pages}{35--59}.
\newblock
\showISSN{0142694X}
\urldef\tempurl%
\url{https://doi.org/10.1016/j.destud.2019.10.003}
\showDOI{\tempurl}


\bibitem[Bannert et~al\mbox{.}(2014)]%
        {bannert_process_2014}
\bibfield{author}{\bibinfo{person}{Maria Bannert}, \bibinfo{person}{Peter Reimann}, {and} \bibinfo{person}{Christoph Sonnenberg}.} \bibinfo{year}{2014}\natexlab{}.
\newblock \showarticletitle{Process Mining Techniques for Analysing Patterns and Strategies in Students' Self-Regulated Learning}.
\newblock \bibinfo{journal}{\emph{Metacognition and Learning}} \bibinfo{volume}{9}, \bibinfo{number}{2} (\bibinfo{date}{Aug.} \bibinfo{year}{2014}), \bibinfo{pages}{161--185}.
\newblock
\showISSN{1556-1623, 1556-1631}
\urldef\tempurl%
\url{https://doi.org/10.1007/s11409-013-9107-6}
\showDOI{\tempurl}


\bibitem[Baumer and Tomlinson(2011)]%
        {baumer_comparing_2011}
\bibfield{author}{\bibinfo{person}{Eric~P.S. Baumer} {and} \bibinfo{person}{Bill Tomlinson}.} \bibinfo{year}{2011}\natexlab{}.
\newblock \showarticletitle{Comparing {{Activity Theory}} with {{Distributed Cognition}} for {{Video Analysis}}: {{Beyond}} "{{Kicking}} the {{Tires}}"}. In \bibinfo{booktitle}{\emph{Proceedings of the {{SIGCHI Conference}} on {{Human Factors}} in {{Computing Systems}}}} \emph{(\bibinfo{series}{{{CHI}} '11})}. \bibinfo{publisher}{Association for Computing Machinery}, \bibinfo{address}{New York, NY, USA}, \bibinfo{pages}{133--142}.
\newblock
\showISBNx{978-1-4503-0228-9}
\urldef\tempurl%
\url{https://doi.org/10.1145/1978942.1978962}
\showDOI{\tempurl}


\bibitem[Berger(2014)]%
        {berger_more_2014}
\bibfield{author}{\bibinfo{person}{Warren Berger}.} \bibinfo{year}{2014}\natexlab{}.
\newblock \bibinfo{booktitle}{\emph{A More Beautiful Question: The Power of Inquiry to Spark Breakthrough Ideas}}.
\newblock \bibinfo{publisher}{Bloomsbury USA}, \bibinfo{address}{New York, NY}.
\newblock
\showISBNx{978-1-62040-145-3}
\showLCCN{HD53 .B448 2014}


\bibitem[Berry and Broadbent(1984)]%
        {berry_relationship_1984}
\bibfield{author}{\bibinfo{person}{Dianne~C. Berry} {and} \bibinfo{person}{Donald~E. Broadbent}.} \bibinfo{year}{1984}\natexlab{}.
\newblock \showarticletitle{On the {{Relationship}} between {{Task Performance}} and {{Associated Verbalizable Knowledge}}}.
\newblock \bibinfo{journal}{\emph{The Quarterly Journal of Experimental Psychology Section A}} \bibinfo{volume}{36}, \bibinfo{number}{2} (\bibinfo{date}{May} \bibinfo{year}{1984}), \bibinfo{pages}{209--231}.
\newblock
\showISSN{0272-4987, 1464-0740}
\urldef\tempurl%
\url{https://doi.org/10.1080/14640748408402156}
\showDOI{\tempurl}


\bibitem[Boehner et~al\mbox{.}(2007)]%
        {boehner_how_2007b}
\bibfield{author}{\bibinfo{person}{Kirsten Boehner}, \bibinfo{person}{Janet Vertesi}, \bibinfo{person}{Phoebe Sengers}, {and} \bibinfo{person}{Paul Dourish}.} \bibinfo{year}{2007}\natexlab{}.
\newblock \showarticletitle{How {{HCI}} Interprets the Probes}. In \bibinfo{booktitle}{\emph{Proceedings of the {{SIGCHI Conference}} on {{Human Factors}} in {{Computing Systems}}}} \emph{(\bibinfo{series}{{{CHI}} '07})}. \bibinfo{publisher}{Association for Computing Machinery}, \bibinfo{address}{New York, NY, USA}, \bibinfo{pages}{1077--1086}.
\newblock
\showISBNx{978-1-59593-593-9}
\urldef\tempurl%
\url{https://doi.org/10.1145/1240624.1240789}
\showDOI{\tempurl}


\bibitem[Borchers et~al\mbox{.}(2024)]%
        {borchers_using_2024}
\bibfield{author}{\bibinfo{person}{Conrad Borchers}, \bibinfo{person}{Jiayi Zhang}, \bibinfo{person}{Ryan~S. Baker}, {and} \bibinfo{person}{Vincent Aleven}.} \bibinfo{year}{2024}\natexlab{}.
\newblock \showarticletitle{Using {{Think-Aloud Data}} to {{Understand Relations}} between {{Self-Regulation Cycle Characteristics}} and {{Student Performance}} in {{Intelligent Tutoring Systems}}}. In \bibinfo{booktitle}{\emph{Proceedings of the 14th {{Learning Analytics}} and {{Knowledge Conference}}}}. \bibinfo{pages}{529--539}.
\newblock
\urldef\tempurl%
\url{https://doi.org/10.1145/3636555.3636911}
\showDOI{\tempurl}
\showeprint[arxiv]{2312.05675}~[cs]


\bibitem[Braun and Clarke(2019)]%
        {braun_reflecting_2019}
\bibfield{author}{\bibinfo{person}{Virginia Braun} {and} \bibinfo{person}{Victoria Clarke}.} \bibinfo{year}{2019}\natexlab{}.
\newblock \showarticletitle{Reflecting on Reflexive Thematic Analysis}.
\newblock \bibinfo{journal}{\emph{Qualitative Research in Sport, Exercise and Health}} \bibinfo{volume}{11}, \bibinfo{number}{4} (\bibinfo{date}{Aug.} \bibinfo{year}{2019}), \bibinfo{pages}{589--597}.
\newblock
\showISSN{2159-676X, 2159-6778}
\urldef\tempurl%
\url{https://doi.org/10.1080/2159676X.2019.1628806}
\showDOI{\tempurl}


\bibitem[Bulathwela et~al\mbox{.}(2023)]%
        {bulathwela_scalable_2023}
\bibfield{author}{\bibinfo{person}{Sahan Bulathwela}, \bibinfo{person}{Hamze Muse}, {and} \bibinfo{person}{Emine Yilmaz}.} \bibinfo{year}{2023}\natexlab{}.
\newblock \showarticletitle{Scalable {{Educational Question Generation}} with {{Pre-trained Language Models}}}.
\newblock In \bibinfo{booktitle}{\emph{Artificial {{Intelligence}} in {{Education}}}}, \bibfield{editor}{\bibinfo{person}{Ning Wang}, \bibinfo{person}{Genaro {Rebolledo-Mendez}}, \bibinfo{person}{Noboru Matsuda}, \bibinfo{person}{Olga~C. Santos}, {and} \bibinfo{person}{Vania Dimitrova}} (Eds.). Vol.~\bibinfo{volume}{13916}. \bibinfo{publisher}{Springer Nature Switzerland}, \bibinfo{address}{Cham}, \bibinfo{pages}{327--339}.
\newblock
\showISBNx{978-3-031-36271-2 978-3-031-36272-9}
\urldef\tempurl%
\url{https://doi.org/10.1007/978-3-031-36272-9_27}
\showDOI{\tempurl}


\bibitem[{{\c C}etin T{\"u}nger} et~al\mbox{.}(2020)]%
        {cetintunger_comparison_2020}
\bibfield{author}{\bibinfo{person}{{{\c C}etin T{\"u}nger}}, \bibinfo{person}{{\c C}etin T{\"u}nger}, \bibinfo{person}{{{\c S}ule Ta{\c s}l{\i} Pekta{\c s}}}, {and} \bibinfo{person}{Sule~Tasli Pektas}.} \bibinfo{year}{2020}\natexlab{}.
\newblock \showarticletitle{A Comparison of the Cognitive Actions of Designers in Geometry-Based and Parametric Design Environments}.
\newblock \bibinfo{journal}{\emph{Open House International}}  \bibinfo{volume}{45} (\bibinfo{date}{June} \bibinfo{year}{2020}), \bibinfo{pages}{87--101}.
\newblock
\urldef\tempurl%
\url{https://doi.org/10.1108/ohi-04-2020-0008}
\showDOI{\tempurl}


\bibitem[Cardoso et~al\mbox{.}(2014)]%
        {cardoso_question_2014}
\bibfield{author}{\bibinfo{person}{Carlos Cardoso}, \bibinfo{person}{Ozgur Eris}, \bibinfo{person}{Petra {Badke-Schaub}}, {and} \bibinfo{person}{Marco Aurisicchio}.} \bibinfo{year}{2014}\natexlab{}.
\newblock \showarticletitle{Question Asking in Design Reviews: How Does Inquiry Facilitate the Learning Interaction?}. In \bibinfo{booktitle}{\emph{Proceedings of the 10th {{Design}} Thinking Research Symposium ({{DTRS}})}}. \bibinfo{publisher}{Purdue University}, \bibinfo{pages}{18}.
\newblock


\bibitem[{Castro-Alonso} and Sweller(2020)]%
        {castro_modality_2020}
\bibfield{author}{\bibinfo{person}{Juan~C. {Castro-Alonso}} {and} \bibinfo{person}{John Sweller}.} \bibinfo{year}{2020}\natexlab{}.
\newblock \showarticletitle{The Modality Effect of Cognitive Load Theory}. In \bibinfo{booktitle}{\emph{Advances in Human Factors in Training, Education, and Learning Sciences}}, \bibfield{editor}{\bibinfo{person}{Waldemar Karwowski}, \bibinfo{person}{Tareq Ahram}, {and} \bibinfo{person}{Salman Nazir}} (Eds.). \bibinfo{publisher}{Springer International Publishing}, \bibinfo{address}{Cham}, \bibinfo{pages}{75--84}.
\newblock
\showISBNx{978-3-030-20135-7}


\bibitem[Chase et~al\mbox{.}(2015)]%
        {chase_development_2015}
\bibfield{author}{\bibinfo{person}{Catherine~C. Chase}, \bibinfo{person}{Jenna Marks}, \bibinfo{person}{Deena Bernett}, \bibinfo{person}{Melissa Bradley}, {and} \bibinfo{person}{Vincent Aleven}.} \bibinfo{year}{2015}\natexlab{}.
\newblock \showarticletitle{Towards the {{Development}} of the {{Invention Coach}}: A {{Naturalistic Study}} of {{Teacher Guidance}} for an {{Exploratory Learning Task}}}.
\newblock In \bibinfo{booktitle}{\emph{Artificial {{Intelligence}} in {{Education}}}}, \bibfield{editor}{\bibinfo{person}{Cristina Conati}, \bibinfo{person}{Neil Heffernan}, \bibinfo{person}{Antonija Mitrovic}, {and} \bibinfo{person}{M.~Felisa Verdejo}} (Eds.). Vol.~\bibinfo{volume}{9112}. \bibinfo{publisher}{Springer International Publishing}, \bibinfo{address}{Cham}, \bibinfo{pages}{558--561}.
\newblock
\showISBNx{978-3-319-19772-2 978-3-319-19773-9}
\urldef\tempurl%
\url{https://doi.org/10.1007/978-3-319-19773-9_61}
\showDOI{\tempurl}


\bibitem[Chaudhury et~al\mbox{.}(2023)]%
        {chaudhury_exploring_2023}
\bibfield{author}{\bibinfo{person}{Rimika Chaudhury}, \bibinfo{person}{Taha Liaqat}, {and} \bibinfo{person}{Parmit~K. Chilana}.} \bibinfo{year}{2023}\natexlab{}.
\newblock \showarticletitle{Exploring the {{Needs}} of {{Informal Learners}} of {{Computational Skills}}: {{Probe-Based Elicitation}} for the {{Design}} of {{Self-Monitoring Interventions}}}.
\newblock


\bibitem[Chen et~al\mbox{.}(2018)]%
        {chen_forte_2018}
\bibfield{author}{\bibinfo{person}{Xiang~'Anthony' Chen}, \bibinfo{person}{Ye Tao}, \bibinfo{person}{Guanyun Wang}, \bibinfo{person}{Runchang Kang}, \bibinfo{person}{Tovi Grossman}, \bibinfo{person}{Stelian Coros}, {and} \bibinfo{person}{Scott~E. Hudson}.} \bibinfo{year}{2018}\natexlab{}.
\newblock \showarticletitle{Forte: {{User-Driven Generative Design}}}. In \bibinfo{booktitle}{\emph{Proceedings of the 2018 {{CHI Conference}} on {{Human Factors}} in {{Computing Systems}}}}. \bibinfo{publisher}{ACM}, \bibinfo{address}{Montreal QC Canada}, \bibinfo{pages}{1--12}.
\newblock
\showISBNx{978-1-4503-5620-6}
\urldef\tempurl%
\url{https://doi.org/10.1145/3173574.3174070}
\showDOI{\tempurl}


\bibitem[Chi(2006)]%
        {chi_laboratory_2006}
\bibfield{author}{\bibinfo{person}{Michelene T.~H. Chi}.} \bibinfo{year}{2006}\natexlab{}.
\newblock \showarticletitle{Laboratory {{Methods}} for {{Assessing Experts}}' and {{Novices}}' {{Knowledge}}}.
\newblock In \bibinfo{booktitle}{\emph{The {{Cambridge Handbook}} of {{Expertise}} and {{Expert Performance}}}}, \bibfield{editor}{\bibinfo{person}{K.~Anders Ericsson}, \bibinfo{person}{Neil Charness}, \bibinfo{person}{Paul~J. Feltovich}, {and} \bibinfo{person}{Robert~R. Hoffman}} (Eds.). \bibinfo{publisher}{Cambridge University Press}, \bibinfo{address}{Cambridge}, \bibinfo{pages}{167--184}.
\newblock
\showISBNx{978-0-511-81679-6}
\urldef\tempurl%
\url{https://doi.org/10.1017/CBO9780511816796.010}
\showDOI{\tempurl}


\bibitem[Chilana et~al\mbox{.}(2018)]%
        {chilana_supporting_2018}
\bibfield{author}{\bibinfo{person}{Parmit~K. Chilana}, \bibinfo{person}{Nathaniel Hudson}, \bibinfo{person}{Srinjita Bhaduri}, \bibinfo{person}{Prashant Shashikumar}, {and} \bibinfo{person}{Shaun~K. Kane}.} \bibinfo{year}{2018}\natexlab{}.
\newblock \showarticletitle{Supporting {{Remote Real-Time Expert Help}}: {{Opportunities}} and {{Challenges}} for {{Novice 3D Modelers}}}.
\newblock  (\bibinfo{date}{Oct.} \bibinfo{year}{2018}), \bibinfo{pages}{157--166}.
\newblock
\urldef\tempurl%
\url{https://doi.org/10.1109/vlhcc.2018.8506568}
\showDOI{\tempurl}


\bibitem[Coimbra~Cardoso et~al\mbox{.}(2016)]%
        {Coimbra_inflection_2016}
\bibfield{author}{\bibinfo{person}{Carlos Coimbra~Cardoso}, \bibinfo{person}{Petra {Badke-Schaub}}, {and} \bibinfo{person}{Ozgur Eris}.} \bibinfo{year}{2016}\natexlab{}.
\newblock \showarticletitle{Inflection Moments in Design Discourse: {{How}} Questions Drive Problem Framing during Idea Generation}.
\newblock \bibinfo{journal}{\emph{Design Studies}}  \bibinfo{volume}{46} (\bibinfo{date}{Sept.} \bibinfo{year}{2016}), \bibinfo{pages}{59--78}.
\newblock
\showISSN{0142-694X}
\urldef\tempurl%
\url{https://doi.org/10.1016/j.destud.2016.07.002}
\showDOI{\tempurl}


\bibitem[Craig et~al\mbox{.}(2006)]%
        {craig_deeplevelreasoningquestion_2006}
\bibfield{author}{\bibinfo{person}{Scotty~D. Craig}, \bibinfo{person}{Jeremiah Sullins}, \bibinfo{person}{Amy Witherspoon}, {and} \bibinfo{person}{Barry Gholson}.} \bibinfo{year}{2006}\natexlab{}.
\newblock \showarticletitle{The {{Deep-Level-Reasoning-Question Effect}}: {{The Role}} of {{Dialogue}} and {{Deep-Level-Reasoning Questions During Vicarious Learning}}}.
\newblock \bibinfo{journal}{\emph{Cognition and Instruction}} \bibinfo{volume}{24}, \bibinfo{number}{4} (\bibinfo{year}{2006}), \bibinfo{pages}{565--591}.
\newblock
\showISSN{0737-0008}
\showeprint[jstor]{27739846}


\bibitem[Cross(2006)]%
        {cross_designerly_2006}
\bibfield{author}{\bibinfo{person}{Nigel Cross}.} \bibinfo{year}{2006}\natexlab{}.
\newblock \bibinfo{booktitle}{\emph{Designerly Ways of Knowing}}.
\newblock \bibinfo{publisher}{Springer}, \bibinfo{address}{Berlin London}.
\newblock
\showISBNx{978-1-84628-300-0}


\bibitem[Dahlb{\"a}ck et~al\mbox{.}(1993)]%
        {dahlback_wizard_1993}
\bibfield{author}{\bibinfo{person}{Nils Dahlb{\"a}ck}, \bibinfo{person}{Arne J{\"o}nsson}, {and} \bibinfo{person}{Lars Ahrenberg}.} \bibinfo{year}{1993}\natexlab{}.
\newblock \showarticletitle{Wizard of {{Oz}} Studies --- Why and How}.
\newblock \bibinfo{journal}{\emph{Knowledge-Based Systems}} \bibinfo{volume}{6}, \bibinfo{number}{4} (\bibinfo{date}{Dec.} \bibinfo{year}{1993}), \bibinfo{pages}{258--266}.
\newblock
\showISSN{09507051}
\urldef\tempurl%
\url{https://doi.org/10.1016/0950-7051(93)90017-N}
\showDOI{\tempurl}


\bibitem[Danry et~al\mbox{.}(2023)]%
        {danry_dont_2023}
\bibfield{author}{\bibinfo{person}{Valdemar Danry}, \bibinfo{person}{Pat Pataranutaporn}, \bibinfo{person}{Yaoli Mao}, {and} \bibinfo{person}{Pattie Maes}.} \bibinfo{year}{2023}\natexlab{}.
\newblock \showarticletitle{Don't {{Just Tell Me}}, {{Ask Me}}: {{AI Systems}} That {{Intelligently Frame Explanations}} as {{Questions Improve Human Logical Discernment Accuracy}} over {{Causal AI}} Explanations}. In \bibinfo{booktitle}{\emph{Proceedings of the 2023 {{CHI Conference}} on {{Human Factors}} in {{Computing Systems}}}}. \bibinfo{publisher}{ACM}, \bibinfo{address}{Hamburg Germany}, \bibinfo{pages}{1--13}.
\newblock
\showISBNx{978-1-4503-9421-5}
\urldef\tempurl%
\url{https://doi.org/10.1145/3544548.3580672}
\showDOI{\tempurl}


\bibitem[Davis et~al\mbox{.}(2015)]%
        {davis_drawing_2015}
\bibfield{author}{\bibinfo{person}{Nicholas Davis}, \bibinfo{person}{Chih-PIn Hsiao}, \bibinfo{person}{Kunwar~Yashraj Singh}, \bibinfo{person}{Lisa Li}, \bibinfo{person}{Sanat Moningi}, {and} \bibinfo{person}{Brian Magerko}.} \bibinfo{year}{2015}\natexlab{}.
\newblock \showarticletitle{Drawing {{Apprentice}}: {{An Enactive Co-Creative Agent}} for {{Artistic Collaboration}}}. In \bibinfo{booktitle}{\emph{Proceedings of the 2015 {{ACM SIGCHI Conference}} on {{Creativity}} and {{Cognition}}}}. \bibinfo{publisher}{ACM}, \bibinfo{address}{Glasgow United Kingdom}, \bibinfo{pages}{185--186}.
\newblock
\showISBNx{978-1-4503-3598-0}
\urldef\tempurl%
\url{https://doi.org/10.1145/2757226.2764555}
\showDOI{\tempurl}


\bibitem[DeLiema et~al\mbox{.}(2019)]%
        {deliema_debugging_2019}
\bibfield{author}{\bibinfo{person}{David DeLiema}, \bibinfo{person}{Maggie Dahn}, \bibinfo{person}{Virginia~J. Flood}, \bibinfo{person}{Ana Asuncion}, \bibinfo{person}{Dor Abrahamson}, \bibinfo{person}{Noel Enyedy}, {and} \bibinfo{person}{Francis Steen}.} \bibinfo{year}{2019}\natexlab{}.
\newblock \bibinfo{booktitle}{\emph{Debugging as a {{Context}} for {{Fostering Reflection}} on {{Critical Thinking}} and {{Emotion}}} (\bibinfo{edition}{1} ed.)}.
\newblock \bibinfo{publisher}{Routledge}, \bibinfo{address}{London}, \bibinfo{pages}{209--228}.
\newblock
\showISBNx{978-0-429-32305-8}
\urldef\tempurl%
\url{https://doi.org/10.4324/9780429323058-13}
\showDOI{\tempurl}


\bibitem[Dillon(1984)]%
        {dillon_classification_1984}
\bibfield{author}{\bibinfo{person}{J.~T. Dillon}.} \bibinfo{year}{1984}\natexlab{}.
\newblock \showarticletitle{The {{Classification}} of {{Research Questions}}}.
\newblock \bibinfo{journal}{\emph{Review of Educational Research}} \bibinfo{volume}{54}, \bibinfo{number}{3} (\bibinfo{date}{Sept.} \bibinfo{year}{1984}), \bibinfo{pages}{327--361}.
\newblock
\showISSN{0034-6543, 1935-1046}
\urldef\tempurl%
\url{https://doi.org/10.3102/00346543054003327}
\showDOI{\tempurl}


\bibitem[Dorst and Cross(2001)]%
        {dorst_creativity_2001}
\bibfield{author}{\bibinfo{person}{Kees Dorst} {and} \bibinfo{person}{Nigel Cross}.} \bibinfo{year}{2001}\natexlab{}.
\newblock \showarticletitle{Creativity in the Design Process: Co-Evolution of Problem--Solution}.
\newblock \bibinfo{journal}{\emph{Design Studies}} \bibinfo{volume}{22}, \bibinfo{number}{5} (\bibinfo{date}{Sept.} \bibinfo{year}{2001}), \bibinfo{pages}{425--437}.
\newblock
\showISSN{0142694X}
\urldef\tempurl%
\url{https://doi.org/10.1016/S0142-694X(01)00009-6}
\showDOI{\tempurl}


\bibitem[Dove et~al\mbox{.}(2016)]%
        {dove_argument_2016a}
\bibfield{author}{\bibinfo{person}{Graham Dove}, \bibinfo{person}{Nicolai~Brodersen Hansen}, {and} \bibinfo{person}{Kim Halskov}.} \bibinfo{year}{2016}\natexlab{}.
\newblock \showarticletitle{An {{Argument For Design Space Reflection}}}. In \bibinfo{booktitle}{\emph{Proceedings of the 9th {{Nordic Conference}} on {{Human-Computer Interaction}}}} \emph{(\bibinfo{series}{{{NordiCHI}} '16})}. \bibinfo{publisher}{Association for Computing Machinery}, \bibinfo{address}{New York, NY, USA}, \bibinfo{pages}{1--10}.
\newblock
\showISBNx{978-1-4503-4763-1}
\urldef\tempurl%
\url{https://doi.org/10.1145/2971485.2971528}
\showDOI{\tempurl}


\bibitem[Drosos et~al\mbox{.}(2024)]%
        {drosos_its_2024}
\bibfield{author}{\bibinfo{person}{Ian Drosos}, \bibinfo{person}{Advait Sarkar}, \bibinfo{person}{Xiaotong Xu}, \bibinfo{person}{Carina Negreanu}, \bibinfo{person}{Sean Rintel}, {and} \bibinfo{person}{Lev Tankelevitch}.} \bibinfo{year}{2024}\natexlab{}.
\newblock \showarticletitle{"{{It}}'s like a Rubber Duck That Talks Back": {{Understanding Generative AI-Assisted Data Analysis Workflows}} through a {{Participatory Prompting Study}}}. In \bibinfo{booktitle}{\emph{Proceedings of the 3rd {{Annual Meeting}} of the {{Symposium}} on {{Human-Computer Interaction}} for {{Work}}}} \emph{(\bibinfo{series}{{{CHIWORK}} '24})}. \bibinfo{publisher}{Association for Computing Machinery}, \bibinfo{address}{New York, NY, USA}, \bibinfo{pages}{1--21}.
\newblock
\showISBNx{979-8-4007-1017-9}
\urldef\tempurl%
\url{https://doi.org/10.1145/3663384.3663389}
\showDOI{\tempurl}


\bibitem[E et~al\mbox{.}(2024)]%
        {e_when_2024}
\bibfield{author}{\bibinfo{person}{Jane~L. E}, \bibinfo{person}{Yu-Chun~Grace Yen}, \bibinfo{person}{Isabelle~Yan Pan}, \bibinfo{person}{Grace Lin}, \bibinfo{person}{Mingyi Li}, \bibinfo{person}{Hyoungwook Jin}, \bibinfo{person}{Mengyi Chen}, \bibinfo{person}{Haijun Xia}, {and} \bibinfo{person}{Steven~P. Dow}.} \bibinfo{year}{2024}\natexlab{}.
\newblock \showarticletitle{When to {{Give Feedback}}: {{Exploring Tradeoffs}} in the {{Timing}} of {{Design Feedback}}}. In \bibinfo{booktitle}{\emph{Proceedings of the 16th {{Conference}} on {{Creativity}} \& {{Cognition}}}} \emph{(\bibinfo{series}{C\&amp;{{C}} '24})}. \bibinfo{publisher}{Association for Computing Machinery}, \bibinfo{address}{New York, NY, USA}, \bibinfo{pages}{292--310}.
\newblock
\showISBNx{979-8-4007-0485-7}
\urldef\tempurl%
\url{https://doi.org/10.1145/3635636.3656183}
\showDOI{\tempurl}


\bibitem[Elder and Paul(2016)]%
        {elder_thinkers_2016}
\bibfield{author}{\bibinfo{person}{Linda Elder} {and} \bibinfo{person}{Richard Paul}.} \bibinfo{year}{2016}\natexlab{}.
\newblock \bibinfo{booktitle}{\emph{The {{Thinker}}'s {{Guide}} to {{The Art}} of {{Socratic Questioning}}}}.
\newblock \bibinfo{publisher}{Foundation for Critical Thinking Press}.
\newblock


\bibitem[Eris(2004)]%
        {eris_effective_2004}
\bibfield{author}{\bibinfo{person}{Ozgur Eris}.} \bibinfo{year}{2004}\natexlab{}.
\newblock \bibinfo{booktitle}{\emph{Effective {{Inquiry}} for {{Innovative Engineering Design}}}}.
\newblock \bibinfo{publisher}{Springer US}, \bibinfo{address}{Boston, MA}.
\newblock
\showISBNx{978-1-4613-4729-3 978-1-4419-8943-7}
\urldef\tempurl%
\url{https://doi.org/10.1007/978-1-4419-8943-7}
\showDOI{\tempurl}


\bibitem[Flavell(1979)]%
        {flavell_metacognition_1979}
\bibfield{author}{\bibinfo{person}{John~H. Flavell}.} \bibinfo{year}{1979}\natexlab{}.
\newblock \showarticletitle{Metacognition and Cognitive Monitoring: {{A}} New Area of Cognitive--Developmental Inquiry.}
\newblock \bibinfo{journal}{\emph{American Psychologist}} \bibinfo{volume}{34}, \bibinfo{number}{10} (\bibinfo{date}{Oct.} \bibinfo{year}{1979}), \bibinfo{pages}{906--911}.
\newblock
\showISSN{1935-990X, 0003-066X}
\urldef\tempurl%
\url{https://doi.org/10.1037/0003-066X.34.10.906}
\showDOI{\tempurl}


\bibitem[{formlabs}(2020)]%
        {formlabs_generative_2020}
\bibfield{author}{\bibinfo{person}{{formlabs}}.} \bibinfo{year}{2020}\natexlab{}.
\newblock \bibinfo{title}{Generative {{Design}} 101}.
\newblock \bibinfo{howpublished}{https://formlabs.com/blog/generative-design/}.
\newblock


\bibitem[Gmeiner et~al\mbox{.}(2023)]%
        {gmeiner_exploring_2023a}
\bibfield{author}{\bibinfo{person}{Frederic Gmeiner}, \bibinfo{person}{Humphrey Yang}, \bibinfo{person}{Lining Yao}, \bibinfo{person}{Kenneth Holstein}, {and} \bibinfo{person}{Nikolas Martelaro}.} \bibinfo{year}{2023}\natexlab{}.
\newblock \showarticletitle{Exploring {{Challenges}} and {{Opportunities}} to {{Support Designers}} in {{Learning}} to {{Co-create}} with {{AI-based Manufacturing Design Tools}}}. In \bibinfo{booktitle}{\emph{Proceedings of the 2023 {{CHI Conference}} on {{Human Factors}} in {{Computing Systems}}}}. \bibinfo{publisher}{ACM}, \bibinfo{address}{Hamburg Germany}, \bibinfo{pages}{1--20}.
\newblock
\showISBNx{978-1-4503-9421-5}
\urldef\tempurl%
\url{https://doi.org/10.1145/3544548.3580999}
\showDOI{\tempurl}


\bibitem[Goldschmidt et~al\mbox{.}(2010)]%
        {goldschmidt_design_2010}
\bibfield{author}{\bibinfo{person}{Gabriela Goldschmidt}, \bibinfo{person}{Hagay Hochman}, {and} \bibinfo{person}{Itay Dafni}.} \bibinfo{year}{2010}\natexlab{}.
\newblock \showarticletitle{The Design Studio ``Crit'': {{Teacher}}--Student Communication}.
\newblock \bibinfo{journal}{\emph{Artificial Intelligence for Engineering Design, Analysis and Manufacturing}} \bibinfo{volume}{24}, \bibinfo{number}{3} (\bibinfo{date}{Aug.} \bibinfo{year}{2010}), \bibinfo{pages}{285--302}.
\newblock
\showISSN{0890-0604, 1469-1760}
\urldef\tempurl%
\url{https://doi.org/10.1017/S089006041000020X}
\showDOI{\tempurl}


\bibitem[Graesser and Person(1994)]%
        {graesser_question_1994}
\bibfield{author}{\bibinfo{person}{Arthur~C. Graesser} {and} \bibinfo{person}{Natalie~K. Person}.} \bibinfo{year}{1994}\natexlab{}.
\newblock \showarticletitle{Question {{Asking During Tutoring}}}.
\newblock \bibinfo{journal}{\emph{American Educational Research Journal}} \bibinfo{volume}{31}, \bibinfo{number}{1} (\bibinfo{date}{March} \bibinfo{year}{1994}), \bibinfo{pages}{104--137}.
\newblock
\showISSN{0002-8312, 1935-1011}
\urldef\tempurl%
\url{https://doi.org/10.3102/00028312031001104}
\showDOI{\tempurl}


\bibitem[Hacker(1998)]%
        {hacker_metacognition_1998}
\bibfield{editor}{\bibinfo{person}{Douglas~J. Hacker}} (Ed.). \bibinfo{year}{1998}\natexlab{}.
\newblock \bibinfo{booktitle}{\emph{Metacognition in Educational Theory and Practice}}.
\newblock \bibinfo{publisher}{Erlbaum}, \bibinfo{address}{Mahwah, NJ}.
\newblock
\showISBNx{978-0-8058-2482-7 978-0-8058-2481-0}


\bibitem[Hausmann and Vanlehn(2010)]%
        {hausmann_effect_2010}
\bibfield{author}{\bibinfo{person}{Robert Hausmann} {and} \bibinfo{person}{Kurt Vanlehn}.} \bibinfo{year}{2010}\natexlab{}.
\newblock \showarticletitle{The {{Effect}} of {{Self-Explaining}} on {{Robust Learning}}.}
\newblock \bibinfo{journal}{\emph{I. J. Artificial Intelligence in Education}}  \bibinfo{volume}{20} (\bibinfo{date}{Jan.} \bibinfo{year}{2010}), \bibinfo{pages}{303--332}.
\newblock
\urldef\tempurl%
\url{https://doi.org/10.3233/JAI-2010-010}
\showDOI{\tempurl}


\bibitem[Heirweg et~al\mbox{.}(2020)]%
        {heirweg_mine_2020}
\bibfield{author}{\bibinfo{person}{Sofie Heirweg}, \bibinfo{person}{Mona De~Smul}, \bibinfo{person}{Emmelien Merchie}, \bibinfo{person}{Geert Devos}, {and} \bibinfo{person}{Hilde Keer}.} \bibinfo{year}{2020}\natexlab{}.
\newblock \showarticletitle{Mine the Process: Investigating the Cyclical Nature of Upper Primary School Students' Self-Regulated Learning}.
\newblock \bibinfo{journal}{\emph{Instructional Science}}  \bibinfo{volume}{48} (\bibinfo{date}{Aug.} \bibinfo{year}{2020}).
\newblock
\urldef\tempurl%
\url{https://doi.org/10.1007/s11251-020-09519-0}
\showDOI{\tempurl}


\bibitem[Ho et~al\mbox{.}(2022)]%
        {ho_imagen_2022}
\bibfield{author}{\bibinfo{person}{Jonathan Ho}, \bibinfo{person}{William Chan}, \bibinfo{person}{Chitwan Saharia}, \bibinfo{person}{Jay Whang}, \bibinfo{person}{Ruiqi Gao}, \bibinfo{person}{Alexey Gritsenko}, \bibinfo{person}{Diederik~P. Kingma}, \bibinfo{person}{Ben Poole}, \bibinfo{person}{Mohammad Norouzi}, \bibinfo{person}{David~J. Fleet}, {and} \bibinfo{person}{Tim Salimans}.} \bibinfo{year}{2022}\natexlab{}.
\newblock \bibinfo{title}{Imagen {{Video}}: {{High Definition Video Generation}} with {{Diffusion Models}}}.
\newblock \bibinfo{howpublished}{https://arxiv.org/abs/2210.02303v1}.
\newblock


\bibitem[Ho et~al\mbox{.}(2023)]%
        {ho_thinking_2023}
\bibfield{author}{\bibinfo{person}{Yueh-Ren Ho}, \bibinfo{person}{Bao-Yu Chen}, {and} \bibinfo{person}{Chien-Ming Li}.} \bibinfo{year}{2023}\natexlab{}.
\newblock \showarticletitle{Thinking More Wisely: Using the {{Socratic}} Method to Develop Critical Thinking Skills amongst Healthcare Students}.
\newblock \bibinfo{journal}{\emph{BMC Medical Education}} \bibinfo{volume}{23}, \bibinfo{number}{1} (\bibinfo{date}{March} \bibinfo{year}{2023}), \bibinfo{pages}{173}.
\newblock
\showISSN{1472-6920}
\urldef\tempurl%
\url{https://doi.org/10.1186/s12909-023-04134-2}
\showDOI{\tempurl}


\bibitem[Hong et~al\mbox{.}(2023)]%
        {hong_cogagent_2023}
\bibfield{author}{\bibinfo{person}{Wenyi Hong}, \bibinfo{person}{Weihan Wang}, \bibinfo{person}{Qingsong Lv}, \bibinfo{person}{Jiazheng Xu}, \bibinfo{person}{Wenmeng Yu}, \bibinfo{person}{Junhui Ji}, \bibinfo{person}{Yan Wang}, \bibinfo{person}{Zihan Wang}, \bibinfo{person}{Yuxuan Zhang}, \bibinfo{person}{Juanzi Li}, \bibinfo{person}{Bin Xu}, \bibinfo{person}{Yuxiao Dong}, \bibinfo{person}{Ming Ding}, {and} \bibinfo{person}{Jie Tang}.} \bibinfo{year}{2023}\natexlab{}.
\newblock \bibinfo{title}{{{CogAgent}}: {{A Visual Language Model}} for {{GUI Agents}}}.
\newblock
\newblock
\showeprint[arxiv]{2312.08914}


\bibitem[Hurst et~al\mbox{.}(2023)]%
        {hurst_comparing_2023}
\bibfield{author}{\bibinfo{person}{Ada Hurst}, \bibinfo{person}{Shirley Lin}, \bibinfo{person}{Claire Treacy}, \bibinfo{person}{Oscar~G. Nespoli}, {and} \bibinfo{person}{John~S. Gero}.} \bibinfo{year}{2023}\natexlab{}.
\newblock \showarticletitle{Comparing {{Academics}} and {{Practitioners Q}}\&{{A Tutoring}} in the {{Engineering Design Studio}}}.
\newblock \bibinfo{journal}{\emph{Proceedings of the Design Society}}  \bibinfo{volume}{3} (\bibinfo{date}{July} \bibinfo{year}{2023}), \bibinfo{pages}{997--1006}.
\newblock
\showISSN{2732-527X}
\urldef\tempurl%
\url{https://doi.org/10.1017/pds.2023.100}
\showDOI{\tempurl}


\bibitem[Joshi et~al\mbox{.}(2020)]%
        {joshi_micromentor_2020}
\bibfield{author}{\bibinfo{person}{Nikhita Joshi}, \bibinfo{person}{Justin Matejka}, \bibinfo{person}{Fraser Anderson}, \bibinfo{person}{Tovi Grossman}, {and} \bibinfo{person}{George Fitzmaurice}.} \bibinfo{year}{2020}\natexlab{}.
\newblock \showarticletitle{{{MicroMentor}}: {{Peer-to-Peer Software Help Sessions}} in {{Three Minutes}} or {{Less}}}.
\newblock  (\bibinfo{date}{April} \bibinfo{year}{2020}), \bibinfo{pages}{1--13}.
\newblock
\urldef\tempurl%
\url{https://doi.org/10.1145/3313831.3376230}
\showDOI{\tempurl}


\bibitem[Kavousi et~al\mbox{.}(2020)]%
        {kavousi_role_2020}
\bibfield{author}{\bibinfo{person}{Shabnam Kavousi}, \bibinfo{person}{Patrick~A. Miller}, {and} \bibinfo{person}{Patricia~A. Alexander}.} \bibinfo{year}{2020}\natexlab{}.
\newblock \showarticletitle{The Role of Metacognition in the First-Year Design Lab}.
\newblock \bibinfo{journal}{\emph{Educational Technology Research and Development}} \bibinfo{volume}{68}, \bibinfo{number}{6} (\bibinfo{date}{Dec.} \bibinfo{year}{2020}), \bibinfo{pages}{3471--3494}.
\newblock
\showISSN{1042-1629, 1556-6501}
\urldef\tempurl%
\url{https://doi.org/10.1007/s11423-020-09848-4}
\showDOI{\tempurl}


\bibitem[Kazi et~al\mbox{.}(2017)]%
        {kazi_dreamsketch_2017}
\bibfield{author}{\bibinfo{person}{Rubaiat~Habib Kazi}, \bibinfo{person}{Tovi Grossman}, \bibinfo{person}{Hyunmin Cheong}, \bibinfo{person}{Ali Hashemi}, {and} \bibinfo{person}{George Fitzmaurice}.} \bibinfo{year}{2017}\natexlab{}.
\newblock \showarticletitle{{{DreamSketch}}: {{Early Stage 3D Design Explorations}} with {{Sketching}} and {{Generative Design}}}. In \bibinfo{booktitle}{\emph{Proceedings of the 30th {{Annual ACM Symposium}} on {{User Interface Software}} and {{Technology}}}}. \bibinfo{publisher}{ACM}, \bibinfo{address}{Qu{\'e}bec City QC Canada}, \bibinfo{pages}{401--414}.
\newblock
\showISBNx{978-1-4503-4981-9}
\urldef\tempurl%
\url{https://doi.org/10.1145/3126594.3126662}
\showDOI{\tempurl}


\bibitem[Khan and Tun{\c c}er(2019)]%
        {khan_gesture_2019}
\bibfield{author}{\bibinfo{person}{Sumbul Khan} {and} \bibinfo{person}{Bige Tun{\c c}er}.} \bibinfo{year}{2019}\natexlab{}.
\newblock \showarticletitle{Gesture and Speech Elicitation for {{3D CAD}} Modeling in Conceptual Design}.
\newblock \bibinfo{journal}{\emph{Automation in Construction}}  \bibinfo{volume}{106} (\bibinfo{date}{Oct.} \bibinfo{year}{2019}), \bibinfo{pages}{102847}.
\newblock
\showISSN{0926-5805}
\urldef\tempurl%
\url{https://doi.org/10.1016/j.autcon.2019.102847}
\showDOI{\tempurl}


\bibitem[Kim et~al\mbox{.}(2016)]%
        {kim_interaxis_2016}
\bibfield{author}{\bibinfo{person}{Hannah Kim}, \bibinfo{person}{Jaegul Choo}, \bibinfo{person}{Haesun Park}, {and} \bibinfo{person}{Alex Endert}.} \bibinfo{year}{2016}\natexlab{}.
\newblock \showarticletitle{{{InterAxis}}: {{Steering Scatterplot Axes}} via {{Observation-Level Interaction}}}.
\newblock \bibinfo{journal}{\emph{IEEE Transactions on Visualization and Computer Graphics}} \bibinfo{volume}{22}, \bibinfo{number}{1} (\bibinfo{date}{Jan.} \bibinfo{year}{2016}), \bibinfo{pages}{131--140}.
\newblock
\showISSN{1077-2626}
\urldef\tempurl%
\url{https://doi.org/10.1109/TVCG.2015.2467615}
\showDOI{\tempurl}


\bibitem[Ko and Myers(2004)]%
        {ko_designing_2004}
\bibfield{author}{\bibinfo{person}{Amy~J. Ko} {and} \bibinfo{person}{Brad~A. Myers}.} \bibinfo{year}{2004}\natexlab{}.
\newblock \showarticletitle{Designing the Whyline: A Debugging Interface for Asking Questions about Program Behavior}. In \bibinfo{booktitle}{\emph{Proceedings of the {{SIGCHI Conference}} on {{Human Factors}} in {{Computing Systems}}}}. \bibinfo{publisher}{ACM}, \bibinfo{address}{Vienna Austria}, \bibinfo{pages}{151--158}.
\newblock
\showISBNx{978-1-58113-702-6}
\urldef\tempurl%
\url{https://doi.org/10.1145/985692.985712}
\showDOI{\tempurl}


\bibitem[Kolko(2010)]%
        {kolko_abductive_2010}
\bibfield{author}{\bibinfo{person}{Jon Kolko}.} \bibinfo{year}{2010}\natexlab{}.
\newblock \showarticletitle{Abductive {{Thinking}} and {{Sensemaking}}: {{The Drivers}} of {{Design Synthesis}}}.
\newblock \bibinfo{journal}{\emph{Design Issues}} \bibinfo{volume}{26}, \bibinfo{number}{1} (\bibinfo{date}{Jan.} \bibinfo{year}{2010}), \bibinfo{pages}{15--28}.
\newblock
\showISSN{0747-9360}
\urldef\tempurl%
\url{https://doi.org/10.1162/desi.2010.26.1.15}
\showDOI{\tempurl}


\bibitem[Krosnick et~al\mbox{.}(2021)]%
        {krosnick_thinkaloud_2021}
\bibfield{author}{\bibinfo{person}{Rebecca Krosnick}, \bibinfo{person}{Fraser Anderson}, \bibinfo{person}{Justin Matejka}, \bibinfo{person}{Steve Oney}, \bibinfo{person}{Walter~S. Lasecki}, \bibinfo{person}{Tovi Grossman}, {and} \bibinfo{person}{George Fitzmaurice}.} \bibinfo{year}{2021}\natexlab{}.
\newblock \showarticletitle{Think-{{Aloud Computing}}: {{Supporting Rich}} and {{Low-Effort Knowledge Capture}}}.
\newblock \bibinfo{journal}{\emph{International Conference on Human Factors in Computing Systems}} (\bibinfo{date}{May} \bibinfo{year}{2021}).
\newblock
\urldef\tempurl%
\url{https://doi.org/10.1145/3411764.3445066}
\showDOI{\tempurl}


\bibitem[Ku and Ho(2010)]%
        {ku_metacognitive_2010}
\bibfield{author}{\bibinfo{person}{Kelly Y.~L. Ku} {and} \bibinfo{person}{Irene~T. Ho}.} \bibinfo{year}{2010}\natexlab{}.
\newblock \showarticletitle{Metacognitive Strategies That Enhance Critical Thinking}.
\newblock \bibinfo{journal}{\emph{Metacognition and Learning}} \bibinfo{volume}{5}, \bibinfo{number}{3} (\bibinfo{date}{Dec.} \bibinfo{year}{2010}), \bibinfo{pages}{251--267}.
\newblock
\showISSN{1556-1631}
\urldef\tempurl%
\url{https://doi.org/10.1007/s11409-010-9060-6}
\showDOI{\tempurl}


\bibitem[Kulesza et~al\mbox{.}(2015)]%
        {kulesza_principles_2015}
\bibfield{author}{\bibinfo{person}{Todd Kulesza}, \bibinfo{person}{Margaret Burnett}, \bibinfo{person}{Weng-Keen Wong}, {and} \bibinfo{person}{Simone Stumpf}.} \bibinfo{year}{2015}\natexlab{}.
\newblock \showarticletitle{Principles of {{Explanatory Debugging}} to {{Personalize Interactive Machine Learning}}}. In \bibinfo{booktitle}{\emph{Proceedings of the 20th {{International Conference}} on {{Intelligent User Interfaces}}}}. \bibinfo{publisher}{ACM}, \bibinfo{address}{Atlanta Georgia USA}, \bibinfo{pages}{126--137}.
\newblock
\showISBNx{978-1-4503-3306-1}
\urldef\tempurl%
\url{https://doi.org/10.1145/2678025.2701399}
\showDOI{\tempurl}


\bibitem[Kurt and Kurt(2017)]%
        {kurt_improving_2017}
\bibfield{author}{\bibinfo{person}{Mustafa Kurt} {and} \bibinfo{person}{Sevinc Kurt}.} \bibinfo{year}{2017}\natexlab{}.
\newblock \showarticletitle{Improving {{Design Understandings}} and {{Skills}} through {{Enhanced Metacognition}}: {{Reflective Design Journals}}}.
\newblock \bibinfo{journal}{\emph{International Journal of Art \& Design Education}} \bibinfo{volume}{36}, \bibinfo{number}{2} (\bibinfo{year}{2017}), \bibinfo{pages}{226--238}.
\newblock
\showISSN{1476-8070}
\urldef\tempurl%
\url{https://doi.org/10.1111/jade.12094}
\showDOI{\tempurl}


\bibitem[Kuusela and Paul(2000)]%
        {kuusela_comparison_2000}
\bibfield{author}{\bibinfo{person}{Hannu Kuusela} {and} \bibinfo{person}{Pallab Paul}.} \bibinfo{year}{2000}\natexlab{}.
\newblock \showarticletitle{A {{Comparison}} of {{Concurrent}} and {{Retrospective Verbal Protocol Analysis}}}.
\newblock \bibinfo{journal}{\emph{The American Journal of Psychology}} \bibinfo{volume}{113}, \bibinfo{number}{3} (\bibinfo{year}{2000}), \bibinfo{pages}{387}.
\newblock
\showISSN{00029556}
\urldef\tempurl%
\url{https://doi.org/10.2307/1423365}
\showDOI{\tempurl}
\showeprint[jstor]{1423365}


\bibitem[Lee and See(2004)]%
        {lee_trust_2004}
\bibfield{author}{\bibinfo{person}{J.~D. Lee} {and} \bibinfo{person}{K.~A. See}.} \bibinfo{year}{2004}\natexlab{}.
\newblock \showarticletitle{Trust in {{Automation}}: {{Designing}} for {{Appropriate Reliance}}}.
\newblock \bibinfo{journal}{\emph{Human Factors: The Journal of the Human Factors and Ergonomics Society}} \bibinfo{volume}{46}, \bibinfo{number}{1} (\bibinfo{date}{Jan.} \bibinfo{year}{2004}), \bibinfo{pages}{50--80}.
\newblock
\showISSN{0018-7208}
\urldef\tempurl%
\url{https://doi.org/10.1518/hfes.46.1.50_30392}
\showDOI{\tempurl}


\bibitem[Lehnert(1978)]%
        {lehnert_process_1978}
\bibfield{author}{\bibinfo{person}{Wendy~G. Lehnert}.} \bibinfo{year}{1978}\natexlab{}.
\newblock \bibinfo{booktitle}{\emph{The {{Process}} of {{Question Answering}}: {{A Computer Simulation}} of {{Cognition}}} (\bibinfo{edition}{1} ed.)}.
\newblock \bibinfo{publisher}{Routledge}, \bibinfo{address}{London}.
\newblock
\showISBNx{978-1-003-31681-7}
\urldef\tempurl%
\url{https://doi.org/10.4324/9781003316817}
\showDOI{\tempurl}


\bibitem[Matejka et~al\mbox{.}(2018)]%
        {matejka_dream_2018}
\bibfield{author}{\bibinfo{person}{Justin Matejka}, \bibinfo{person}{Michael Glueck}, \bibinfo{person}{Erin Bradner}, \bibinfo{person}{Ali Hashemi}, \bibinfo{person}{Tovi Grossman}, {and} \bibinfo{person}{George Fitzmaurice}.} \bibinfo{year}{2018}\natexlab{}.
\newblock \showarticletitle{Dream {{Lens}}: {{Exploration}} and {{Visualization}} of {{Large-Scale Generative Design Datasets}}}. In \bibinfo{booktitle}{\emph{Proceedings of the 2018 {{CHI Conference}} on {{Human Factors}} in {{Computing Systems}}}}. \bibinfo{publisher}{ACM}, \bibinfo{address}{Montreal QC Canada}, \bibinfo{pages}{1--12}.
\newblock
\showISBNx{978-1-4503-5620-6}
\urldef\tempurl%
\url{https://doi.org/10.1145/3173574.3173943}
\showDOI{\tempurl}


\bibitem[Matejka et~al\mbox{.}(2011)]%
        {matejka_ambient_2011}
\bibfield{author}{\bibinfo{person}{Justin Matejka}, \bibinfo{person}{Tovi Grossman}, {and} \bibinfo{person}{George Fitzmaurice}.} \bibinfo{year}{2011}\natexlab{}.
\newblock \showarticletitle{Ambient Help}. In \bibinfo{booktitle}{\emph{Proceedings of the {{SIGCHI Conference}} on {{Human Factors}} in {{Computing Systems}}}} \emph{(\bibinfo{series}{{{CHI}} '11})}. \bibinfo{publisher}{Association for Computing Machinery}, \bibinfo{address}{New York, NY, USA}, \bibinfo{pages}{2751--2760}.
\newblock
\showISBNx{978-1-4503-0228-9}
\urldef\tempurl%
\url{https://doi.org/10.1145/1978942.1979349}
\showDOI{\tempurl}


\bibitem[Mehrish et~al\mbox{.}(2023)]%
        {mehrish_review_2023}
\bibfield{author}{\bibinfo{person}{Ambuj Mehrish}, \bibinfo{person}{Navonil Majumder}, \bibinfo{person}{Rishabh Bhardwaj}, \bibinfo{person}{Rada Mihalcea}, {and} \bibinfo{person}{Soujanya Poria}.} \bibinfo{year}{2023}\natexlab{}.
\newblock \bibinfo{title}{A {{Review}} of {{Deep Learning Techniques}} for {{Speech Processing}}}.
\newblock
\newblock
\showeprint[arxiv]{2305.00359}


\bibitem[Menges and Ahlquist(2011)]%
        {menges_computational_2011}
\bibfield{editor}{\bibinfo{person}{Achim Menges} {and} \bibinfo{person}{Sean Ahlquist}} (Eds.). \bibinfo{year}{2011}\natexlab{}.
\newblock \bibinfo{booktitle}{\emph{Computational Design Thinking} (\bibinfo{edition}{1. publ} ed.)}.
\newblock \bibinfo{publisher}{Wiley}, \bibinfo{address}{Chichester}.
\newblock
\showISBNx{978-0-470-66570-1 978-0-470-66565-7}


\bibitem[Niu et~al\mbox{.}(2022)]%
        {niu_multimodal_2022}
\bibfield{author}{\bibinfo{person}{Hongwei Niu}, \bibinfo{person}{Cees Van~Leeuwen}, \bibinfo{person}{Jia Hao}, \bibinfo{person}{Guoxin Wang}, {and} \bibinfo{person}{Thomas Lachmann}.} \bibinfo{year}{2022}\natexlab{}.
\newblock \showarticletitle{Multimodal {{Natural Human}}--{{Computer Interfaces}} for {{Computer-Aided Design}}: {{A Review Paper}}}.
\newblock \bibinfo{journal}{\emph{Applied Sciences}} \bibinfo{volume}{12}, \bibinfo{number}{13} (\bibinfo{date}{June} \bibinfo{year}{2022}), \bibinfo{pages}{6510}.
\newblock
\showISSN{2076-3417}
\urldef\tempurl%
\url{https://doi.org/10.3390/app12136510}
\showDOI{\tempurl}


\bibitem[Niu et~al\mbox{.}(2024)]%
        {niu_screenagent_2024}
\bibfield{author}{\bibinfo{person}{Runliang Niu}, \bibinfo{person}{Jindong Li}, \bibinfo{person}{Shiqi Wang}, \bibinfo{person}{Yali Fu}, \bibinfo{person}{Xiyu Hu}, \bibinfo{person}{Xueyuan Leng}, \bibinfo{person}{He Kong}, \bibinfo{person}{Yi Chang}, {and} \bibinfo{person}{Qi Wang}.} \bibinfo{year}{2024}\natexlab{}.
\newblock \showarticletitle{{{ScreenAgent}}: {{A Vision Language Model-driven Computer Control Agent}}}.
\newblock  (\bibinfo{year}{2024}).
\newblock
\urldef\tempurl%
\url{https://doi.org/10.48550/ARXIV.2402.07945}
\showDOI{\tempurl}


\bibitem[Oh et~al\mbox{.}(2013)]%
        {oh_theoretical_2013}
\bibfield{author}{\bibinfo{person}{Yeonjoo Oh}, \bibinfo{person}{Suguru Ishizaki}, \bibinfo{person}{Mark~D. Gross}, {and} \bibinfo{person}{Ellen {Yi-Luen Do}}.} \bibinfo{year}{2013}\natexlab{}.
\newblock \showarticletitle{A Theoretical Framework of Design Critiquing in Architecture Studios}.
\newblock \bibinfo{journal}{\emph{Design Studies}} \bibinfo{volume}{34}, \bibinfo{number}{3} (\bibinfo{date}{May} \bibinfo{year}{2013}), \bibinfo{pages}{302--325}.
\newblock
\showISSN{0142-694X}
\urldef\tempurl%
\url{https://doi.org/10.1016/j.destud.2012.08.004}
\showDOI{\tempurl}


\bibitem[Panadero(2017)]%
        {panadero_review_2017}
\bibfield{author}{\bibinfo{person}{Ernesto Panadero}.} \bibinfo{year}{2017}\natexlab{}.
\newblock \showarticletitle{A {{Review}} of {{Self-regulated Learning}}: {{Six Models}} and {{Four Directions}} for {{Research}}}.
\newblock \bibinfo{journal}{\emph{Frontiers in Psychology}}  \bibinfo{volume}{8} (\bibinfo{date}{April} \bibinfo{year}{2017}), \bibinfo{pages}{422}.
\newblock
\showISSN{1664-1078}
\urldef\tempurl%
\url{https://doi.org/10.3389/fpsyg.2017.00422}
\showDOI{\tempurl}


\bibitem[Park et~al\mbox{.}(2024)]%
        {park_thinking_2024}
\bibfield{author}{\bibinfo{person}{Soya Park}, \bibinfo{person}{Hari Subramonyam}, {and} \bibinfo{person}{Chinmay Kulkarni}.} \bibinfo{year}{2024}\natexlab{}.
\newblock \bibinfo{title}{Thinking {{Assistants}}: {{LLM-Based Conversational Assistants}} That {{Help Users Think By Asking}} Rather than {{Answering}}}.
\newblock
\newblock
\urldef\tempurl%
\url{https://doi.org/10.48550/arXiv.2312.06024}
\showDOI{\tempurl}
\showeprint[arxiv]{2312.06024}~[cs]


\bibitem[Parreira et~al\mbox{.}(2023)]%
        {parreira_robot_2023}
\bibfield{author}{\bibinfo{person}{Maria~Teresa Parreira}, \bibinfo{person}{Sarah Gillet}, {and} \bibinfo{person}{Iolanda Leite}.} \bibinfo{year}{2023}\natexlab{}.
\newblock \showarticletitle{Robot {{Duck Debugging}}: {{Can Attentive Listening Improve Problem Solving}}?}. In \bibinfo{booktitle}{\emph{Proceedings of the 25th {{International Conference}} on {{Multimodal Interaction}}}} \emph{(\bibinfo{series}{{{ICMI}} '23})}. \bibinfo{publisher}{Association for Computing Machinery}, \bibinfo{address}{New York, NY, USA}, \bibinfo{pages}{527--536}.
\newblock
\showISBNx{979-8-4007-0055-2}
\urldef\tempurl%
\url{https://doi.org/10.1145/3577190.3614160}
\showDOI{\tempurl}


\bibitem[Plumb et~al\mbox{.}(2018)]%
        {plumb_measuring_2018}
\bibfield{author}{\bibinfo{person}{Carolyn Plumb}, \bibinfo{person}{Rose Marra}, \bibinfo{person}{Douglas Hacker}, {and} \bibinfo{person}{John Dunlosky}.} \bibinfo{year}{2018}\natexlab{}.
\newblock \showarticletitle{Measuring {{Engineering Students}}' {{Metacognition}} with a {{Think-Aloud Protocol}}}. In \bibinfo{booktitle}{\emph{2018 {{ASEE Annual Conference}} \& {{Exposition}} {{Proceedings}}}}. \bibinfo{publisher}{ASEE Conferences}, \bibinfo{address}{Salt Lake City, Utah}, \bibinfo{pages}{30796}.
\newblock
\urldef\tempurl%
\url{https://doi.org/10.18260/1-2--30796}
\showDOI{\tempurl}


\bibitem[Plumed et~al\mbox{.}(2021)]%
        {plumed_voicebased_2021}
\bibfield{author}{\bibinfo{person}{Raquel Plumed}, \bibinfo{person}{Carmen {Gonz{\'a}lez-Lluch}}, \bibinfo{person}{David {P{\'e}rez-L{\'o}pez}}, \bibinfo{person}{Manuel Contero}, {and} \bibinfo{person}{Jorge~D Camba}.} \bibinfo{year}{2021}\natexlab{}.
\newblock \showarticletitle{A Voice-Based Annotation System for Collaborative Computer-Aided Design}.
\newblock \bibinfo{journal}{\emph{Journal of Computational Design and Engineering}} \bibinfo{volume}{8}, \bibinfo{number}{2} (\bibinfo{date}{April} \bibinfo{year}{2021}), \bibinfo{pages}{536--546}.
\newblock
\showISSN{2288-5048}
\urldef\tempurl%
\url{https://doi.org/10.1093/jcde/qwaa092}
\showDOI{\tempurl}


\bibitem[{Pontificia Universidad Javeriana} et~al\mbox{.}(2019)]%
        {pontificiauniversidadjaveriana_metacognition_2019}
\bibfield{author}{\bibinfo{person}{{Pontificia Universidad Javeriana}}, \bibinfo{person}{Juanita Tob{\'o}n}, \bibinfo{person}{Fabio Tellez}, {and} \bibinfo{person}{Oscar Alzate}.} \bibinfo{year}{2019}\natexlab{}.
\newblock \showarticletitle{Metacognition in the {{Wild}}: {{Metacognitive Studies}} in {{Design Education}}}. In \bibinfo{booktitle}{\emph{Insider {{Knowledge}} - {{Proceedings}} of the {{Design Research Society Learn X Design Conference}}, 2019}}. \bibinfo{publisher}{Design Research Society}.
\newblock
\showISBNx{978-1-912294-00-8}
\urldef\tempurl%
\url{https://doi.org/10.21606/learnxdesign.2019.09128}
\showDOI{\tempurl}


\bibitem[Price and Lloyd(2022)]%
        {price_asking_2022}
\bibfield{author}{\bibinfo{person}{Rebecca~Anne Price} {and} \bibinfo{person}{Peter Lloyd}.} \bibinfo{year}{2022}\natexlab{}.
\newblock \showarticletitle{Asking {{Effective Questions}}: {{Awareness}} of {{Bias}} in {{Designerly Thinking}}}.
\newblock In \bibinfo{booktitle}{\emph{Handbook of {{Engineering Systems Design}}}}, \bibfield{editor}{\bibinfo{person}{Anja Maier}, \bibinfo{person}{Josef Oehmen}, {and} \bibinfo{person}{Pieter~E. Vermaas}} (Eds.). \bibinfo{publisher}{Springer International Publishing}, \bibinfo{address}{Cham}, \bibinfo{pages}{1--16}.
\newblock
\showISBNx{978-3-030-46054-9}
\urldef\tempurl%
\url{https://doi.org/10.1007/978-3-030-46054-9_24-3}
\showDOI{\tempurl}


\bibitem[Ren et~al\mbox{.}(2000)]%
        {ren_experimental_2000}
\bibfield{author}{\bibinfo{person}{Xiangshi Ren}, \bibinfo{person}{Gao Zhang}, {and} \bibinfo{person}{Guozhong Dai}.} \bibinfo{year}{2000}\natexlab{}.
\newblock \showarticletitle{An {{Experimental Study}} of {{Input Modes}} for {{Multimodal Human-Computer Interaction}}}. In \bibinfo{booktitle}{\emph{Advances in {{Multimodal Interfaces}} --- {{ICMI}} 2000}} \emph{(\bibinfo{series}{Lecture {{Notes}} in {{Computer Science}}})}, \bibfield{editor}{\bibinfo{person}{Tieniu Tan}, \bibinfo{person}{Yuanchun Shi}, {and} \bibinfo{person}{Wen Gao}} (Eds.). \bibinfo{publisher}{Springer}, \bibinfo{address}{Berlin, Heidelberg}, \bibinfo{pages}{49--56}.
\newblock
\showISBNx{978-3-540-40063-9}
\urldef\tempurl%
\url{https://doi.org/10.1007/3-540-40063-X_7}
\showDOI{\tempurl}


\bibitem[Risko and Gilbert(2016)]%
        {risko_cognitive_2016}
\bibfield{author}{\bibinfo{person}{Evan~F. Risko} {and} \bibinfo{person}{Sam~J. Gilbert}.} \bibinfo{year}{2016}\natexlab{}.
\newblock \showarticletitle{Cognitive {{Offloading}}}.
\newblock \bibinfo{journal}{\emph{Trends in Cognitive Sciences}} \bibinfo{volume}{20}, \bibinfo{number}{9} (\bibinfo{date}{Sept.} \bibinfo{year}{2016}), \bibinfo{pages}{676--688}.
\newblock
\showISSN{1364-6613}
\urldef\tempurl%
\url{https://doi.org/10.1016/j.tics.2016.07.002}
\showDOI{\tempurl}


\bibitem[Roy et~al\mbox{.}(2024)]%
        {roy_designing_2024}
\bibfield{author}{\bibinfo{person}{Debrina Roy}, \bibinfo{person}{Nicole Calpin}, \bibinfo{person}{Kathy Cheng}, \bibinfo{person}{Alison Olechowski}, \bibinfo{person}{Andrea~P. Arg{\"u}elles}, \bibinfo{person}{Nicol{\'a}s~F. Soria~Zurita}, {and} \bibinfo{person}{Jessica Menold}.} \bibinfo{year}{2024}\natexlab{}.
\newblock \showarticletitle{Designing {{Together}}: {{Exploring Collaborative Dynamics}} of {{Multi-Objective Design Problems}} in {{Virtual Environments}}}.
\newblock \bibinfo{journal}{\emph{Journal of Mechanical Design}} \bibinfo{volume}{146}, \bibinfo{number}{3} (\bibinfo{date}{March} \bibinfo{year}{2024}), \bibinfo{pages}{031702}.
\newblock
\showISSN{1050-0472, 1528-9001}
\urldef\tempurl%
\url{https://doi.org/10.1115/1.4063658}
\showDOI{\tempurl}


\bibitem[Royo et~al\mbox{.}(2021)]%
        {royo_guiding_2021}
\bibfield{author}{\bibinfo{person}{Marta Royo}, \bibinfo{person}{Elena Mulet}, \bibinfo{person}{Vicente Chulvi}, {and} \bibinfo{person}{Francisco Felip}.} \bibinfo{year}{2021}\natexlab{}.
\newblock \showarticletitle{Guiding Questions for Increasing the Generation of Product Ideas to Meet Changing Needs ({{QuChaNe}})}.
\newblock \bibinfo{journal}{\emph{Research in Engineering Design}} \bibinfo{volume}{32}, \bibinfo{number}{3} (\bibinfo{date}{July} \bibinfo{year}{2021}), \bibinfo{pages}{411--430}.
\newblock
\showISSN{0934-9839, 1435-6066}
\urldef\tempurl%
\url{https://doi.org/10.1007/s00163-021-00364-x}
\showDOI{\tempurl}


\bibitem[Saharia et~al\mbox{.}(2022)]%
        {saharia_photorealistic_2022}
\bibfield{author}{\bibinfo{person}{Chitwan Saharia}, \bibinfo{person}{William Chan}, \bibinfo{person}{Saurabh Saxena}, \bibinfo{person}{Lala Li}, \bibinfo{person}{Jay Whang}, \bibinfo{person}{Emily Denton}, \bibinfo{person}{Seyed Kamyar~Seyed Ghasemipour}, \bibinfo{person}{Burcu~Karagol Ayan}, \bibinfo{person}{S.~Sara Mahdavi}, \bibinfo{person}{Rapha~Gontijo Lopes}, \bibinfo{person}{Tim Salimans}, \bibinfo{person}{Jonathan Ho}, \bibinfo{person}{David~J. Fleet}, {and} \bibinfo{person}{Mohammad Norouzi}.} \bibinfo{year}{2022}\natexlab{}.
\newblock \bibinfo{title}{Photorealistic {{Text-to-Image Diffusion Models}} with {{Deep Language Understanding}}}.
\newblock \bibinfo{howpublished}{https://arxiv.org/abs/2205.11487v1}.
\newblock


\bibitem[Sch{\"o}n(1983)]%
        {schon_reflective_1983}
\bibfield{author}{\bibinfo{person}{Donald~A. Sch{\"o}n}.} \bibinfo{year}{1983}\natexlab{}.
\newblock \bibinfo{booktitle}{\emph{The Reflective Practitioner: How Professionals Think in Action}}.
\newblock \bibinfo{publisher}{Basic Books}, \bibinfo{address}{New York}.
\newblock
\showISBNx{978-0-465-06878-4 978-0-465-06874-6}


\bibitem[Sharmin and Bailey(2011)]%
        {sharmin_reflect_2011}
\bibfield{author}{\bibinfo{person}{Moushumi Sharmin} {and} \bibinfo{person}{Brian~P. Bailey}.} \bibinfo{year}{2011}\natexlab{}.
\newblock \showarticletitle{"{{I}} Reflect to Improve My Design": Investigating the Role and Process of Reflection in Creative Design}. In \bibinfo{booktitle}{\emph{Proceedings of the 8th {{ACM}} Conference on {{Creativity}} and Cognition}} \emph{(\bibinfo{series}{C\&amp;{{C}} '11})}. \bibinfo{publisher}{Association for Computing Machinery}, \bibinfo{address}{New York, NY, USA}, \bibinfo{pages}{389--390}.
\newblock
\showISBNx{978-1-4503-0820-5}
\urldef\tempurl%
\url{https://doi.org/10.1145/2069618.2069710}
\showDOI{\tempurl}


\bibitem[Shridhar et~al\mbox{.}(2022)]%
        {shridhar_automatic_2022}
\bibfield{author}{\bibinfo{person}{Kumar Shridhar}, \bibinfo{person}{Jakub Macina}, \bibinfo{person}{Mennatallah {El-Assady}}, \bibinfo{person}{Tanmay Sinha}, \bibinfo{person}{Manu Kapur}, {and} \bibinfo{person}{Mrinmaya Sachan}.} \bibinfo{year}{2022}\natexlab{}.
\newblock \showarticletitle{Automatic {{Generation}} of {{Socratic Subquestions}} for {{Teaching Math Word Problems}}}. In \bibinfo{booktitle}{\emph{Proceedings of the 2022 {{Conference}} on {{Empirical Methods}} in {{Natural Language Processing}}}}, \bibfield{editor}{\bibinfo{person}{Yoav Goldberg}, \bibinfo{person}{Zornitsa Kozareva}, {and} \bibinfo{person}{Yue Zhang}} (Eds.). \bibinfo{publisher}{Association for Computational Linguistics}, \bibinfo{address}{Abu Dhabi, United Arab Emirates}, \bibinfo{pages}{4136--4149}.
\newblock
\urldef\tempurl%
\url{https://doi.org/10.18653/v1/2022.emnlp-main.277}
\showDOI{\tempurl}


\bibitem[Stenning et~al\mbox{.}(2016)]%
        {stenning_socratic_2016}
\bibfield{author}{\bibinfo{person}{Keith Stenning}, \bibinfo{person}{Alexander Schmoelz}, \bibinfo{person}{Heather Wren}, \bibinfo{person}{Elias Stouraitis}, \bibinfo{person}{Theodore Scaltsas}, \bibinfo{person}{Constantine Alexopoulos}, {and} \bibinfo{person}{Amelie Aichhorn}.} \bibinfo{year}{2016}\natexlab{}.
\newblock \showarticletitle{Socratic Dialogue as a Teaching and Research Method for Co-Creativity?}
\newblock \bibinfo{journal}{\emph{Digital Culture and Education}} \bibinfo{volume}{8}, \bibinfo{number}{2} (\bibinfo{year}{2016}), \bibinfo{pages}{13}.
\newblock


\bibitem[Subramonyam et~al\mbox{.}(2024)]%
        {subramonyam_bridging_2024}
\bibfield{author}{\bibinfo{person}{Hari Subramonyam}, \bibinfo{person}{Roy Pea}, \bibinfo{person}{Christopher Pondoc}, \bibinfo{person}{Maneesh Agrawala}, {and} \bibinfo{person}{Colleen Seifert}.} \bibinfo{year}{2024}\natexlab{}.
\newblock \showarticletitle{Bridging the {{Gulf}} of {{Envisioning}}: {{Cognitive Challenges}} in {{Prompt Based Interactions}} with {{LLMs}}}. In \bibinfo{booktitle}{\emph{Proceedings of the 2024 {{CHI Conference}} on {{Human Factors}} in {{Computing Systems}}}} \emph{(\bibinfo{series}{{{CHI}} '24})}. \bibinfo{publisher}{Association for Computing Machinery}, \bibinfo{address}{New York, NY, USA}, \bibinfo{pages}{1--19}.
\newblock
\showISBNx{979-8-4007-0330-0}
\urldef\tempurl%
\url{https://doi.org/10.1145/3613904.3642754}
\showDOI{\tempurl}


\bibitem[Tamang et~al\mbox{.}(2021)]%
        {tamang_comparative_2021}
\bibfield{author}{\bibinfo{person}{Lasang~Jimba Tamang}, \bibinfo{person}{Zeyad Alshaikh}, \bibinfo{person}{Nisrine~Ait Khayi}, \bibinfo{person}{Priti Oli}, {and} \bibinfo{person}{Vasile Rus}.} \bibinfo{year}{2021}\natexlab{}.
\newblock \showarticletitle{A {{Comparative Study}} of {{Free Self-Explanations}} and {{Socratic Tutoring Explanations}} for {{Source Code Comprehension}}}.
\newblock \bibinfo{journal}{\emph{Proceedings of the 52nd ACM Technical Symposium on Computer Science Education}} (\bibinfo{date}{March} \bibinfo{year}{2021}), \bibinfo{pages}{219--225}.
\newblock
\showISBNx{9781450380621}
\urldef\tempurl%
\url{https://doi.org/10.1145/3408877.3432423}
\showDOI{\tempurl}


\bibitem[Tankelevitch et~al\mbox{.}(2023)]%
        {tankelevitch_metacognitive_2023}
\bibfield{author}{\bibinfo{person}{Lev Tankelevitch}, \bibinfo{person}{Viktor Kewenig}, \bibinfo{person}{Auste Simkute}, \bibinfo{person}{Ava~Elizabeth Scott}, \bibinfo{person}{Advait Sarkar}, \bibinfo{person}{Abigail Sellen}, {and} \bibinfo{person}{Sean Rintel}.} \bibinfo{year}{2023}\natexlab{}.
\newblock \bibinfo{title}{The {{Metacognitive Demands}} and {{Opportunities}} of {{Generative AI}}}.
\newblock
\newblock
\urldef\tempurl%
\url{https://doi.org/10.48550/arXiv.2312.10893}
\showDOI{\tempurl}
\showeprint[arxiv]{2312.10893}~[cs]


\bibitem[Tawfik et~al\mbox{.}(2020)]%
        {tawfik_role_2020}
\bibfield{author}{\bibinfo{person}{Andrew~A. Tawfik}, \bibinfo{person}{Arthur Graesser}, \bibinfo{person}{Jessica Gatewood}, {and} \bibinfo{person}{Jaclyn Gishbaugher}.} \bibinfo{year}{2020}\natexlab{}.
\newblock \showarticletitle{Role of Questions in Inquiry-Based Instruction: Towards a Design Taxonomy for Question-Asking and Implications for Design}.
\newblock \bibinfo{journal}{\emph{Educational Technology Research and Development}} \bibinfo{volume}{68}, \bibinfo{number}{2} (\bibinfo{date}{April} \bibinfo{year}{2020}), \bibinfo{pages}{653--678}.
\newblock
\showISSN{1042-1629, 1556-6501}
\urldef\tempurl%
\url{https://doi.org/10.1007/s11423-020-09738-9}
\showDOI{\tempurl}


\bibitem[Tolmeijer et~al\mbox{.}(2021)]%
        {tolmeijer_female_2021}
\bibfield{author}{\bibinfo{person}{Suzanne Tolmeijer}, \bibinfo{person}{Naim Zierau}, \bibinfo{person}{Andreas Janson}, \bibinfo{person}{Jalil~Sebastian Wahdatehagh}, \bibinfo{person}{Jan Marco~Marco Leimeister}, {and} \bibinfo{person}{Abraham Bernstein}.} \bibinfo{year}{2021}\natexlab{}.
\newblock \showarticletitle{Female by {{Default}}? -- {{Exploring}} the {{Effect}} of {{Voice Assistant Gender}} and {{Pitch}} on {{Trait}} and {{Trust Attribution}}}. In \bibinfo{booktitle}{\emph{Extended {{Abstracts}} of the 2021 {{CHI Conference}} on {{Human Factors}} in {{Computing Systems}}}} \emph{(\bibinfo{series}{{{CHI EA}} '21})}. \bibinfo{publisher}{Association for Computing Machinery}, \bibinfo{address}{New York, NY, USA}, \bibinfo{pages}{1--7}.
\newblock
\showISBNx{978-1-4503-8095-9}
\urldef\tempurl%
\url{https://doi.org/10.1145/3411763.3451623}
\showDOI{\tempurl}


\bibitem[Triantafyllopoulos et~al\mbox{.}(2023)]%
        {triantafyllopoulos_overview_2023}
\bibfield{author}{\bibinfo{person}{Andreas Triantafyllopoulos}, \bibinfo{person}{Bj{\"o}rn~W. Schuller}, \bibinfo{person}{G{\"o}k{\c c}e {\.I}ymen}, \bibinfo{person}{Metin Sezgin}, \bibinfo{person}{Xiangheng He}, \bibinfo{person}{Zijiang Yang}, \bibinfo{person}{Panagiotis Tzirakis}, \bibinfo{person}{Shuo Liu}, \bibinfo{person}{Silvan Mertes}, \bibinfo{person}{Elisabeth Andr{\'e}}, \bibinfo{person}{Ruibo Fu}, {and} \bibinfo{person}{Jianhua Tao}.} \bibinfo{year}{2023}\natexlab{}.
\newblock \showarticletitle{An {{Overview}} of {{Affective Speech Synthesis}} and {{Conversion}} in the {{Deep Learning Era}}}.
\newblock \bibinfo{journal}{\emph{Proc. IEEE}} \bibinfo{volume}{111}, \bibinfo{number}{10} (\bibinfo{date}{Oct.} \bibinfo{year}{2023}), \bibinfo{pages}{1355--1381}.
\newblock
\showISSN{0018-9219, 1558-2256}
\urldef\tempurl%
\url{https://doi.org/10.1109/JPROC.2023.3250266}
\showDOI{\tempurl}


\bibitem[Van Den~Haak et~al\mbox{.}(2003)]%
        {vandenhaak_retrospective_2003}
\bibfield{author}{\bibinfo{person}{Maaike Van Den~Haak}, \bibinfo{person}{Menno De~Jong}, {and} \bibinfo{person}{Peter Jan~Schellens}.} \bibinfo{year}{2003}\natexlab{}.
\newblock \showarticletitle{Retrospective vs. Concurrent Think-Aloud Protocols: {{Testing}} the Usability of an Online Library Catalogue}.
\newblock \bibinfo{journal}{\emph{Behaviour \& Information Technology}} \bibinfo{volume}{22}, \bibinfo{number}{5} (\bibinfo{date}{Sept.} \bibinfo{year}{2003}), \bibinfo{pages}{339--351}.
\newblock
\showISSN{0144-929X, 1362-3001}
\urldef\tempurl%
\url{https://doi.org/10.1080/0044929031000}
\showDOI{\tempurl}


\bibitem[VanLehn et~al\mbox{.}(1992)]%
        {vanlehn_model_1992}
\bibfield{author}{\bibinfo{person}{Kurt VanLehn}, \bibinfo{person}{Randolph~M. Jones}, {and} \bibinfo{person}{Michelene~T.H. Chi}.} \bibinfo{year}{1992}\natexlab{}.
\newblock \showarticletitle{A {{Model}} of the {{Self-Explanation Effect}}}.
\newblock \bibinfo{journal}{\emph{Journal of the Learning Sciences}} \bibinfo{volume}{2}, \bibinfo{number}{1} (\bibinfo{date}{Jan.} \bibinfo{year}{1992}), \bibinfo{pages}{1--59}.
\newblock
\showISSN{1050-8406, 1532-7809}
\urldef\tempurl%
\url{https://doi.org/10.1207/s15327809jls0201_1}
\showDOI{\tempurl}


\bibitem[Vaughan(2022)]%
        {vaughan_metacognition_2022}
\bibfield{author}{\bibinfo{person}{Geoffrey Vaughan}.} \bibinfo{year}{2022}\natexlab{}.
\newblock \showarticletitle{Metacognition and {{Self-Regulated Learning}}: {{Recent Perspectives}} for an {{International Context}}}.
\newblock \bibinfo{journal}{\emph{Opus et Educatio}} \bibinfo{volume}{9}, \bibinfo{number}{2} (\bibinfo{date}{Aug.} \bibinfo{year}{2022}).
\newblock
\showISSN{2064-9908}
\urldef\tempurl%
\url{https://doi.org/10.3311/ope.501}
\showDOI{\tempurl}


\bibitem[Veuskens et~al\mbox{.}(2022)]%
        {veuskens_identifying_2022}
\bibfield{author}{\bibinfo{person}{Tom Veuskens}, \bibinfo{person}{Danny Leen}, {and} \bibinfo{person}{Raf Ramakers}.} \bibinfo{year}{2022}\natexlab{}.
\newblock \showarticletitle{Identifying {{Opportunities}} to {{Reimagine Parametric Modeling}} for {{Makers}}}.
\newblock  (\bibinfo{year}{2022}).
\newblock


\bibitem[Wang and Zeng(2009)]%
        {wang_asking_2009}
\bibfield{author}{\bibinfo{person}{Min Wang} {and} \bibinfo{person}{Yong Zeng}.} \bibinfo{year}{2009}\natexlab{}.
\newblock \showarticletitle{Asking the Right Questions to Elicit Product Requirements}.
\newblock \bibinfo{journal}{\emph{International Journal of Computer Integrated Manufacturing}} \bibinfo{volume}{22}, \bibinfo{number}{4} (\bibinfo{date}{April} \bibinfo{year}{2009}), \bibinfo{pages}{283--298}.
\newblock
\showISSN{0951-192X, 1362-3052}
\urldef\tempurl%
\url{https://doi.org/10.1080/09511920802232902}
\showDOI{\tempurl}


\bibitem[Wilson(1987)]%
        {wilson_socratic_1987}
\bibfield{author}{\bibinfo{person}{Judith~D. Wilson}.} \bibinfo{year}{1987}\natexlab{}.
\newblock \showarticletitle{A {{Socratic}} Approach to Helping Novice Programmers Debug Programs}. In \bibinfo{booktitle}{\emph{Proceedings of the Eighteenth {{SIGCSE}} Technical Symposium on {{Computer}} Science Education - {{SIGCSE}} '87}}. \bibinfo{publisher}{ACM Press}, \bibinfo{address}{St. Louis, Missouri, United States}, \bibinfo{pages}{179--182}.
\newblock
\showISBNx{978-0-89791-217-4}
\urldef\tempurl%
\url{https://doi.org/10.1145/31820.31755}
\showDOI{\tempurl}


\bibitem[Woodbury(2010)]%
        {woodbury_elements_2010a}
\bibfield{author}{\bibinfo{person}{Robert Woodbury}.} \bibinfo{year}{2010}\natexlab{}.
\newblock \bibinfo{booktitle}{\emph{Elements of Parametric Design}}.
\newblock \bibinfo{publisher}{Routledge}, \bibinfo{address}{London}.
\newblock
\showISBNx{978-0-415-77986-9 978-0-415-77987-6}


\bibitem[Yang et~al\mbox{.}(2020)]%
        {yang_simulearn_2020}
\bibfield{author}{\bibinfo{person}{Humphrey Yang}, \bibinfo{person}{Kuanren Qian}, \bibinfo{person}{Haolin Liu}, \bibinfo{person}{Yuxuan Yu}, \bibinfo{person}{Jianzhe Gu}, \bibinfo{person}{Matthew McGehee}, \bibinfo{person}{Yongjie~Jessica Zhang}, {and} \bibinfo{person}{Lining Yao}.} \bibinfo{year}{2020}\natexlab{}.
\newblock \showarticletitle{{{SimuLearn}}: {{Fast}} and {{Accurate Simulator}} to {{Support Morphing Materials Design}} and {{Workflows}}}. In \bibinfo{booktitle}{\emph{Proceedings of the 33rd {{Annual ACM Symposium}} on {{User Interface Software}} and {{Technology}}}}. \bibinfo{publisher}{ACM}, \bibinfo{address}{Virtual Event USA}, \bibinfo{pages}{71--84}.
\newblock
\showISBNx{978-1-4503-7514-6}
\urldef\tempurl%
\url{https://doi.org/10.1145/3379337.3415867}
\showDOI{\tempurl}


\bibitem[Young(2009)]%
        {young_direct_2009}
\bibfield{author}{\bibinfo{person}{Kirsty Young}.} \bibinfo{year}{2009}\natexlab{}.
\newblock \showarticletitle{Direct from the Source: The Value of 'think-Aloud' Data in Understanding Learning}.
\newblock \bibinfo{journal}{\emph{The Journal of Educational Enquiry}}  \bibinfo{volume}{6} (\bibinfo{year}{2009}).
\newblock


\bibitem[Zaman et~al\mbox{.}(2015)]%
        {zaman_gemni_2015}
\bibfield{author}{\bibinfo{person}{Loutfouz Zaman}, \bibinfo{person}{Wolfgang Stuerzlinger}, \bibinfo{person}{Christian Neugebauer}, \bibinfo{person}{Rob Woodbury}, \bibinfo{person}{Maher Elkhaldi}, \bibinfo{person}{Naghmi Shireen}, {and} \bibinfo{person}{Michael Terry}.} \bibinfo{year}{2015}\natexlab{}.
\newblock \showarticletitle{{{{\emph{GEM-NI}}}}: {{A System}} for {{Creating}} and {{Managing Alternatives In Generative Design}}}. In \bibinfo{booktitle}{\emph{Proceedings of the 33rd {{Annual ACM Conference}} on {{Human Factors}} in {{Computing Systems}}}}. \bibinfo{publisher}{ACM}, \bibinfo{address}{Seoul Republic of Korea}, \bibinfo{pages}{1201--1210}.
\newblock
\showISBNx{978-1-4503-3145-6}
\urldef\tempurl%
\url{https://doi.org/10.1145/2702123.2702398}
\showDOI{\tempurl}


\bibitem[{Zamfirescu-Pereira} et~al\mbox{.}(2021)]%
        {zamfirescu-pereira_fake_2021}
\bibfield{author}{\bibinfo{person}{J.D. {Zamfirescu-Pereira}}, \bibinfo{person}{David Sirkin}, \bibinfo{person}{David Goedicke}, \bibinfo{person}{Ray LC}, \bibinfo{person}{Natalie Friedman}, \bibinfo{person}{Ilan Mandel}, \bibinfo{person}{Nikolas Martelaro}, {and} \bibinfo{person}{Wendy Ju}.} \bibinfo{year}{2021}\natexlab{}.
\newblock \showarticletitle{Fake {{It}} to {{Make It}}: {{Exploratory Prototyping}} in {{HRI}}}. In \bibinfo{booktitle}{\emph{Companion of the 2021 {{ACM}}/{{IEEE International Conference}} on {{Human-Robot Interaction}}}} \emph{(\bibinfo{series}{{{HRI}} '21 {{Companion}}})}. \bibinfo{publisher}{Association for Computing Machinery}, \bibinfo{address}{New York, NY, USA}, \bibinfo{pages}{19--28}.
\newblock
\showISBNx{978-1-4503-8290-8}
\urldef\tempurl%
\url{https://doi.org/10.1145/3434074.3446909}
\showDOI{\tempurl}


\bibitem[{Zamfirescu-Pereira} et~al\mbox{.}(2023)]%
        {zamfirescu-pereira_why_2023}
\bibfield{author}{\bibinfo{person}{J.D. {Zamfirescu-Pereira}}, \bibinfo{person}{Richmond~Y. Wong}, \bibinfo{person}{Bjoern Hartmann}, {and} \bibinfo{person}{Qian Yang}.} \bibinfo{year}{2023}\natexlab{}.
\newblock \showarticletitle{Why {{Johnny Can}}'t {{Prompt}}: {{How Non-AI Experts Try}} (and {{Fail}}) to {{Design LLM Prompts}}}. In \bibinfo{booktitle}{\emph{Proceedings of the 2023 {{CHI Conference}} on {{Human Factors}} in {{Computing Systems}}}}. \bibinfo{publisher}{ACM}, \bibinfo{address}{Hamburg Germany}, \bibinfo{pages}{1--21}.
\newblock
\showISBNx{978-1-4503-9421-5}
\urldef\tempurl%
\url{https://doi.org/10.1145/3544548.3581388}
\showDOI{\tempurl}


\bibitem[Zhang et~al\mbox{.}(2021)]%
        {zhang_cautionary_2021}
\bibfield{author}{\bibinfo{person}{Guanglu Zhang}, \bibinfo{person}{Ayush Raina}, \bibinfo{person}{Jonathan Cagan}, {and} \bibinfo{person}{Christopher McComb}.} \bibinfo{year}{2021}\natexlab{}.
\newblock \showarticletitle{A Cautionary Tale about the Impact of {{AI}} on Human Design Teams}.
\newblock \bibinfo{journal}{\emph{Design Studies}}  \bibinfo{volume}{72} (\bibinfo{date}{Jan.} \bibinfo{year}{2021}), \bibinfo{pages}{100990}.
\newblock
\showISSN{0142694X}
\urldef\tempurl%
\url{https://doi.org/10.1016/j.destud.2021.100990}
\showDOI{\tempurl}


\bibitem[Zhang et~al\mbox{.}(2024)]%
        {zhang_using_2024}
\bibfield{author}{\bibinfo{person}{Jiayi Zhang}, \bibinfo{person}{Conrad Borchers}, \bibinfo{person}{Vincent Aleven}, {and} \bibinfo{person}{Ryan~Shaun Baker}.} \bibinfo{year}{2024}\natexlab{}.
\newblock \bibinfo{title}{Using {{Large Language Models}} to {{Detect Self-Regulated Learning}} in {{Think-Aloud Protocols}}}.
\newblock
\newblock
\urldef\tempurl%
\url{https://doi.org/10.35542/osf.io/hrtz6}
\showDOI{\tempurl}


\end{thebibliography}

\appendix
\newpage
\onecolumn
\pagestyle{empty}

\section{Additional Materials}

\begin{table*}[h]
\caption{Overview of study participants.}
\Description{Study participant demographics. This table is machine-readable.}
\begin{tabular}{llclllllll}
\toprule
\textbf{ID } & 
\begin{tabular}[c]{@{}l@{}}\textbf{Agent} \\ \textbf{Group} \end{tabular} & 
\textbf{Age} & 
\textbf{Role}                                &
\begin{tabular}[c]{@{}l@{}}\textbf{MechDes} \\ \textbf{Exp.} \\ \textbf{Years}\end{tabular} & 
\begin{tabular}[c]{@{}l@{}}\textbf{Indus.} \\ \textbf{Exp.} \\ \textbf{Years}\end{tabular} & 
\begin{tabular}[c]{@{}l@{}}\textbf{CAD} \\ \textbf{Exp.} \\ \textbf{Years}\end{tabular}& 
\begin{tabular}[c]{@{}l@{}}\textbf{FEA } \\ \textbf{Prof.}\end{tabular}&  
\begin{tabular}[c]{@{}l@{}}\textbf{DFM } \\ \textbf{Prof.}\end{tabular}\\

\midrule
B1 & No Support      & 22  & Student, MA Mechanical Engineering  & 3–5                         & 0                   & 2–4           & 5               & 2                            \\
B2 & No Support      & 28  & Student, PhD Mechanical Engineering & 3–5                         & 1–2                 & 5+            & 6               & 2                          \\
B3 & No Support      & 27  & Researcher, Mechanical Engineering  & 6–10                        & 0                   & 5+            & 7               & 5                          \\
B4 & No Support      & 23  & Student, MA Mechanical Engineering  & 3–5                         & 3–5                 & 2–4           & 7               & 5                         \\
B5 & No Support      & 39  & Researcher, Mechanical Engineering  & 3–5                         & 0                   & 5+            & 1               & 1                            \\

S1 & SocratAIs       & 26  & Student, BS Mechanical Engineering  & 6–10                        & 0                   & 5+            & 4               & 1                            \\
S2 & SocratAIs       & 23  & Student, MS Mechanical Engineering  & 3–5                         & 0                   & 2–4           & 5               & 5                            \\
S3 & SocratAIs       & 22  & Student, MS Mechanical Engineering  & 1–2                         & 1–2                 & 2–4           & 4               & 1                            \\
S4 & SocratAIs       & 26  & Student, PhD Mechanical Engineering & 6–10                        & 1–2                 & 5+            & 1               & 5                          \\
S5 & SocratAIs       & 20  & Student, BA Mechanical Engineering  & 3–5                         & 0                   & 2–4           & 4               & 6                          \\

H1 & Hephaistus      & 30  & Student, PhD Mechanical Engineering & 3–5                         & 3–5                 & 5+            & 4               & 4                         \\
H2 & Hephaistus      & 42  & Student, PhD Mechanical Engineering & 6–10                        & 6–10                & 5+            & 4               & 1                            \\
H3 & Hephaistus      & 22  & Student, BA Mechanical Engineering  & 1–2                         & 1–2                 & 2–4           & 3               & 2                          \\
H4 & Hephaistus      & 26  & Mechanical Engineer                 & 3–5                         & 3–5                 & 5+            & 2               & 2                            \\
H5 & Hephaistus      & 21  & Student, BA Mechanical Engineering  & 1–2                         & 1–2                 & 2–4           & 3               & 5                          \\

E1 & Expert-Freeform  & 38  & Mechanical Engineer                 & 10+                         & 10+                 & 5+            & 7               & 7                          \\
E2 & Expert-Freeform  & 26  & Mechanical Engineer                 & 1–2                         & 3–5                 & 5+            & 4               & 6                          \\
E3 & Expert-Freeform  & 29  & Mechanical Designer                 & 10+                         & 3–5                 & 5+            & 5               & 7                          \\
E4 & Expert-Freeform  & 29  & Mechanical Engineer                 & 6–10                        & 6–10                & 5+            & 5               & 5                          \\
E5 & Expert-Freeform  & 23  & Student, MS Mechanical Engineering  & 3–5                         & 0                   & 5+            & 2               & 5                        
      \\                                       
\bottomrule
\end{tabular}
\label{tab:participants}
\end{table*}

\begin{table*}[h]
\caption{Overview of demographics of Autodesk Fusion360 Generative Design experts who acted as wizards in the Expert-Freeform condition. Fusion360 Generative Design software proficiency was self-rated on a 1–7 scale. }
\Description{ This table is machine-readable.}
\begin{tabular}{lllllll}
\toprule
\textbf{ID} & 
\textbf{Age} &  
\textbf{Role} & 
\begin{tabular}[c]{@{}l@{}}\textbf{MechDes} \\ \textbf{Exp. Years}\end{tabular} & 

\begin{tabular}[c]{@{}l@{}}\textbf{F360} \\ 
\textbf{GenDes} \\ \textbf{Prof.}   \end{tabular} & 
\begin{tabular}[c]{@{}l@{}}\textbf{F360} \\ 
\textbf{GenDes} \\ \textbf{Training Exp.}\end{tabular}   &
\begin{tabular}[c]{@{}l@{}}\textbf{Paired} \\ \textbf{with}\end{tabular} 
\\
\midrule
Expert 1  & 31   & Senior Research Engineer                                    & 3 – 5                        & 6/7               & Trained customers, students, colleagues & E2, E5\\

Expert 2  & 35   & Sr. Research \& Design Engineer                             & 6 – 10                       & 7/7               & Taught lectures, trained colleagues  & E1     \\

Expert 3  & 47  & Principal Research Engineer                                 & 15+                          & 7/7               & Trained customer support teams   & E4             \\

Expert 4  & 27  & Research and Design Engineer                                & 3 – 5                        & 6/7               & Trained customers and colleagues  & E3\\  

\bottomrule
\end{tabular}
\label{tab:experts}
\end{table*}

\newpage

\begin{figure*}[t]
  \centering
  \includegraphics[width=\linewidth]{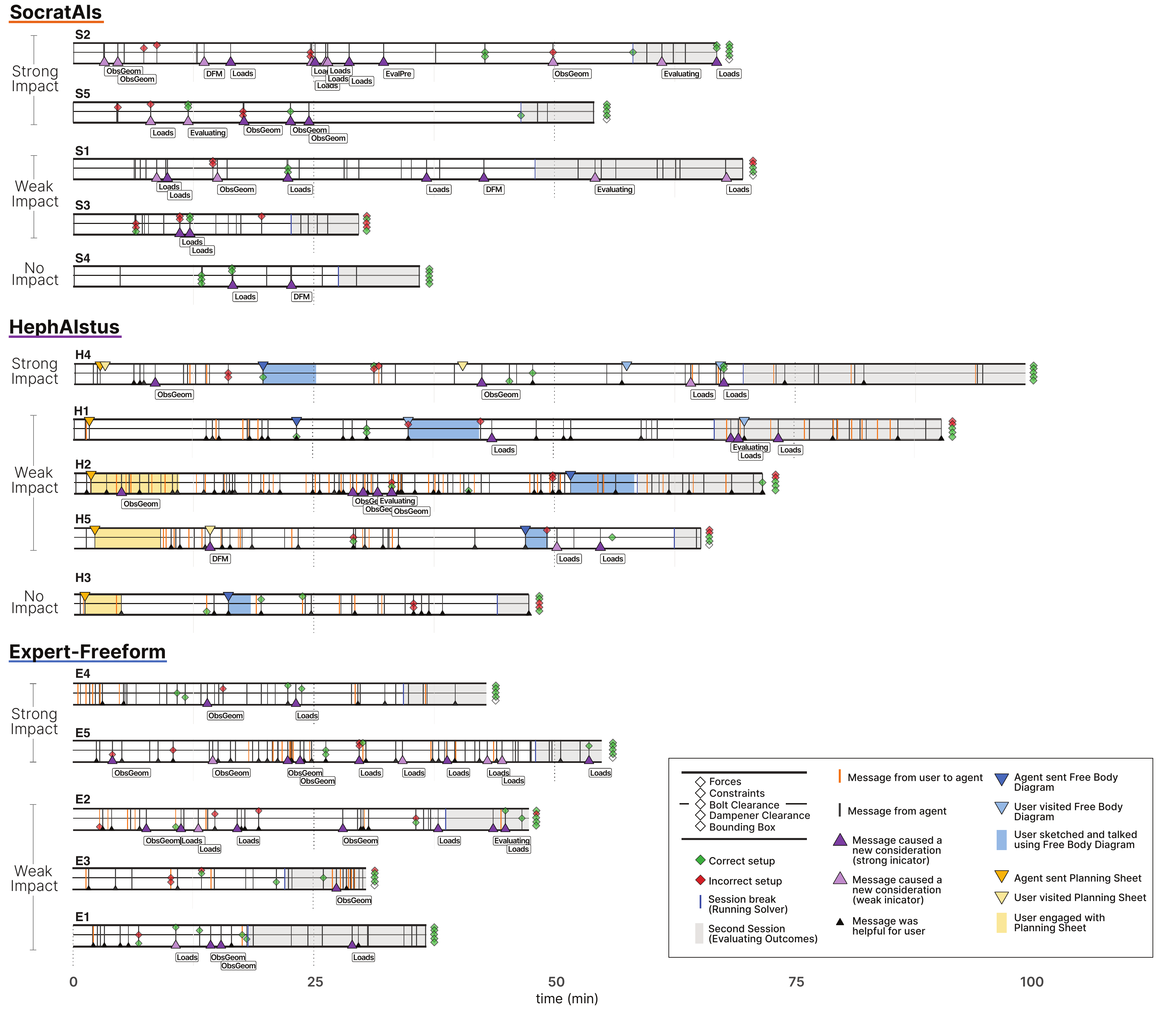}
  \caption{Timeline plots visualizing participant and agent interactions throughout the design task; timelines are divided into lanes, each showing (in)correct GenAI input specifications (diamond shapes) for 
  (1) forces, (2) constraints, (3) bolt and, (4) dampener clearances, (5) bounding box  (from top to bottom); black and orange vertical lines represent exchanged agent and user messages with purple and black triangles indicating an observable impact on the design process.}
  
  \Description{The figure presents timeline plots that visualize participant and agent interactions throughout the design task. Timelines are grouped by agent type (SocratAIs, HephAIstus, Expert) and categorized by their observed impact levels: Strong Impact, Weak Impact, and No Impact. Each timeline is divided into lanes showing (in)correct GenAI input specifications for forces, constraints, bolt clearance, dampener clearance, and bounding box. Diamond shapes indicate (in)correct setups, while vertical lines represent exchanged agent and user messages. Purple and black triangles mark messages that caused observable new considerations or were helpful to the user.

For SocratAIs, timelines such as S2 and S5 show strong impacts where agent messages led to significant design improvements. In contrast, sessions like S1 and S3 demonstrate weaker impacts, where the agent’s messages were less effective in overcoming cognitive challenges. 

For HephAIstus, strong impacts are observed in cases like H4, where the free-body diagram and agent suggestions helped correct errors. Weaker impacts, as in H2 and H5, show limited user response to agent interventions. Expert timelines highlight varying degrees of support, with E4 showing strong impacts through effective interventions and E1 displaying weaker outcomes despite agent efforts.

The plots offer a detailed temporal view of interactions, illustrating the effectiveness of different agents and strategies in influencing the design process. The legend clarifies the annotations, such as correct and incorrect setups, session breaks, and tool usage like the free-body diagram or planning sheet.}

    \label{fig:appendix_timelines}
 
\end{figure*}

\newpage

\subsection{Wizard Guidelines} \label{wizard-guides}

\rev{The agents \highsocratais{\textit{SocratAIs}} and \highhephaistus{\textit{HephAIstus}} were facilitated by the first author, with experience in mechanical engineering, Fusion360, and Generative Design. 
In some sessions, a second research team member with experience in mechanical engineering and Generative Design was co-present to provide additional verbal support for the main wizard.}

\subsubsection{\textbf{Guidelines for the SocratAIs and HephAIstus wizard}} \hfill \\
\rev{The wizard of \highsocratais{\textit{SocratAIs}} and \highhephaistus{\textit{HephAIstus}}  followed these general guidelines:
\begin{enumerate}
  \item[1)] Follow the designer’s verbalizations and screen actions and pay close attention to the task-specific design steps and challenges as outlined in Section \ref{case-study}, such as specifying the bracket’s load cases (forces and structural constraints), modeling appropriate geometry for keeping bolts and dampeners free of material (obstacle geometry), defining DFM parameters such as materials and manufacturing options, and also to support users in evaluating the design previews and generated outcomes.
    \item[2)] Pay close attention to inconsistencies between the requirements stated in the design brief and the input parameters set by the designer. Such requirements could be explicit (e.g., the force the bracket needs to hold) or implicit features, such as bolt clearances, which were not explicitly mentioned in the design brief.
  \item[3)] Never directly tell the participant what to do, but rather provide supportive questions, hints, or suggestions (depending on the enacted agent type). 
  \item[4)] You are free to send messages whenever and how often you consider it helpful to the designer. However, pay special attention to moments in which designers transition between design sub-tasks (such as from specifying obstacle geometry to specifying loads), as well as when designers show hesitation or use hedging expressions (e.g., \textit{“I am unsure if...”}).
  \item[5)] You are free to formulate the messages in a way you consider to be most helpful, while adhering to the agent's support strategy (e.g., only asking questions).
\end{enumerate}}

\subsubsection{\textbf{Guidelines for the Expert-Freeform wizards}} \hfill \\
The \highexpert{\textit{Expert-Freeform}} wizards \rev{(external experts not part of the research team)} received fewer instructions since we wanted to observe their natural support behavior. 
However, experts were told not to directly tell the participant what to do, but rather to help them work on the design task and with the GenAI system.

\subsubsection{\textbf{SocratAIs Agent Introduction }} \hfill

\begin{itemize}
\setlength{\leftskip}{1cm}
\setlength{\rightskip}{1cm}
    \ttfamily
  \item[\highsocratais{\textit{SocratAIs}}:] Hey! I am a voice agent here to support you during the design task. I can hear what you are saying, and I can see your screen and follow along with you while you work on the task. 
  From time to time, I will ask you questions that are supposed to help you think through the design task. You can also ask me questions at any time.
\end{itemize}

\subsubsection{\textbf{HephAIstus}} \hfill \\
\label{appendix-hephaistus}

{\raggedright
\textbf{Agent introduction:}
}

\begin{itemize}
\setlength{\leftskip}{1cm}
\setlength{\rightskip}{1cm}
    \ttfamily
  \item[\highhephaistus{\textit{HephAIstus}} :] Hey! I am a voice agent here to support you during the design task. I can hear what you are saying, and I can see your screen and follow along with you while you work on the task. 
  Feel free to ask me any questions, and I'll do my best to provide you with answers. From time to time, I'll also chime in with tips and guidance to help you along the way.
\end{itemize}

\vspace{8pt}

{\raggedright
\textbf{Project Planning Activity:}

HephAIstus provided metacognitive support by suggesting the user engage in a \textbf{project planning activity} by sharing a pre-generated text document outlining critical project-relevant aspects with the user:
}

\begin{itemize}
\setlength{\leftskip}{1cm}
\setlength{\rightskip}{1cm}
    \ttfamily
  \item[\highhephaistus{\textit{HephAIstus}} :] Hey! I noticed you are working on designing a ship engine mounting bracket and that you are planning to use Autodesk Fusion 360 Generative Design to explore different materials and manufacturing options. Before you start working on the task, I suggest having a planning session together to help you walk through the steps and design considerations in designing the bracket. What do you think?
    \item[] \textit{(waiting for confirmation from user)}
    \item[\highhephaistus{\textit{HephAIstus}} :] I created a shared doc with a high-level breakdown of the steps and design considerations to be discussed. You can find the link in the chat.

\end{itemize}

\newpage
\textbf{Planning Sheet Document:}

\begin{framed}
\textbf{1) Project Scope and Requirements}

\begin{itemize}
\item
  Define the objectives of the bracket design.
\item
  Understand the operational conditions of the ship engine.
\item
  Identify load types (static, dynamic, thermal) and magnitudes.
\item
  Clarify space constraints and installation considerations.
\end{itemize}

\textbf{2) Material Selection}

\begin{itemize}
\item
  Discuss different material options (metal alloys, composites, etc.).
\item
  Consider material properties such as strength, weight, corrosion
  resistance, and cost.
\item
  Review the material performance under extreme marine conditions.
\end{itemize}

\textbf{3) Manufacturing Method}

\begin{itemize}
\item
  Determine feasible manufacturing methods (casting, machining, additive
  manufacturing, etc.).
\item
  Evaluate the pros and cons of each method concerning the design
  objectives.
\item
  Discuss generative design constraints for each manufacturing process.
\end{itemize}

\textbf{4) Generative Design Parameters}

\begin{itemize}
\item
  Set up load cases and boundary conditions in Fusion 360.
\item
  Define the design space and apply necessary constraints and
  conditions.
\item
  Choose the resolution of the generative design mesh.
\end{itemize}

\textbf{5) Design Constraints and Criteria}

\begin{itemize}
\item
  Set criteria for minimum safety factors.
\item
  Define requirements for vibration dampening.
\item
  Consider access for maintenance and installation.
\end{itemize}

\textbf{6) Simulation and Analysis}

\begin{itemize}
\item
  Plan for simulations to predict performance under various loads.
\item
  Analyze stress distribution, deformation, and fatigue life.
\item
  Review thermal and fluid flow analysis if necessary.
\end{itemize}

\textbf{7) Optimization Objectives}

\begin{itemize}
\item
  Establish the optimization goals, such as weight reduction, strength
  optimization, cost efficiency, etc.
\item
  Discuss trade-offs between different optimization objectives.
\end{itemize}

\textbf{8) Compliance and Standards}

\begin{itemize}
\item
  Ensure the design meets marine industry standards and regulatory
  compliance.
\item
  Consider classification society requirements and certifications.
\end{itemize}

\textbf{9) Integration with Existing Systems}

\begin{itemize}
\item
  Discuss how the bracket will integrate with the ship\textquotesingle s
  engine and surrounding structures.
\item
  Plan for interfaces with other systems and parts.
\end{itemize}

\textbf{10) Lifecycle Considerations}

\begin{itemize}
\item
  Consider the lifecycle impacts, such as ease of manufacture,
  sustainability, recyclability, and end-of-life disposal.
\item
  Maintenance.
\end{itemize}

\end{framed}

\newpage
{\raggedright
\textbf{Free-body Diagram Sketching Activity:}

The agent can suggest that the designer sketch out load case-relevant forces and constraints as a free-body diagram by sharing a link to a 2D drawing canvas containing the side and top view of the bracket as a starting point:
}

\vspace{4pt}

\begin{figure}[H]
    \makebox[\textwidth]{
        \includegraphics[width=0.9\textwidth]{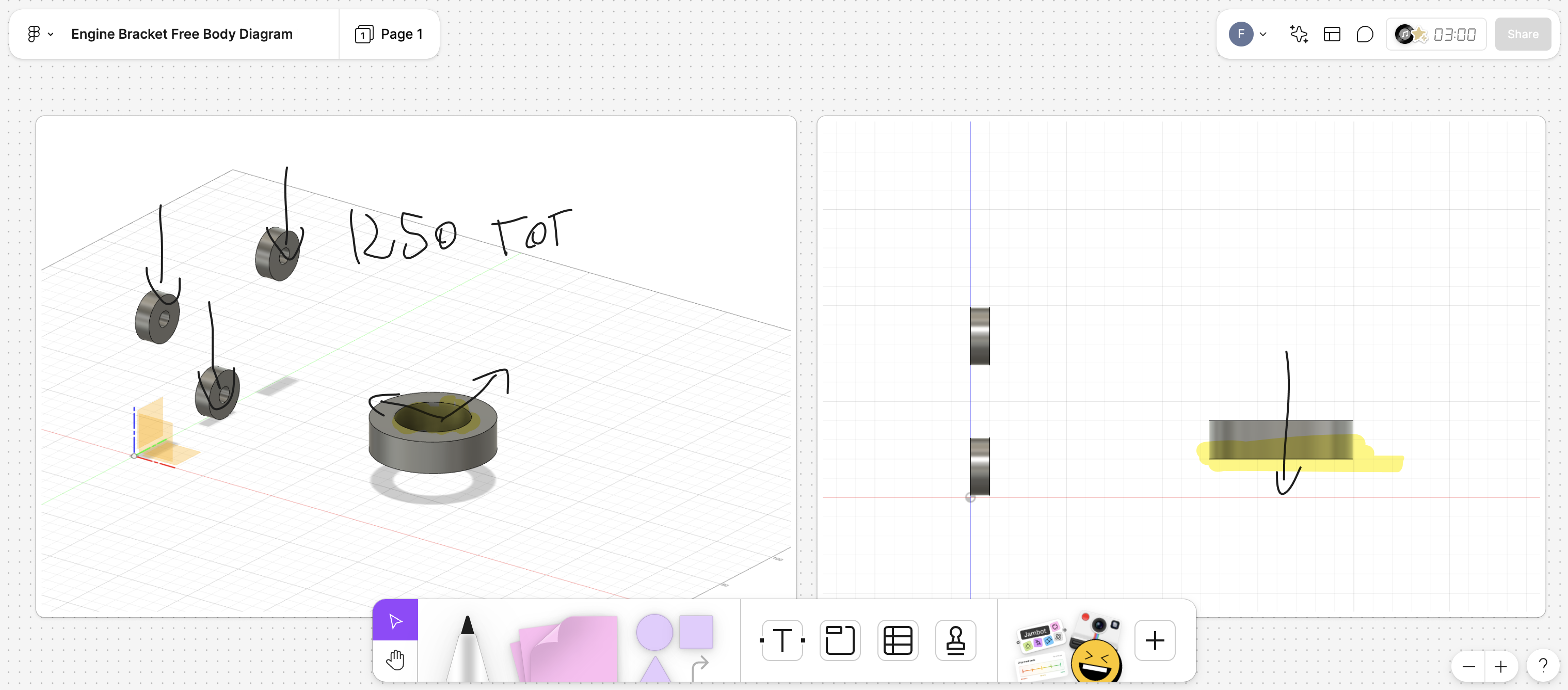}
    }
    \caption{Screenshot of the sketching board  \highhephaistus{\textit{HephAIstus}}  sent to users (with scribbles from H3 on it).  }
    \label{fig:appendix_sketching_board}
    \Description{The figure shows a screenshot of the sketching board interface used by HephAIstus to communicate with a user, including scribbles from participant H3. The left pane displays a 3D visualization of an engine bracket free-body diagram, annotated with arrows indicating force directions and the label “1250 TOT.” The right pane shows a side view of the bracket, with a highlighted area in yellow and an arrow pointing to it, representing an area of interest. The interface includes a toolbar at the bottom, featuring tools for drawing, annotation, and other collaborative features. This visualization demonstrates how HephAIstus used the sketching board to provide interactive feedback and guide the user through the design process.}
    
\end{figure}

\begin{itemize}
\setlength{\leftskip}{1cm}
\setlength{\rightskip}{1cm}
    \ttfamily
  \item[\highhephaistus{\textit{HephAIstus}} :] Can you walk me through your load cases and constraints by sketching a free-body diagram? I shared a link to a board for you to sketch on in the chat (see Fig. \ref{fig:appendix_sketching_board}).
\end{itemize}

\subsubsection{\textbf{Expert-Freeform Agent Introduction}}\hfill

\begin{itemize}
\setlength{\leftskip}{1cm}
\setlength{\rightskip}{1cm}
    \ttfamily
  \item[\highexpert{Expert Agent}:] Hey! I am a voice agent here to support you during the design task. I can hear what you are saying, and I can see your screen and follow along with you while you work on the task. Feel free to ask me any questions, and I'll do my best to provide you with answers. From time to time, I'll also chime in with tips and guidance to help you along the way.
\end{itemize}

\begin{table*}
  \caption{Interview protocol with questions of the semi-structured post-task interview.}
  \includegraphics[width=0.9\linewidth]{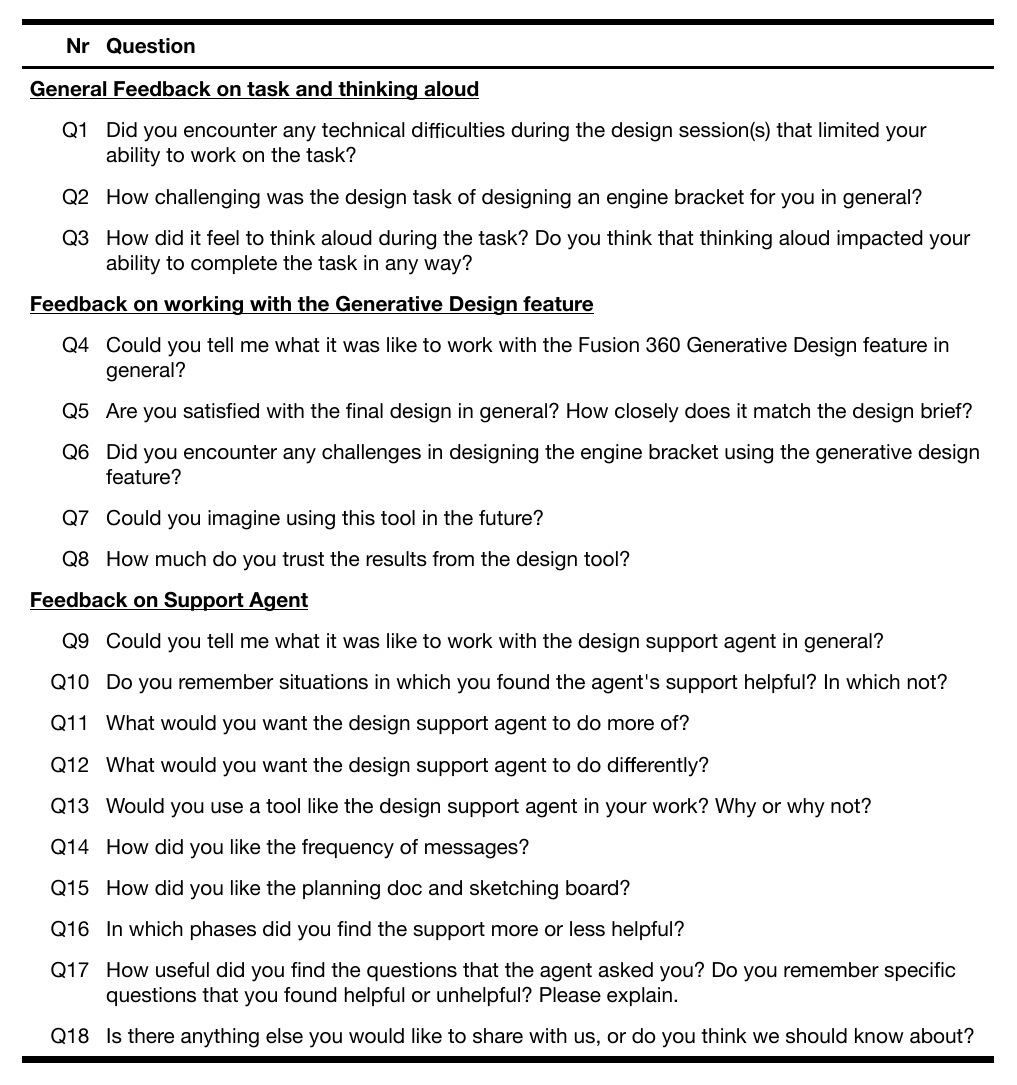}
  \Description{Interview protocol with 18 questions of the semi-structured post-task interview. This table is machine-readable.}
\end{table*}
 
\end{document}